\documentclass[10pt]{feynman}
\usepackage{epsfig}
\usepackage{psfig}
\usepackage{graphics}
\begin{document}

\newcommand{\chapterheader}{CRYSTALLINE COMPUTATION}
\newcommand{\chapterauthor}{NORMAN H. MARGOLUS}

\newcommand{\fref}[1]{Figure~\ref{fig.#1}}
\newcommand{\byby}[0]{$\times$}

\newcommand{\figspace}[3]{
\begin{figure}
    \begin{tt}			 
    \vspace*{#2}		 
    \end{tt}
  \caption{#3}			 
  \label{fig.#1}
\end{figure}}

 \newcommand{\figfig}[3]{
 \begin{figure}
  $$#2$$\relax
  \caption{#3}
 \label{fig.#1}
 \end{figure}                 }

\setlength{\fboxsep}{.1pt}
\setlength{\fboxrule}{.1pt}

\chapter{Crystalline Computation}

\begin{center}
{\bf Norman Margolus}
\begin{picture}(0,0)(0,0)
\put (12,135) {\shortstack[r]{\small
          To appear in {\em Feynman and Computation}\\ 
          (A. {\sc Hey}, ed.),  Perseus Books (1998).} }
\end{picture}

\end{center}

\section*{Abstract}

Discrete lattice systems have had a long and productive history in
physics.  Examples range from exact theoretical models studied in
statistical mechanics to approximate numerical treatments of continuum
models.  There has, however, been relatively little attention paid to
exact lattice models which obey an {\em invertible dynamics}: from any
state of the dynamical system you can infer the previous state.  This
kind of microscopic reversibility is an important property of all
microscopic physical dynamics.  Invertible lattice systems become even
more physically realistic if we impose locality of interaction and
exact conservation laws.  In fact, some invertible and momentum
conserving lattice dynamics---in which discrete particles hop between
neighboring lattice sites at discrete times---accurately reproduce
hydrodynamics in the macroscopic limit.

These kinds of discrete systems not only provide an intriguing
information-dynamics approach to modeling macroscopic physics, but
they may also be supremely practical.  Exactly the same properties
that make these models physically realistic also make them efficiently
realizable.  Algorithms that incorporate constraints such as locality
of interaction and invertibility can be run on microscopic physical
hardware that shares these constraints.  Such hardware can, in
principle, achieve a higher density and rate of computation than any
other kind of computer.

Thus it is interesting to construct discrete lattice dynamics which
are more physics-like both in order to capture more of the richness of
physical dynamics in informational models, and in order to improve our
ability to harness physics for computation.  In this chapter, we
discuss techniques for bringing discrete lattice dynamics closer to
physics, and some of the interesting consequences of doing so.


\section{Introduction}\label{sec.intro}

In 1981, Richard Feynman gave a talk at a conference hosted by the MIT
Information Mechanics Group.  This talk was entitled ``Simulating
Physics with Computers,'' and is reproduced in this volume.

In this talk Feynman asked whether it is possible that, at some
extremely microscopic scale, nature may operate exactly like discrete
computer-logic.  In particular, he discussed whether crystalline
arrays of logic called {\em Cellular Automata} (CA) might be able to
simulate our known laws of physics in a direct fashion.  This question
had been the subject of long and heated debates between him and his
good friend Edward Fredkin (the head of the MIT Group) who has long
maintained that some sort of discrete classical-information model will
eventually replace continuous differential equations as the
mathematical machinery used for describing fundamental physical
dynamics\cite{fredkin-dm,fredkin-nature}.

For classical physics, Feynman could see no fundamental impediment to
a very direct CA simulation.  For quantum physics, he saw serious
difficulties.  In addition to discussing well known issues having to
do with hidden variables and non-separability, Feynman brought up a
new issue: simulation efficiency.  He pointed out that, as far as we
know, the only general way to simulate a lattice of quantum spins on
an ordinary computer takes an exponentially greater number of bits
than the number of spins.  This kind of inefficiency, if unavoidable,
would make it impossible to have a CA simulation of quantum physics in
a very direct manner.

Of course the enormous calculation needed to simulate a spin system on
an ordinary computer gives us the result of not just a single
experiment on the system, but instead approximates the complete
statistical distribution of results for an {\em infinite} number of
repetitions of the experiment.  Feynman made the suggestion that it
might be more efficient to use one quantum system to simulate another.
One could imagine building a new kind of computer, a {\em quantum spin
computer}, that was able to mimic the quantum dynamics of any spin
system using about the same number of spins as the original system.
Each simulation on the quantum computer would then act statistically
like a {\em single} experiment on the original spin system.  This
observation that a quantum computer could do some things easily that
we don't know how to do efficiently classically, stimulated others to
look for and find algorithms for quantum computers that are much
faster than any currently known classical
equivalents\cite{shor,grover}.  In fact, if we restrict our classical
hardware to perform the ``same kind'' of computation as the quantum
hardware---rather than to just solve the same problem---then we can
actually {\em prove} that some quantum computations are faster.  These
fast quantum computations present further challenges to hypothetical
classical-information models of quantum physics\cite{joe-thesis}.

Despite such difficulties, Feynman did not rule out the possibility
that some more subtle approach to the efficient classical
computational modeling 
of physics might yet succeed.  He found something very tantalizing
about the relationship between classical information and quantum
mechanics, and about the fact that in some ways quantum mechanics
seems much {\em more} suited to being economically simulated with bits
than classical mechanics: unlike a continuous classical system, the
entropy of a quantum system is finite.  The informational economy of
quantum systems that Feynman alluded to has of course long been
exploited in statistical mechanics, where classical bits are sometimes
used to provide finite combinatorial models that reproduce some of the
macroscopic {\em equilibrium} properties of quantum
systems\cite{ising,huang}.  It is natural then to ask how much of the
macroscopic {\em dynamical} behavior of physical systems can also be
captured with simple classical information models.  This is an
interesting question even if your objective is not to revolutionize
quantum physics: we can improve our understanding of nature by making
simple discrete models of phenomena.


My own interest in CA modeling of physics stems from exactly this
desire to try to understand nature better by capturing aspects of it
in exact informational models.  This kind of modeling in some ways
resembles numerical computation of differential equation models, where
at each site in a spatial lattice we perform computations that involve
data coming from neighboring lattice sites.  In CA modeling, however,
the conceptual model is not a continuous dynamics which can only be
approximated on a computer, but is instead a finite logical dynamics
that can be simulated {\em exactly} on a digital computer, without
roundoff or truncation errors.  Every CA simulation is an exact
digital integration of the discrete equations of motion, over whatever
length of time is desired.  Conservations can be exact, invertibility
of the dynamics can be exact, and discrete symmetries can be exact.
Continuous behavior, on the other hand, can only emerge in a
large-scale average sense---in the {\em macroscopic limit}.  CA models
have been developed in which realistic classical physics behavior is
recovered in this limit\cite{rothman-book,chopard-book}.

Physics-like CA models are of more than conceptual and pedagogical
interest.  Exactly the same general constraints that we impose on our
CA systems to make them more like physics also make them more
efficiently realizable as physical devices.  CA hardware that matches
the structure and constraints of microscopic physical dynamics can in
principle be more efficient than any other kind of computer: it can
perform more logic operations in less space and less time and with
less energy dissipation\cite{toffoli-nature,frank-thesis}.  It is also
scalable: a crystalline array of processing elements can be
indefinitely extended.  Finally, this kind of uniform computer is
simpler to design, control, build and test than a more randomly
structured machine.  The prospect of efficient large-scale CA hardware
provides a practical impetus for studying CA models.

\section{Modeling dynamics with classical spins}\label{sec.ising}

{From} the point of view of a physicist, a CA model is a fully discrete
classical field theory.  Space is discrete, time is discrete, and the
state at each discrete lattice point has only a finite
number of possible discrete values.  The most essential property of
CA's is that they emulate the spatial locality of physical law: the
state at a given lattice site depends only upon the previous state at
nearby neighboring sites.  You can think of a CA computation as a
regular spacetime crystal of processing events: a regular pattern of
communication and logic events that is repeated in space and in time.
Of course it is only the structure of the computer that is regular,
not the patterns of data that evolve within it! These patterns can
become arbitrarily complicated.

Discrete lattice models have been used in statistical mechanics since
the 1920's\cite{ising,huang}.  In such models, a finite set of
distinct quantum states is replaced by a finite set of distinct
classical states.  Consider, for example, a hypothetical quantum
system consisting of $n$ spin-${1 \over 2}$ particles arranged on a
lattice, interacting locally.  The spin behavior of such a system can
be fully described in terms of $2^n$ distinct (mutually orthogonal)
quantum states.  The {\em Ising model} accurately reproduces essential
aspects of phase-change behavior in such a system using $n$ classical
bits---which give us $2^n$ distinct classical states.  In the Ising
model, at each site in our lattice we put a {\em classical spin:} a
particle that can be in one of two classical states.  We define bond
energies between neighboring spins: we might say, for example, that
two adjacent spins that are parallel (i.e., are in the same state)
have a bond energy of $\epsilon_=$, while two antiparallel (not same)
neighbors have energy $\epsilon_{\neq}$.  This gives us a classical
system which has many possible states, each of which has an energy
associated with it.  In calculating the equilibrium properties of this
system, we simply assume that the dynamics is complicated enough that
all states with the same energy as we started with will appear with
equal probability.  Thus we ignore the actual quantum dynamics of the
original spin system, and instead substitute an energy-conserving
random pseudo-dynamics.


We could equally well substitute any classical dynamics that has a
sufficiently complicated evolution.  We will consider one simple CA
model that has been successfully used in this
manner\cite{vichniac,pomeau-invariant,herrmann,creutz}.  We assume
that we are dealing with an isolated spin system, not in contact with
any heat bath, so that total energy must be exactly conserved.  We
will also impose the realistic constraint that the microscopic
dynamics of an isolated physical system must be exactly invertible:
there must always be enough information in the current state to
recover any previous state.  This constraint helps to ensure that a
deterministic dynamics explores its available state-space thoroughly,
and so can be analyzed statistically---this issue is discussed in
Section~\ref{sec.rev}.

We can construct a simple CA that has these properties, in which the
next value of the spin at each site on a 2D square lattice only
depends upon the current values of its four nearest neighbors.  The
rule is very simple: a given spin changes state if and only if this
doesn't change the total energy associated with its bonds to its four
nearest neighbors.  Equivalently, a given spin (bit) is flipped
(complemented) if exactly two of its four neighbors are zero's, and
two are one's.  This doesn't change its total bond energy: both before
and after the flip, it will be parallel to half of its neighbors
(contributing $2\epsilon_=$ to the total), and antiparallel to
the rest (contributing $2\epsilon_{\neq}$).

The rule as stated above would be fine if we updated just one spin on
the lattice at a time, but we would like to update the lattice in parallel.
To make this work, we will adopt a checkerboard updating scheme: we
imagine that our lattice is a giant black and white checkerboard, and
we alternately hold the bits at all of the black sites fixed while we
update all of the white ones, and then hold the white sublattice fixed
while updating the black.  In this way, the neighbors of a spin that
is changed are not also simultaneously changed, and so our logic about
conserving energy remains valid.

Now how can we add invertibility?  We already have!  If we apply our
rule to the same checkerboard sublattice twice in a row, then each
spin is either flipped twice or not at all---the net effect is no
change.  Thus the most recent step in our time evolution can always be
undone simply by applying the rule a second time to the appropriate
sublattice, and we can recover any earlier state by undoing enough
steps.

This example demonstrates that we can simultaneously capture several
basic aspects of physics in an exact digital model.  First of all, the
finite-state character of a quantum spin system is captured by using
classical bits.  Next, spatial locality is captured by making the rule
at each lattice site depend only on nearby neighbors.  Finally, both
energy conservation and invertibility are captured by splitting the
updating process into two phases, and alternately looking at one half
of the bits while changing the other half.  By making only changes
that conserve a bond energy locally, we conserve energy globally.  By
making the change at each site be a permutation operation that depends
only upon unchanged neighbor information, we can always go backwards
by taking the same neighbor information and performing the inverse
permutation.

\figfig{creutz}{ \hfill \fbox{%
\epsfig{figure=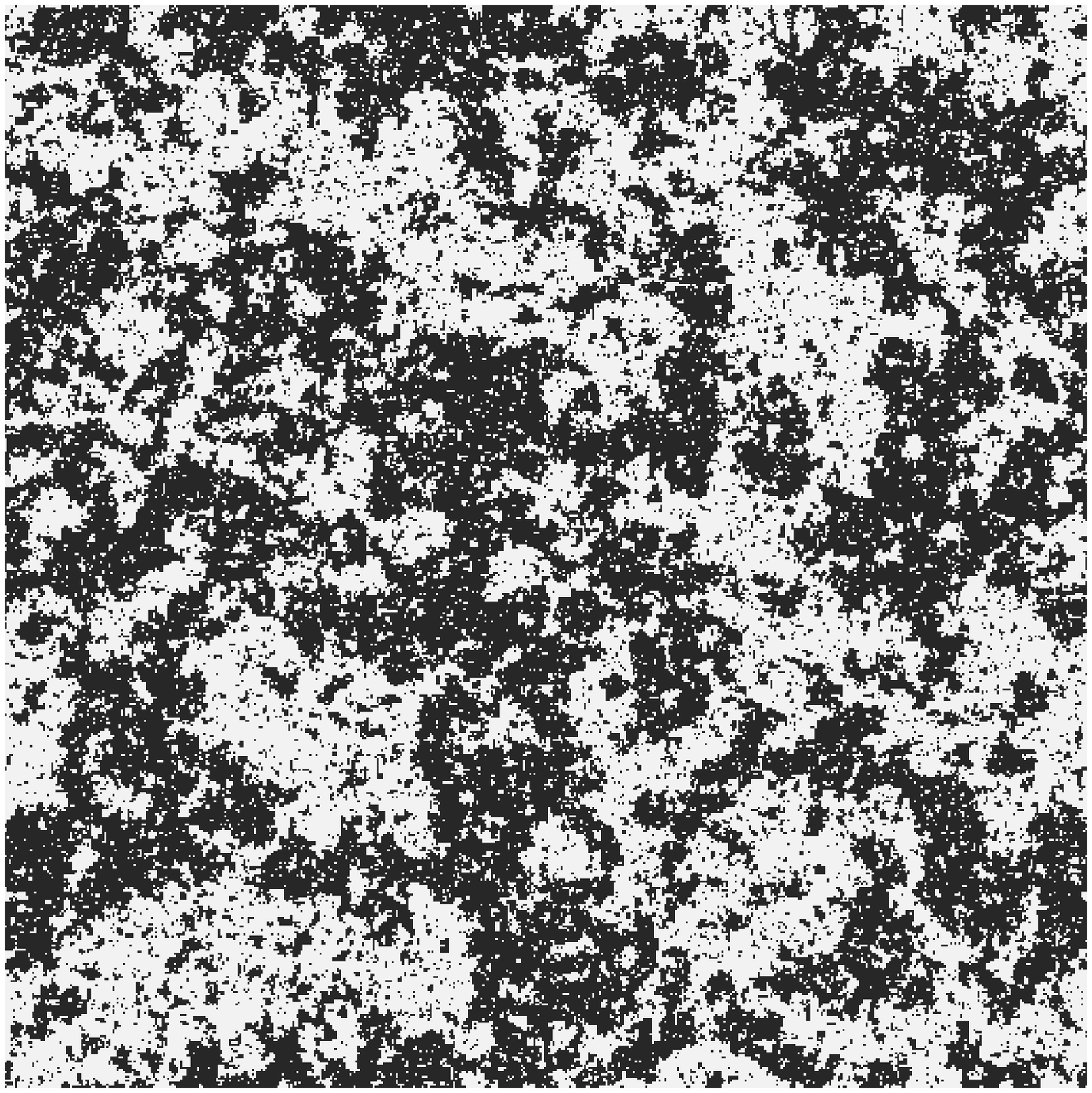,height=1.5in}}\hskip .2in \fbox{%
\epsfig{figure=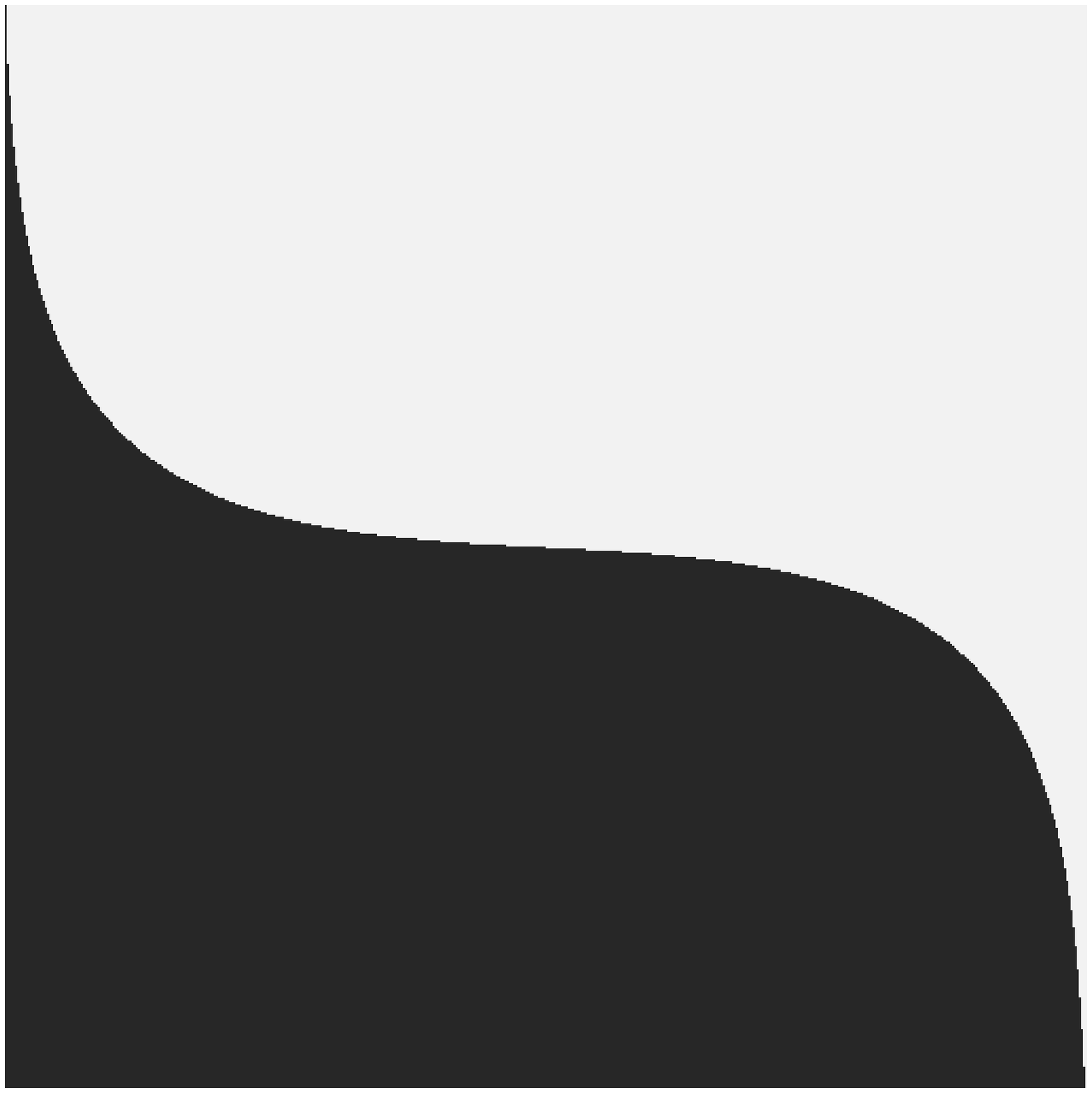,height=1.5in}}\hskip .2in \fbox{%
\epsfig{figure=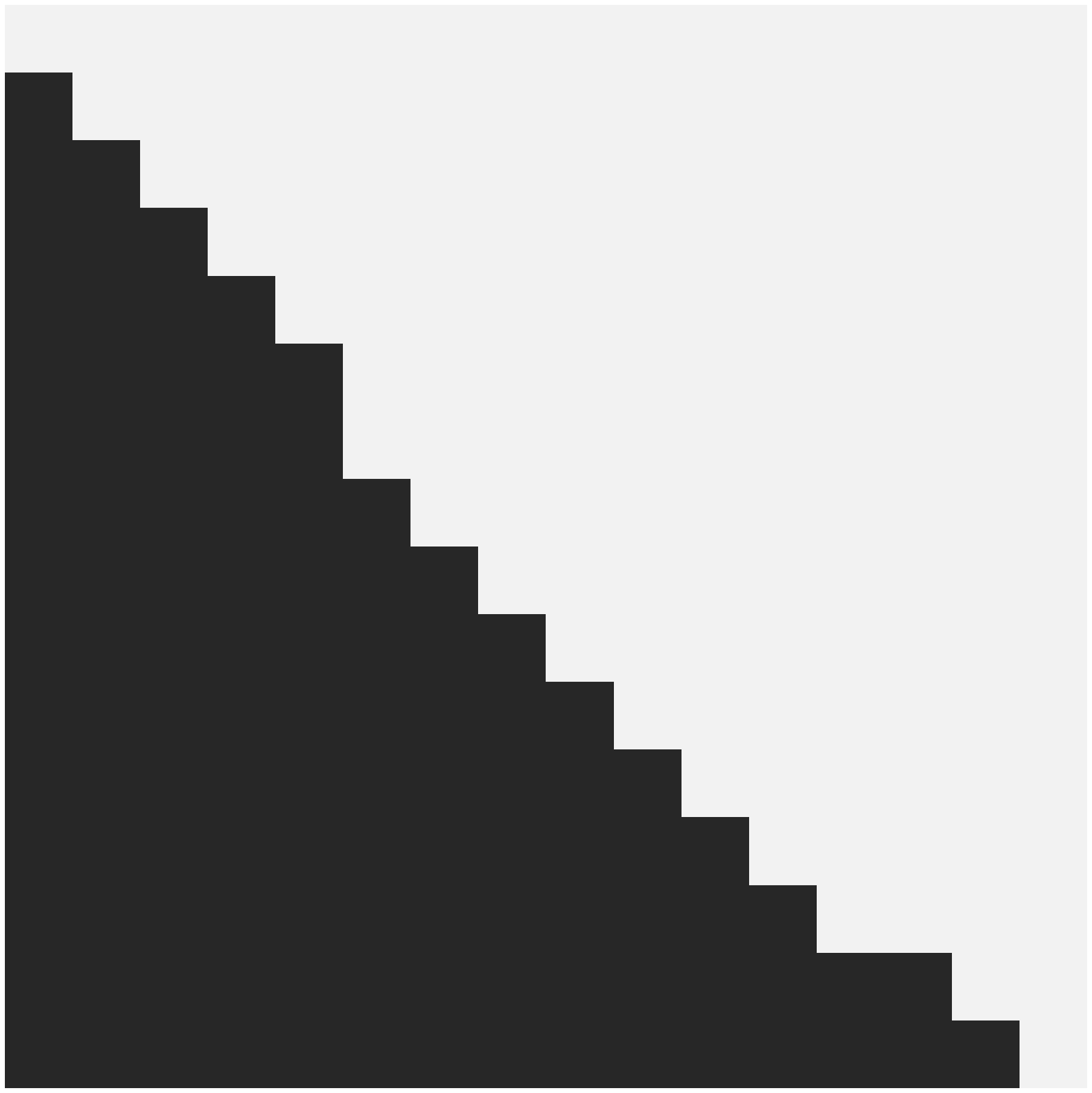,height=1.5in}}\hfill}%
 {An Ising CA
(a)~A state that evolved from a random pattern of 90\% 0's and 10\% 1's.
(b)~Wave on the boundary between a domain of all 1's, and a domain of
all 0's.  (c)~Closeup of a portion of the boundary.}

Figure~\ref{fig.creutz}a shows the state of the Ising CA after 100,000
steps of time evolution on a 512\byby512 lattice, when started from a
randomly generated configuration of site values that consisted of 10\%
1's and 90\% 0's.  This is an {\em equilibrium} configuration: if we
compare this to the configuration after 100,000,000 steps, the picture
looks qualitatively the same.  This equilibrium configuration is
divided about equally between large domains that are mostly 0's, and
large domains that are mostly 1's.  Since bond energy is conserved,
the total length of boundary between regions of 0's and regions of 1's
must be unchanged from that in the initial configuration---the numbers
of 0's and 1's are not themselves conserved.

This CA has some surprising behavior when started from a more ordered
initial state: it supports the continuum wave equation in an exact
fashion.  Figure~\ref{fig.creutz}b illustrates a wave on the boundary
between two {\em pure} domains (all 0's, or all 1's).  If we hold the
values at the edges of the lattice fixed, then we find that the boundary
shown behaves like a standing wave, oscillating in a harmonic fashion
that repeats forever without any damping.  In fact, it is easy to show
that any waveform that we set up along this diagonal boundary---as
long as it isn't too steep---exactly obeys the wave equation ({\em
cf.}  \cite{hrgovcic}).  To see this, notice
(Figure~\ref{fig.creutz}c) that the boundary {\em between} the two
domains consists of a sequence of vertical and horizontal
line-segments each the height or width of one site.  If we number
these segments sequentially along the boundary, then it is easy to
verify that, at each update of the lattice, all of the even-numbered
segments move one position along the boundary in one direction, while
all of the odd-numbered segments move one position in the opposite
direction.  Thus the shape of the boundary is exactly the
superposition of two discrete waveforms moving in opposite directions.

\figfig{other-ising}{ \hfill \hbox{%
\epsfig{figure=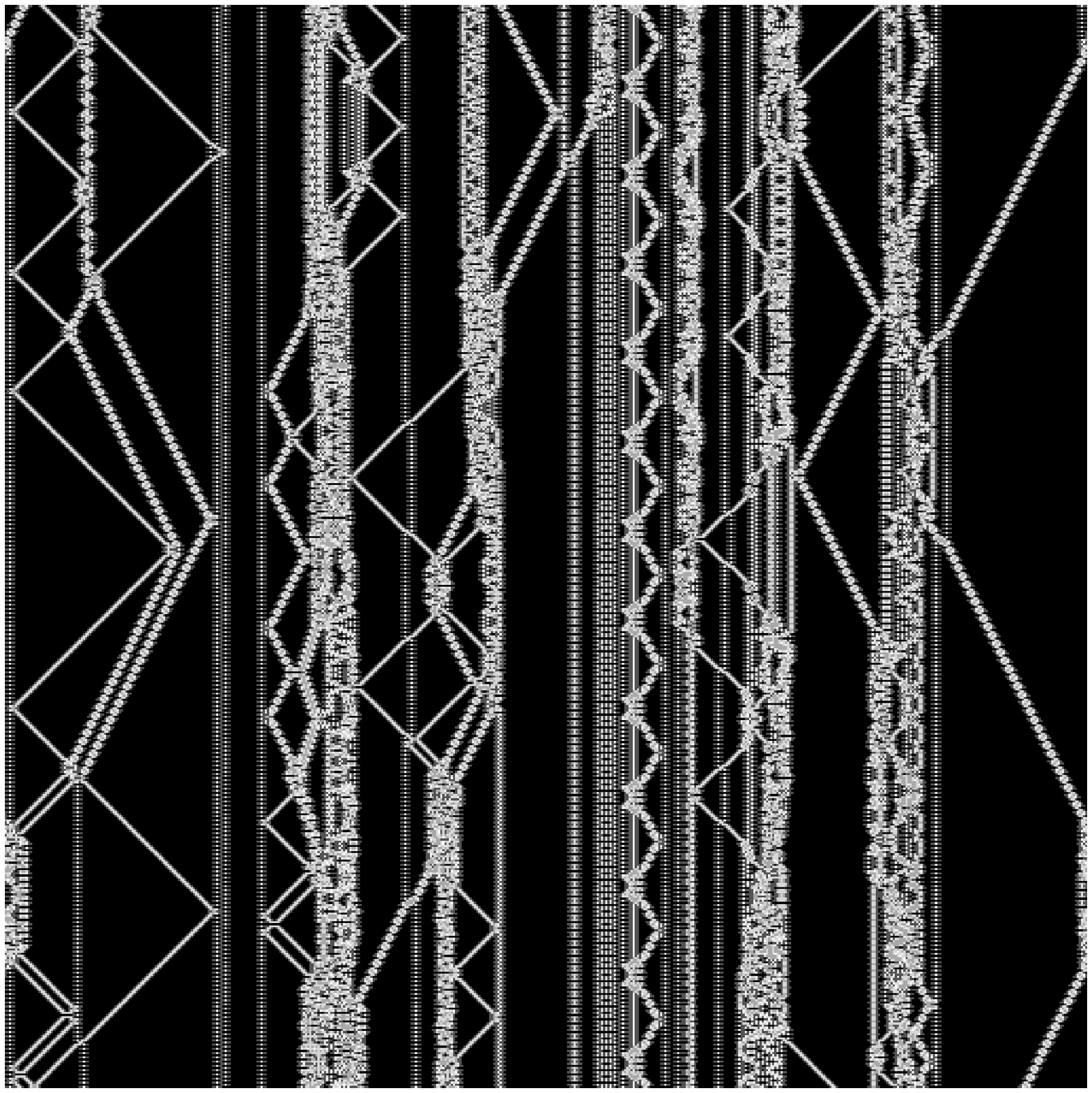,height=2in}}\hskip .6in \hbox{%
\epsfig{figure=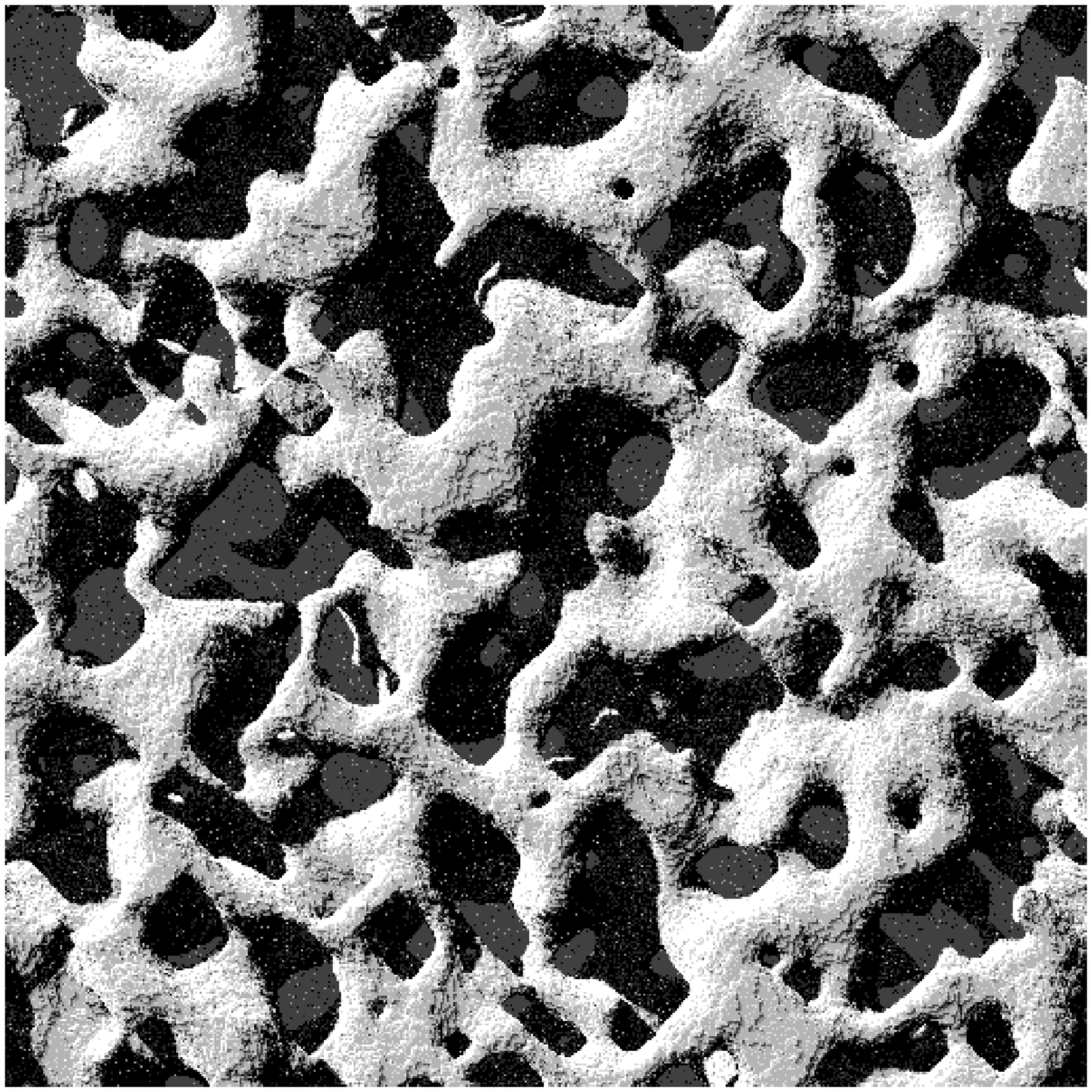,height=2in}}\hfill}%
{Ising-like CA's in 1D and 3D.  (a)~Time history of Bennett's 1D CA.
(b)~A 3D Ising CA cooled with a heatbath.} 

Similar techniques to those used in the Ising CA give a variety of
related CA models\cite{cambook,margolus-thesis,creutz,raissa}.  For
example, in Figure~\ref{fig.other-ising}a we show the time-history of
a 1D rule invented by Charles Bennett that has exactly the same
bond-energy conservation that we've just seen\cite{pomeau-invariant}.
In Bennett's CA, instead of 1-bit at each site we put 2-bits, which
we'll call $A_i$ and $B_i$.  The $A$'s and $B$'s will play the roles
of the two sublattices in the Ising CA.  We first update all of the
$A$'s in parallel, holding the $B$'s fixed, and then vice versa.  For
each $A_i$, it's neighbors along the 1D chain will be the two $B$'s on
either side of it: $B_{i-2}$, $B_{i-1}$, $B_{i+1}$ and $B_{i+2}$.  Our
rule is the same as before: we complement an $A_i$ if exactly half of
its four neighbors are 1's, and half are 0's.  Once we have updated
all of the $A$'s, then we update the $B$'s in the same manner, using
the $A$'s as neighbors.  If we consider that there is a bond between
each ``spin'' and its four neighbors, then we are again flipping the
spin only if it doesn't change the total bond energy.  If we update
the same sublattice twice in a row, the net effect is no change: the
rule is invertible, exactly like the Ising CA.

In the figure, our 1D lattice is 512 sites wide, with periodic
boundaries (joined at the edges).  We started the system with all sites
empty except for a patch of randomly set bits in sites near the
center.  Time advances upward in the figure, and we show a segment of
the evolution after about 100,000 steps.  Rather than show the domains
directly, we show all bonds that join antiparallel spins---the number
of such ``domain boundary'' bonds is not changed by the dynamics.
Note the variety of ``particle'' sizes and speeds.

In Figure~\ref{fig.other-ising}b, we show a 3D Ising dynamics with a
heat bath.  Here the rule is an invertible 3D checkerboard Ising CA
similar to our 2D version, except that at every site in our 3D lattice
we have added a few extra {\em heatbath} bits.  The heatbath bits at
each site record a binary number that is interpreted as an energy.
Now our invertible rule is again ``flip whenever it is energetically
allowed.''  As long as the heatbath energy at a given site is not too
near its maximum, then a spin flip that would lower the bond energy is
allowed, because we can put the energy difference into the heatbath.
Similarly with transitions that would raise the bond energy.  This
heatbath-CA technique is due to Michael Creutz\cite{creutz}.  He
thought of the bond energy as being potential energy, and the heatbath
energy as being the associated kinetic energy.  This heatbath CA is
perfectly invertible, since applying the dynamics twice to the same
sublattice leaves both the spins and the heatbath unchanged.

By adjusting the energy in the heatbath portion of this 3D CA, we can
directly control the {\em temperature} of our system.  We simply stop
the simulation for a moment while we reach into our system and reset
the heatbath values---without changing the spin values.  As we {\em
cool} the system in this way, energy will be extracted from bonds, and
so if (for example) $\epsilon_{\neq} > \epsilon_=$, then
there will be fewer domain boundaries---the domains will grow
larger.  The system shown has been cooled in this manner, and we
render the interface between the up and down spins.

\figfig{same}{ \hfill \fbox{%
\epsfig{figure=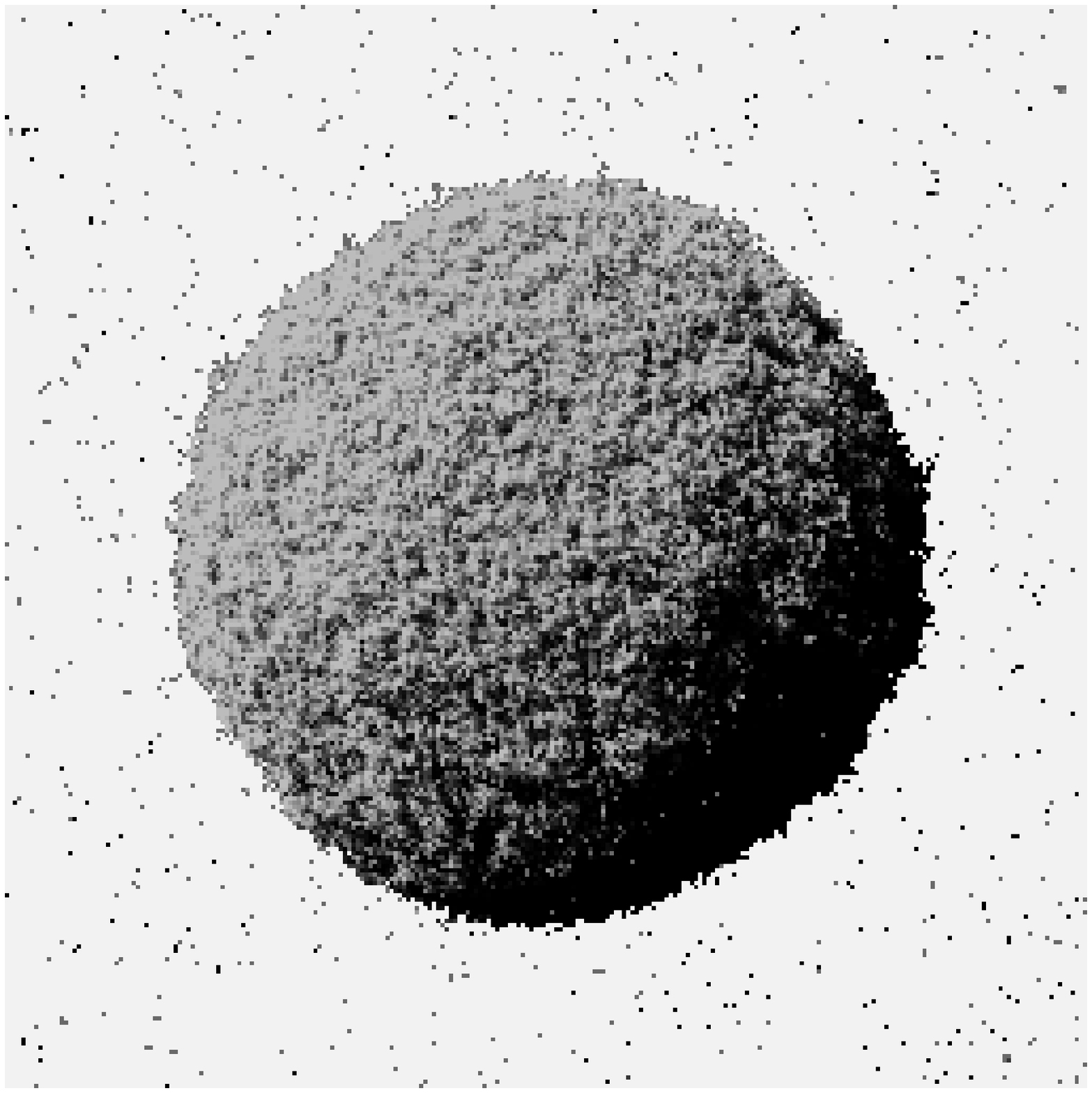,height=2in}}\hskip .6in \fbox{%
\epsfig{figure=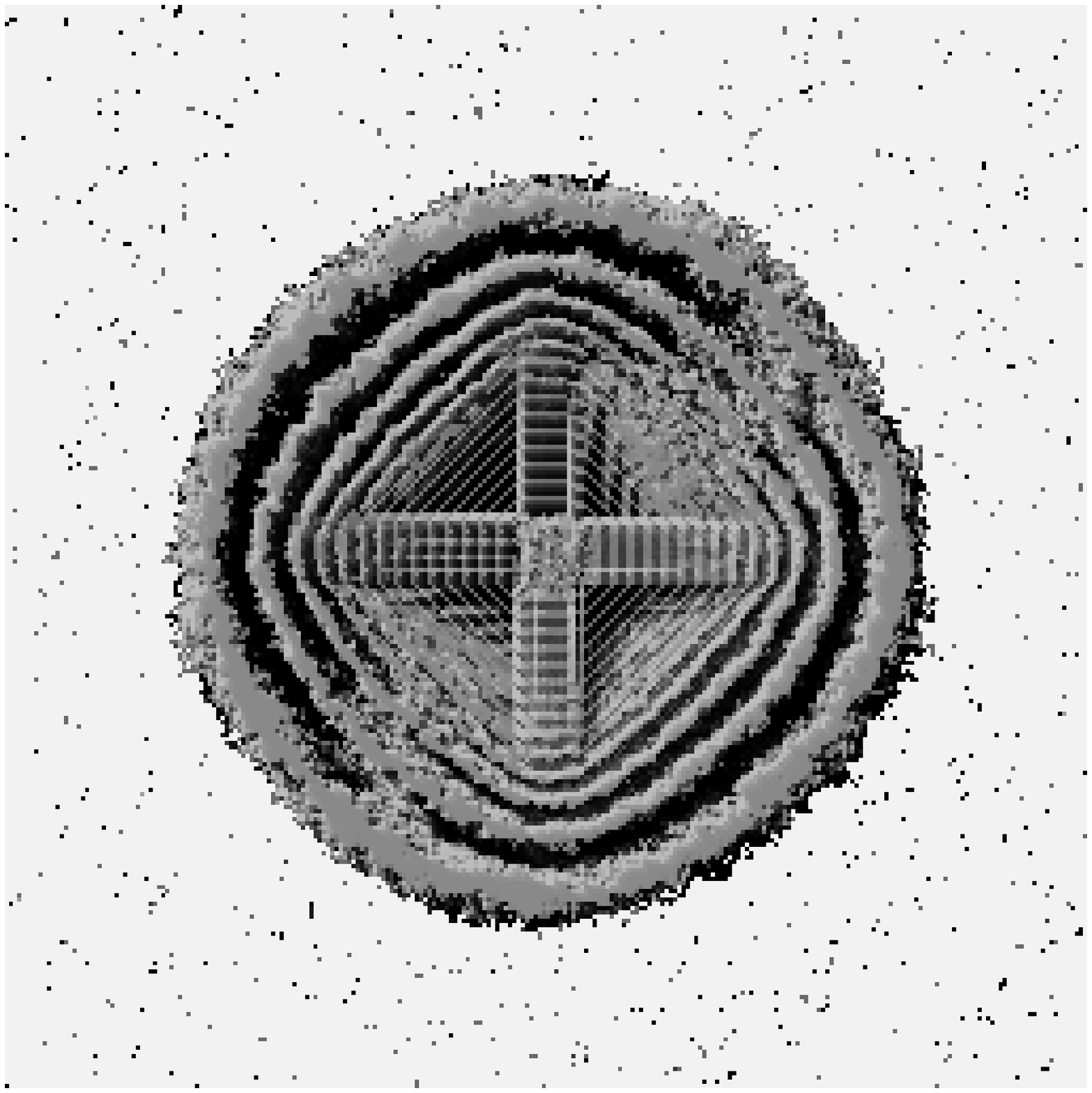,height=2in}}\hfill}%
{An Ising-like 3D CA.  (a)~A macroscopic equilibrium configuration.
(b)~The same configuration, with the front half of the ball removed.}

Figure~\ref{fig.same} shows another Ising-like CA defined on a 3D
cubic lattice.  As in our 2D Ising CA, we have only one bit of state
at each lattice site.  Each site has a bond with each of its six
nearest neighbors, and we perform a 3D checkerboard updating.  This
time our rule is, ``flip a given spin if its six neighbors all have
the same value: six 0's or six 1's.''  We'll call this the ``Same''
rule.  If we label half of the bonds attached to each site as
``antiferromagnetic'' (i.e., the energy values associated with
parallel and antiparallel spins are interchanged for these labeled
bonds), then this rule again conserves the total bond energy.  Notice,
though, that there are many different ways of labeling half of the
bonds, and each way corresponds to a different additively conserved
energy.  We need to use several of these energies simultaneously if we
want to express the Same rule as ``flip whenever permitted by energy
conservation.''

The system in Figure~\ref{fig.same}a is 512$\times$512$\times$512 and
was started from an empty space (all 0's) with a small random block of
spin values in the center.  After about 5000 steps of time evolution,
this invertible system settles into the ball shown, which then doesn't
change further macroscopically.  Microscopically it must keep
changing---otherwise if we ran the system backwards it couldn't tell
when to start changing again and {\em unform} the ball.  The local
density of 1's defines the surface that is being rendered.  In
Figure~\ref{fig.same}b we remove the front half of the ball to show
its interior structure.  The analogous rule in 2D does not form stable
``balls.''


It is easy to define other energy-conserving invertible Ising-like
CA's.  We could, for example, take any model that has a bond energy
defined, find a sublattice of sites that aren't directly connected to
each other by bonds, and update those sites in an invertible and
energy conserving manner, holding their neighbors fixed.  By running
through a sequence of such sublattices, we would eventually update all
sites.  We could also make CA models with the same energy
conservations with just two sublattices, by using the technique
illustrated in Bennett's CA.  Simply duplicate the state at each site
in the original model, calling one copy $A_i$, and the other $B_i$.
If the $A$'s are only bonded to the $B$'s and vice versa, then we can
update half of our system in parallel, while holding all neighbors
that they depend on fixed.  Of course we can construct additional
invertible energy-conserving rules by taking any of these examples and
forbidding some changes that are energetically allowed.

\section{Simple CA's with arbitrarily complex behavior}\label{sec.complex}

When the Ising model was first conceived in the 1920's, it was not
thought of as a computer model: there were no electronic computers
yet!  It was only decades later that it and other discrete lattice
models could begin to be investigated on computers.  One of the first
to think about such models was John von Neumann\cite{burks,ulam}.  He
was particularly interested in using computer ideas to construct a
mechanical model that would capture certain aspects of biology that
are essential for reproduction and evolution.  What he constructed was
a discrete world in which one could arrange patterns of signals that
act much like the logic circuitry in a computer.  Just as computer programs can
be arbitrarily complex, so too could the animated patterns in his CA
world.  Digital ``creatures'' in his digital universe reproduced
themselves by following a digital program.  This work anticipated the
discovery that biological life also uses a digital program (DNA) in
order to reproduce itself.

As we will see below, the level of complexity needed in a CA rule in
order to simulate arbitrary patterns of logic and hence {\em universal
computation} is quite low.  In physics, this same possibility of
building arbitrarily complicated mechanisms out of a fixed set of
components seems to be an essential property of the evolution that
built us.  Is this the {\em only} essential property of evolution?

In a paradoxical sense, computation universality gives us so much that
it really gives us very little.  Once we have computation
universality, we have a CA that is just as powerful as any
conventional computer!  By being able to simulate the logic circuitry
of any computer, given enough time and space, any universal CA can
compute exactly the same things as any ordinary computer.  It can play
chess. It can simulate quantum mechanics.  It can even perform a
simulation of a Pentium Processor running Tom Ray's Tierra
evolutionary-simulation program\cite{ray1}, and thus we know that it
is capable of exhibiting Darwinian evolution.  But if we don't put in
such an unlikely initial state by hand, is evolution of interesting
complexity something that we are ever likely to see?  Is it a {\em
robust} property of the dynamics?

Nature has computation universality along with locality, exact
conservations and many other constraints.  Which of these constraints
are important for promoting evolution, and whether it is possible to
capture all of the important constraints simultaneously in a CA model,
are both interesting questions.  Here we will examine a well-known CA
rule that is universal, and then discuss some physical constraints that
it lacks that might make it a better candidate as a model for
Darwinian evolution.

\figfig{life}{ \hfill \fbox{%
\epsfig{figure=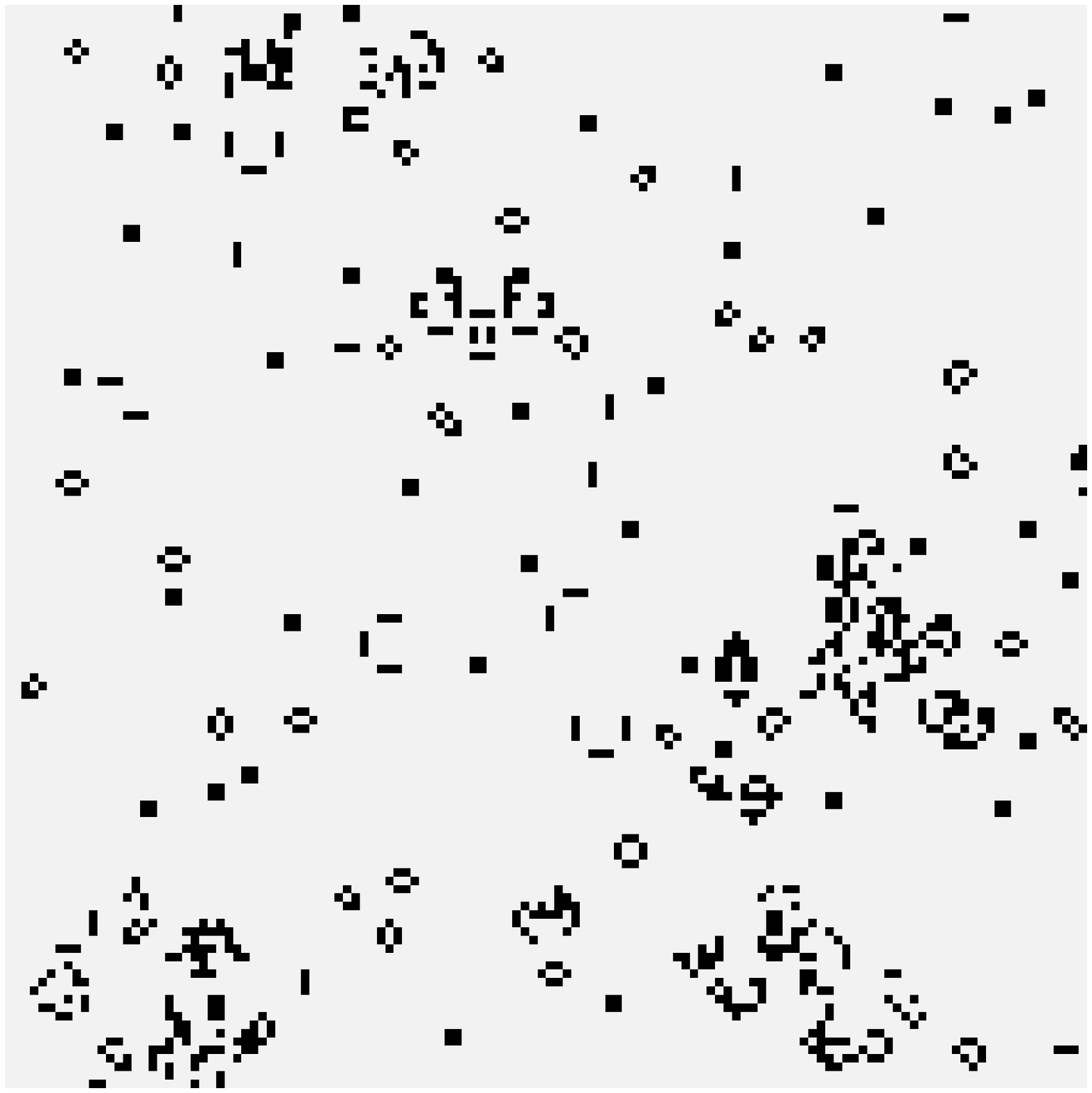,height=1.5in}}\hskip .2in \fbox{%
\epsfig{figure=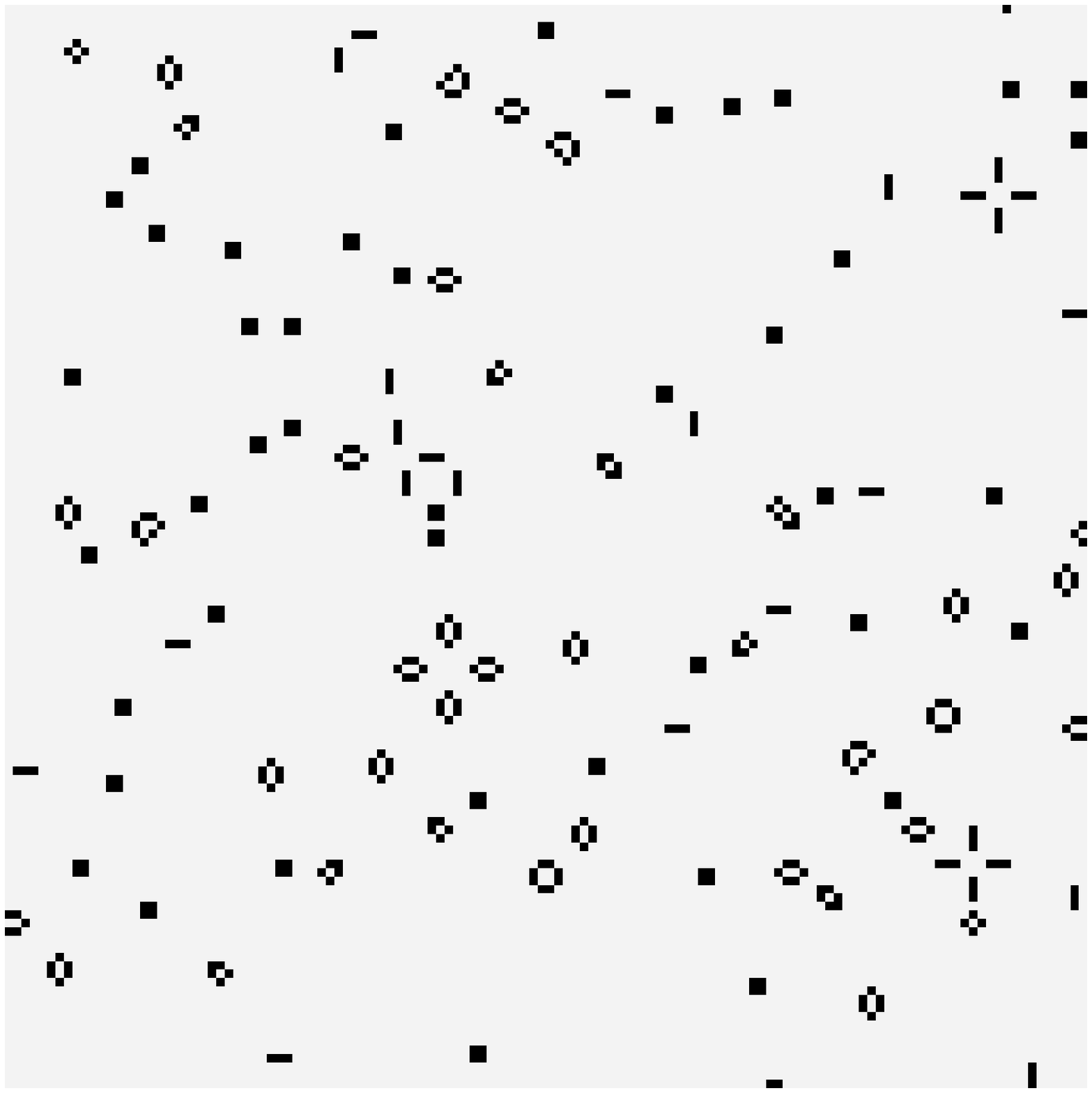,height=1.5in}}\hskip .2in \fbox{%
\epsfig{figure=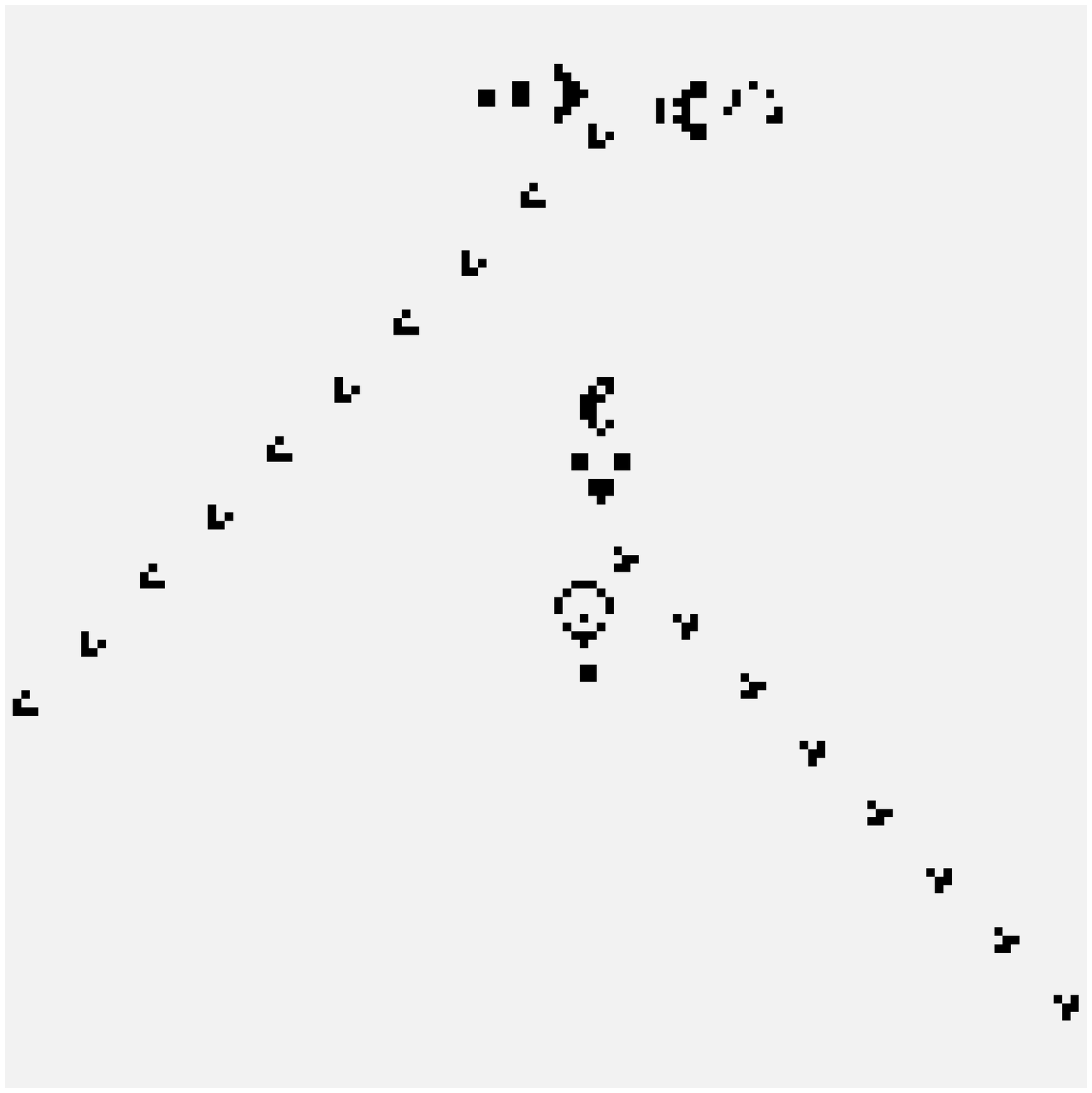,height=1.5in}}\hfill}%
{Conway's non-invertible ``Game of Life'' CA (128\byby128 closeups taken
{from} a 2K$\times$2K space). 
(a)~One thousand steps into an evolution started from a random
configuration of 1's and 0's. (b)~The same region after 16 thousand
steps---the evolution has settled down into small uncoupled repeating
patterns.  (c)~A configuration that started with two ``glider guns.''}

In Figure~\ref{fig.life} we illustrate Conway's {\em Game of Life},
probably the most widely known CA rule\cite{life}.  This is a CA
that involves a 2D square lattice with one bit at each lattice site,
and a rule for updating each site that depends on the total number of
1's present in its eight nearest neighboring sites.  If the total of
its neighbors is 3, a given site becomes a 1, if the total is 2, the
site remains unchanged; in all other cases the site becomes a 0.  This
rule is applied to all sites simultaneously.

The Life rule is clearly non-invertible since, for example, an
isolated 1 surrounded by 0's turns into a 0: you cannot then tell from
the resulting configuration of site values whether that site was a 0
or a 1 in the previous configuration.

If you fill your computer screen with a random pattern of 0 and 1
pixels, and run the Life dynamics on it at video rates, then you see a
lively churning and boiling pattern of activity, dying down in places
to a scattering of small-period oscillating structures, and then being
reignited from adjacent areas (Figure~\ref{fig.life}a).  If you speed
up your simulation to a few hundred frames per second, then typically
after less than a minute for a 2K$\times$2K system all of the
interesting activity has died out, and the pattern has settled down
into a set of isolated small-period oscillators
(Figure~\ref{fig.life}b).

If you watch the initial activity closely, however, and pick out some
of the interesting dynamical structures that arise, you can ``build''
configurations containing constructs such as the ones in
Figure~\ref{fig.life}c.  These are called {\em glider guns}.  When the
Life dynamics is applied to a glider gun, at regular intervals the gun
spits out a small pattern that then goes through a short cycle of
shapes, with the same shape reappearing every few steps in a shifted
position.  These {\em gliders} are the smallest moving objects in the
Life universe.  By putting together such constructs, one can show how
to build arbitrary logic circuits, using sequences of gliders as the
signals that travel around and interact\cite{life,lwod}.

This then is our first example of a universal CA rule.  Many other
non-invertible universal CA rules are known---the simplest is due to
Roger Banks\cite{cambook}.  All of these can support arbitrary
complexity if you rig up a special enough initial state.  Life is
notable because it spontaneously develops interesting complexity
starting from {\em most} initial states.  Small structures that do
something recognizable occasionally appear briefly, before being
sucked back into the digital froth.

\section{Invertible CA's are more interesting}\label{sec.rev}

One problem with Conway's Life as a model of evolution is that it
lasts for such a short time when started from generic initial
conditions.  For a space of 2K$\times$2K bits, there are
$2^{4,194,304}$ distinct possible configurations, and this rule
typically goes through fewer than $2^{14}$ of them before repeating a
configuration and entering a cycle.  This doesn't allow much time for
the evolution of complexity!  Furthermore, useful computing structures
in Life are very fragile: gliders typically vanish as soon as they
touch anything.

The short Life-time problem can be attributed largely to the
non-invertible nature of the Life rule---invertible rules do not
behave like this.  We typically have no idea just how long the cycle
times of our invertible CA's actually are, because we have never seen
them cycle, except from very special initial states or on very tiny
spaces.  The reason that invertible CA's have such long cycle-times is
actually the same as the reason that essentially all invertible
information dynamics have long cycles: {\em an invertible dynamics
cannot repeat any state it has gone through until it first repeats the
state it started in.}  In other words, if we run an invertible rule
for a while, we know what the unique predecessor of every state we
have already seen is, except for the first state---its predecessor is
still coming up!  Thus an invertible system is forced to keep sampling
distinct states until it stumbles onto its initial state.  Since there
is nothing forcing it toward that state as it explores its state
space, the cycle of distinct states is typically enormously long: if
our invertible CA really did sample states at random without
repetition, it would typically have to go through about half of all
possible states before chancing upon its initial
state\cite{toffoli-four}.  A non-invertible system doesn't have this
constraint, and can re-enter its past trajectory at any
point\cite{kauffman}.  The moral here is that if you want to make a
discrete world that lasts long enough to do interesting things, it is
a good idea to make it invertible.  As a bonus, a more thorough
exploration of the available state-space tends to make a system more
amenable to statistical mechanical analysis.

To make it easier to capture physical properties in CA's, we will use
a technique called {\em partitioning}, which was developed
specifically for this purpose\cite{margolus-bbm,cambook}.  This
technique is closely related to the sublattice technique introduced in
Section~\ref{sec.ising}\cite{ica,margolus-thesis}.  The idea of
partitioning is to divide up all of the bits in our CA system into
disjoint local groupings---each bit is part of only one group.  Then
we update all of the groups independently, before changing the
groupings and repeating the process---changing the groupings allows
information to propagate between groups.  If the updating of each
group conserves the number of 1's, for example, then so does the
global dynamics.  If the updating applied independently to each group
is an invertible function, then the global dynamics is also
invertible.  Since all invertible CA's can be
reexpressed isomorphically in a partitioning format---where
conservations and invertibility are manifest---this is a particularly
convenient format to use for our
models\cite{ica,kari-uncomputable,kari-partitioning}.

\figfig{critters-rule}{ \hfill \hbox{%
\epsfig{figure=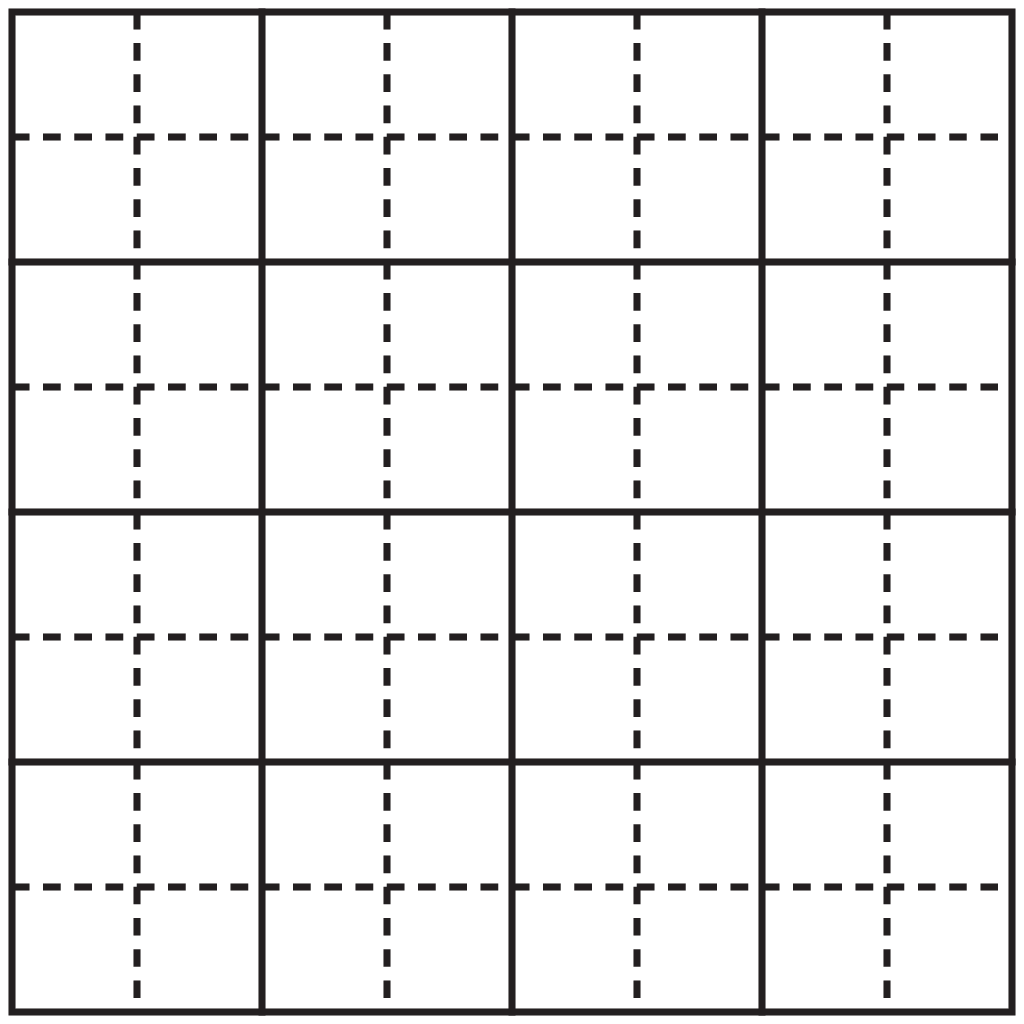,height=2in}}\hskip 1in \hbox{%
\epsfig{figure=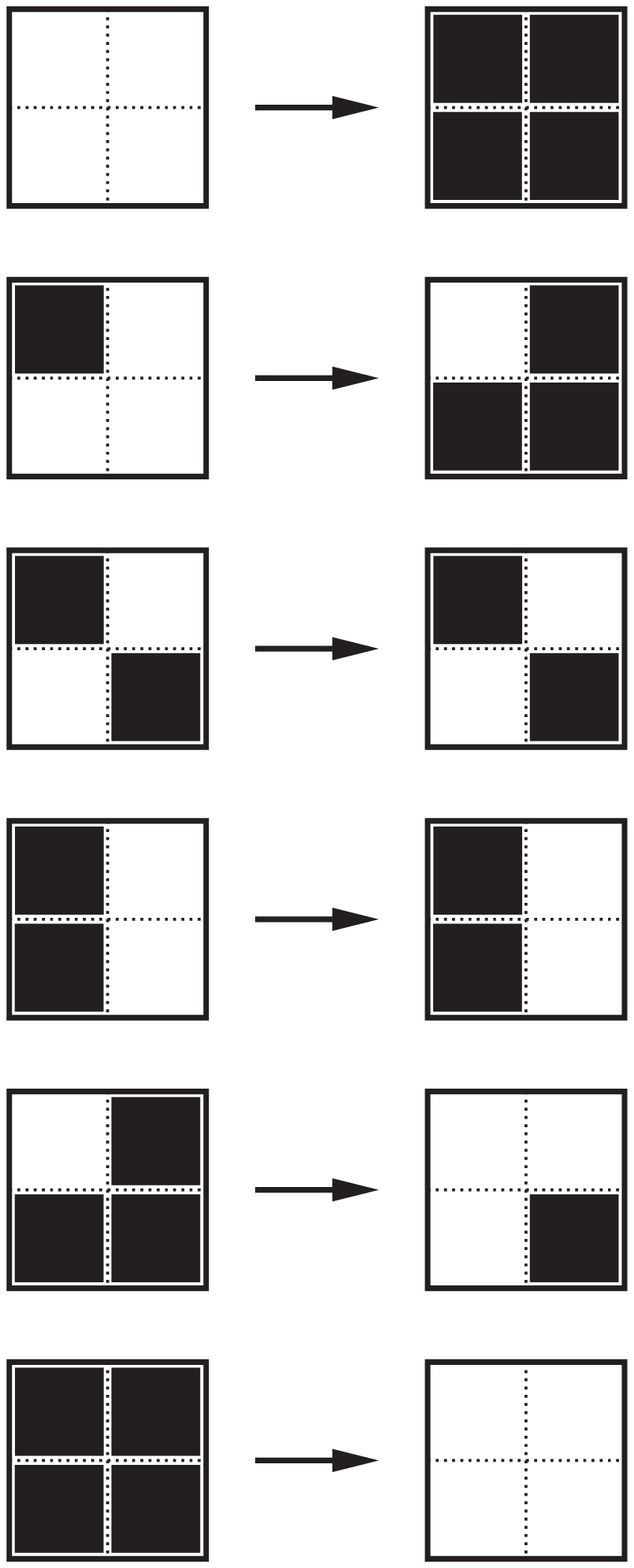,height=2in}}\hfill}%
{The invertible ``Critters'' CA.  (a)~The
solid and dotted blockings are used alternately.  (b)~The Critters rule.}

Our first example of a partitioned CA is called ``Critters.''  This is
a universal invertible CA that evolves interesting complexity.  The
Critters rule uses a 2$\times$2 block partition on a 2D square
lattice.  In Figure~\ref{fig.critters-rule}a we show an 8$\times$8
region of the lattice---each square represents a lattice site that can
hold a 0 or a 1.  The solid lines show the grouping of the bits into
2$\times$2 blocks that is used on the even time-steps, the dotted
lines show the odd-time grouping.  The Critters rule is shown in
Figure~\ref{fig.critters-rule}b.  This same rule is used for both the
even-time grouping and the odd-time grouping.  All possible sets of
initial values for the four bits in a 2$\times$2 block are shown on
the left, the corresponding results are shown on the right.  The rule
is rotationally symmetric, so not all cases need to be shown
explicitly: each of the four discrete rotations of a block that is
shown on the left turns into the same rotation of the corresponding
result-block shown on the right.

Notice that each of the 16 possible initial states of a block is
turned into a distinct result state.  Thus the Critters rule is
invertible.  Notice also that the number of 1's in the initial state
of each block is, in all cases, equal to the number of 0's in the
result.  Thus this property is true for each update of the entire
lattice.  If we call 1's {\em particles} on even steps, and call 0's
{\em particles} on odd steps, then particles are conserved by this
dynamics.  Notice that the Critters rule also conserves the parity
(sum mod 2) along each diagonal of each block, which leads to
conservation of parity along every second diagonal line running across
the entire space.

It is not interesting to run an invertible rule such as Critters
starting from a completely random initial state, as we did in the case
of Life.  This is because the vast majority of all possible states are
random-looking and so, by a simple counting argument, almost all of
them have to turn into other random-looking states.  To see this, note
that any given number of steps of an invertible dynamics must turn
each distinct initial state into a distinct final state.  Since the
set of states with recognizable structure is such a tiny subset of the
set of all possible states, almost every random-looking state must
turn into another random-looking state.

\figfig{critters}{ \hfill \fbox{%
\epsfig{figure=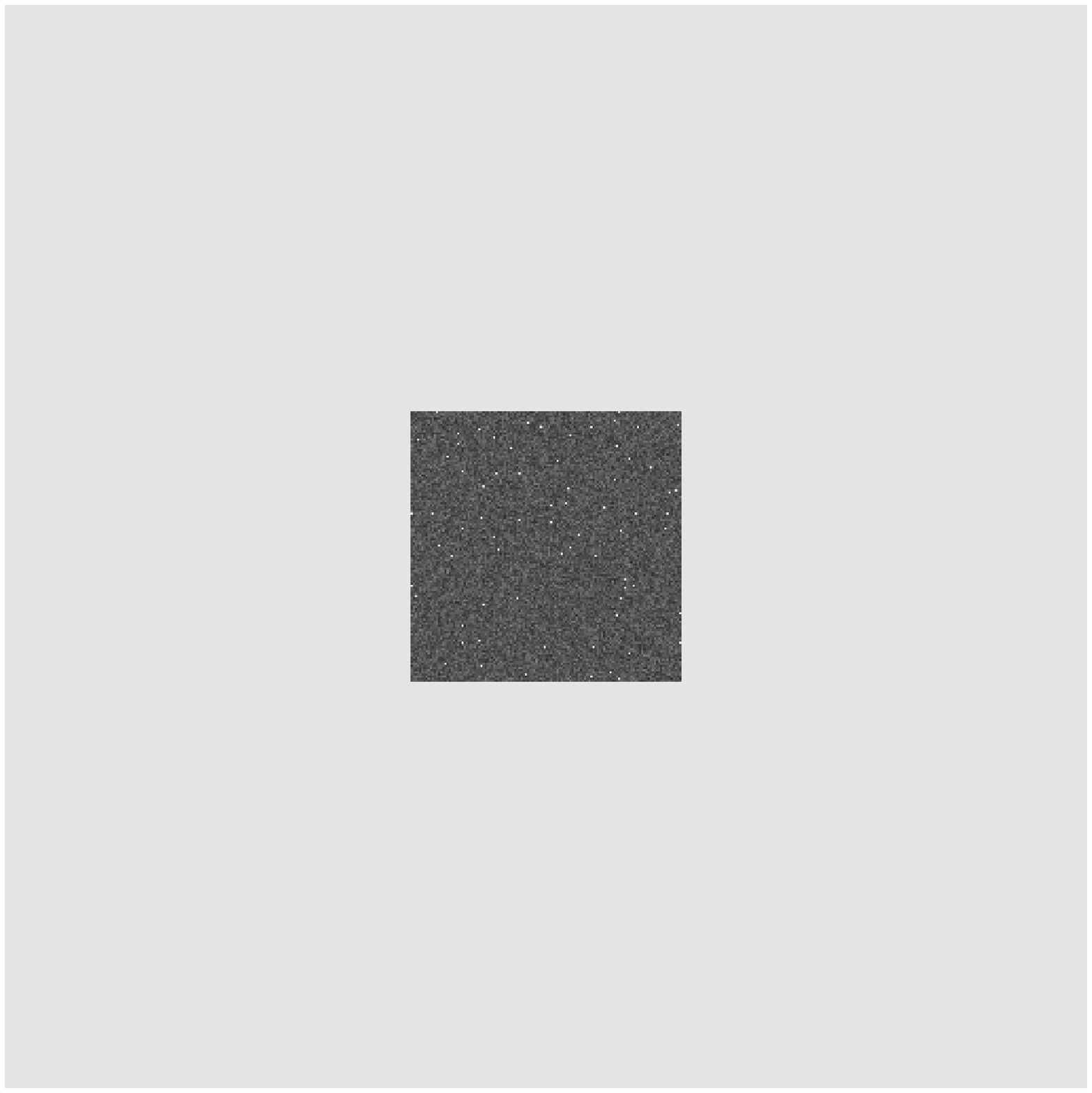,height=1.5in}}\hskip .2in \fbox{%
\epsfig{figure=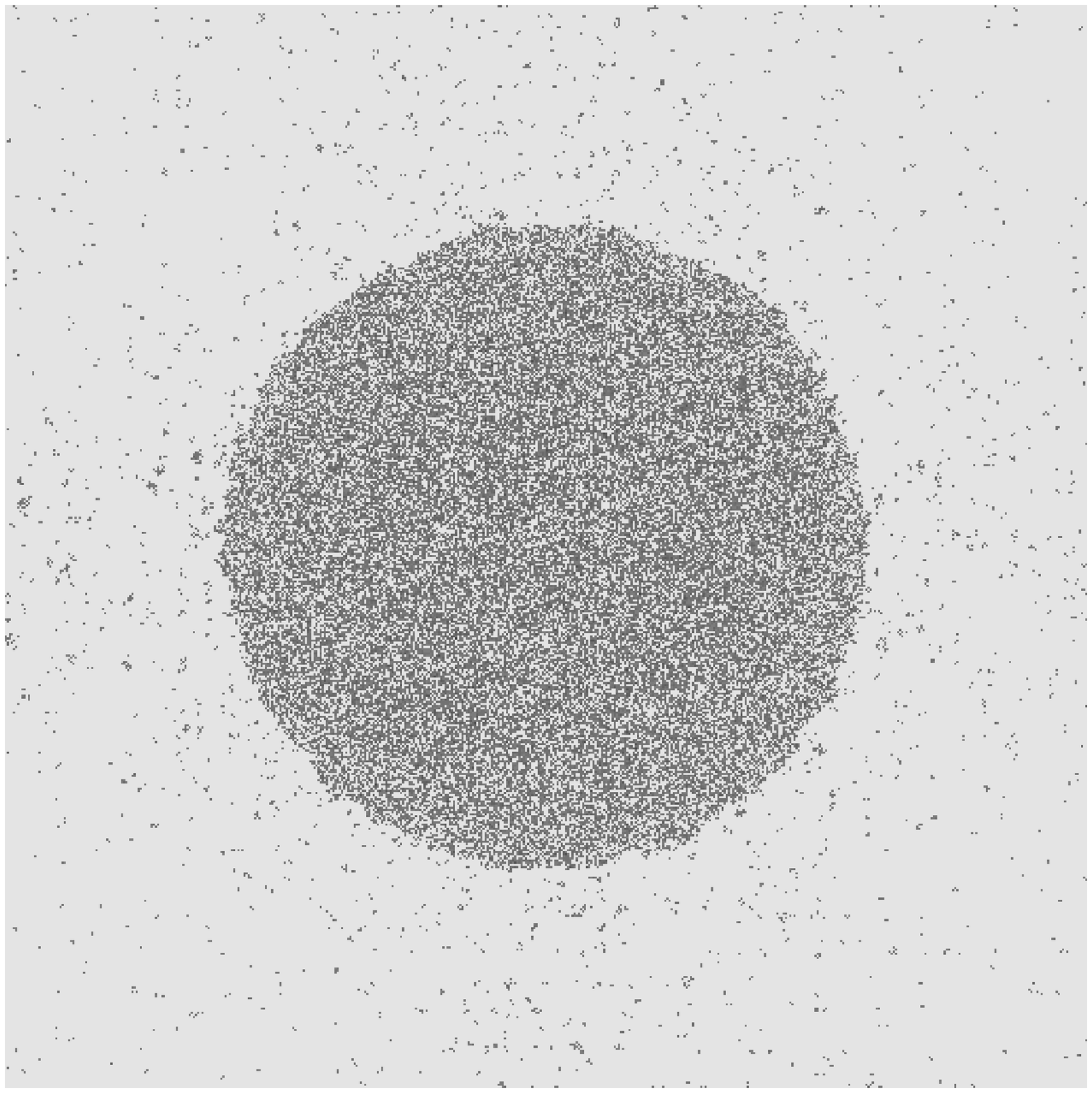,height=1.5in}}\hskip .2in \fbox{%
\epsfig{figure=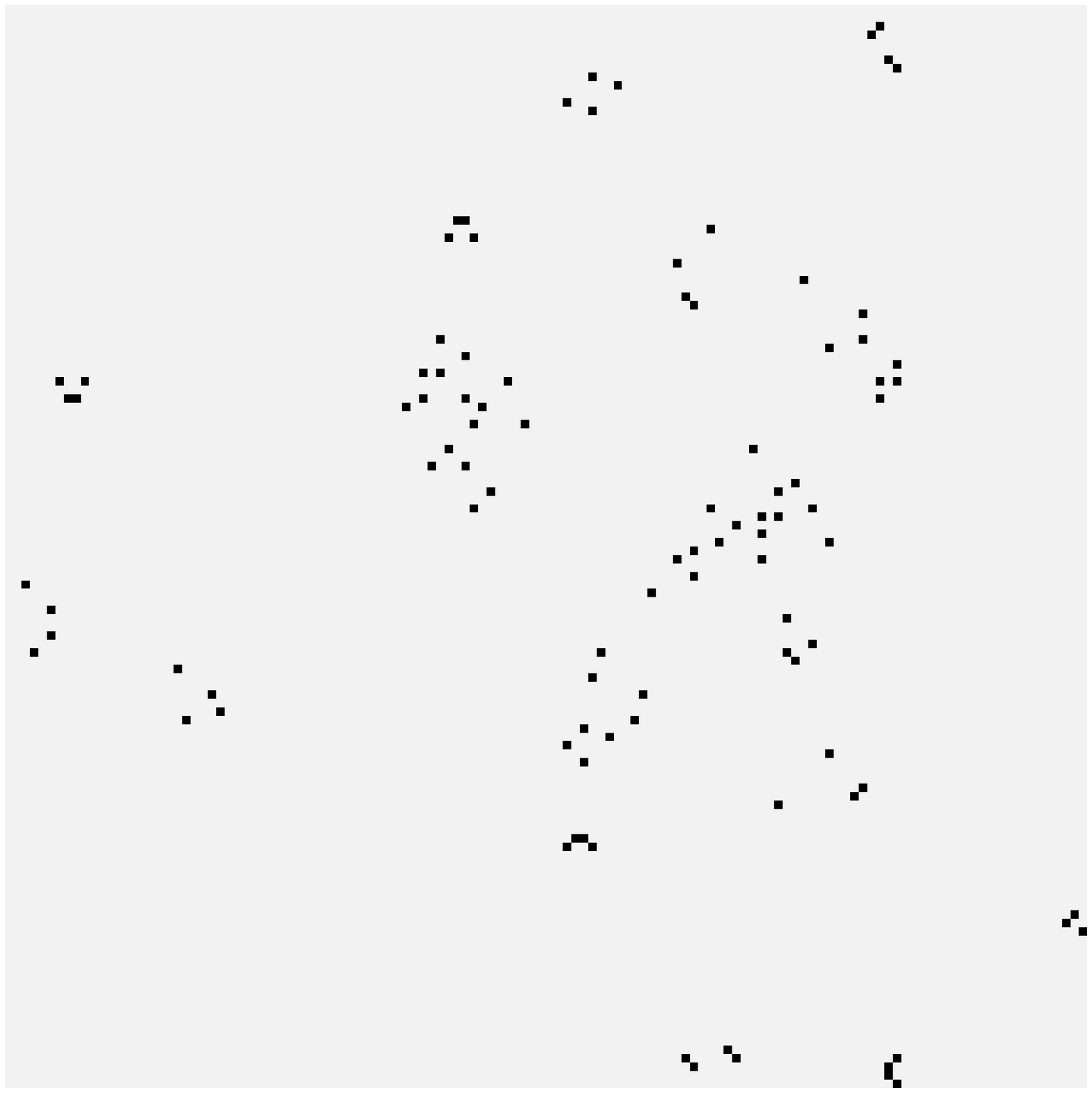,height=1.5in}}\hfill}%
{A Critters simulation.  (a)~The initial state
of the full 2K$\times$2K lattice.  (b)~The state after 1,000,000
steps.  (c)~A closeup of a region on the right.}

Thus instead of a random state, a ``generic'' initial state for an
invertible CA will be some easily generated ``low-entropy'' state---we
saw several examples of invertible evolutions from such states in
Section~\ref{sec.ising}.  For the Critters CA, we show a sample
simulation started from an empty 2K$\times$2K lattice with a randomly
filled 512$\times$512 block of 0's and 1's in the middle
(Figure~\ref{fig.critters}a).  In Figure~\ref{fig.critters}b we see
the state after 1,000,000 updates of the entire space.  In this
simulation, opposite edges of the space have been connected together
(periodic boundaries).  Figure~\ref{fig.critters}c shows a closeup of
a region near the right edge of the space: all of the structure
present has arisen from collisions of small moving objects that
emerged from the central region.  In analogy to Life, we will call
these small moving objects {\em gliders}.  You can see several of
these gliders in various phases of their motion in the closeup: they
are the compact symmetrical structures composed of four particles,
with two particles adjacent to each other, and two slightly separated
(see also Figure~\ref{fig.crit-universal}).  In the Critters dynamics,
a glider goes through a cycle of four configurations each time it
shifts by two positions.

Unlike the gliders in Life, Critters gliders are quite robust.
Consider, for example, what happens when two of these gliders collide
in an empty region.  At first they form a blob of eight particles that
goes through some pattern of contortions.  If nothing hits this blob
for a while, we always see at least one of the gliders emerge.  This
property arises from the combination of conservation and
invertibility: we can prove, from invertibility, that the blob must
break up, but since the only moving objects we can make with eight
particles are one or two gliders, then that's what must come out.  To
see that the blob must break up, we can suppose the opposite.  The
particles that make up the blob can only get so far apart without the
blob breaking up, and so there are only a finite number of possible
configurations of the blob.  The blob cannot repeat any configuration
(and hence enter a cycle of states) because of invertibility: a local
cycle of states would be stable going back in time as well as forward,
but we know that the blob has to break up if we run backwards past the
collision that formed it.  Since the blob runs out of distinct
configurations and cannot repeat, it must break up.  At least one
glider must come out.  If the collision that formed the blob was
rotationally symmetric, then both gliders must come out, since the
dynamics is also rotationally symmetric.  The robustness of particles
that we saw in Figure~\ref{fig.other-ising}a arises in a similar
manner.

The Critters rule is fascinating to watch because of the complicated
structures that form, with swarms of gliders bouncing around within
them and slowly altering them.  Sometimes, for example, a glider will
hit a little flap left from a previous glider collision, flipping it
{from} one diagonal orientation to another.  This will affect what
another glider does when it subsequently hits that flap.  Gliders will
hit obstacles and turn corners, sometimes going one way, sometimes
another, depending on the details of the collisions.  The pattern must
gradually change, because the system as a whole cannot repeat.  After
a little observation it is clear that there are many ways to build
arbitrary logic out of the Critters rule---one simple way is sketched
in the next section in order to demonstrate this rule's universality.

Started from an ordered state, the Critters CA accumulates disorder
for the same reason that a neat room accumulates disorder: most
changes make it messier.  As we have already noted, in an invertible
dynamics, a simple counting argument tells us that most messy states
don't get neater.  Localized patterns of structured activity that may
arise within this CA must deal with an increasingly messy environment.
No such structure in an invertible world can take in inputs that have
statistical properties that are unpredictable by it, and produce
outputs that are less messy, again because of our counting argument.
Thus its fair to call a measure of the messiness of an invertible CA
world the total {\em entropy}.  We can think of this total entropy as
being approximated by the size of the file we would get if we took the
whole array of bits that fills our lattice and used some standard
compression algorithm on it.

It is possible to construct invertible CA's in which a simple initial
state turns into a completely random looking mess very quickly.  While
it is still true that this invertible CA will probably take forever to
cycle, it has found another way to end its interesting activity
quickly---what we might call a rapid {\em heat death}.  Of course heat
death is the inevitable fate of any CA evolution that has a long
enough cycle: since the vast majority of states are random-looking,
very long cycles must consist mostly of such states.  We can, however,
try to put off the inevitable.  In the Critters CA, symmetries and
conservation laws act as constraints on the rate of increase of
entropy, and so make the interesting low-entropy phase of the dynamics
last much longer.  It would be interesting to try to capture within CA
dynamics other mechanisms that occur in real physics that contribute
to metastability and hence delay the heat death.

\section{A bridge to the continuum}




Historically, the partitioning technique used in the previous section
was first de\-vel\-op\-ed\cite{margolus-bbm} for use in the
construction of a very simple universal invertible CA modeled after
Fredkin's Billiard Ball Model (BBM) of
computation\cite{fredkin-bbm}.\footnote{The first universal invertible
CA was actually constructed by Toffoli\cite{toffoli-universal}, who
showed how to take a universal 2D CA that was non-invertible, and add
a third dimension that would keep a complete time history of the
dynamics, thus rendering it invertible.}  The BBM is a beautiful
example of a continuum physical system that is turned into a digital
system simply by constraining its initial conditions and the times at
which we observe it.  This makes it a wonderful bridge between the
tools and concepts of continuum mechanics, and the world of exact
discrete information dynamics.  This model is discussed in
\cite{feynman-comp}, but we will review it very briefly here.

\figfig{fred-coll}{ \hfill \hbox{%
\epsfig{figure=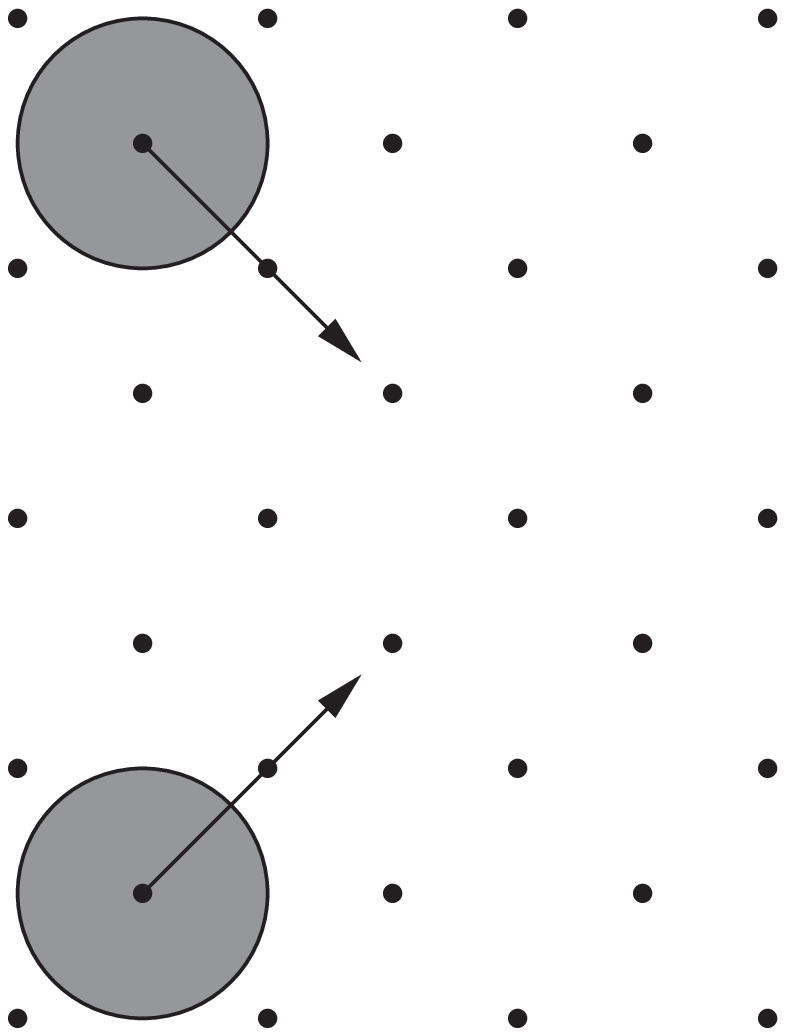,height=2in}}\hskip .6in \hbox{%
\epsfig{figure=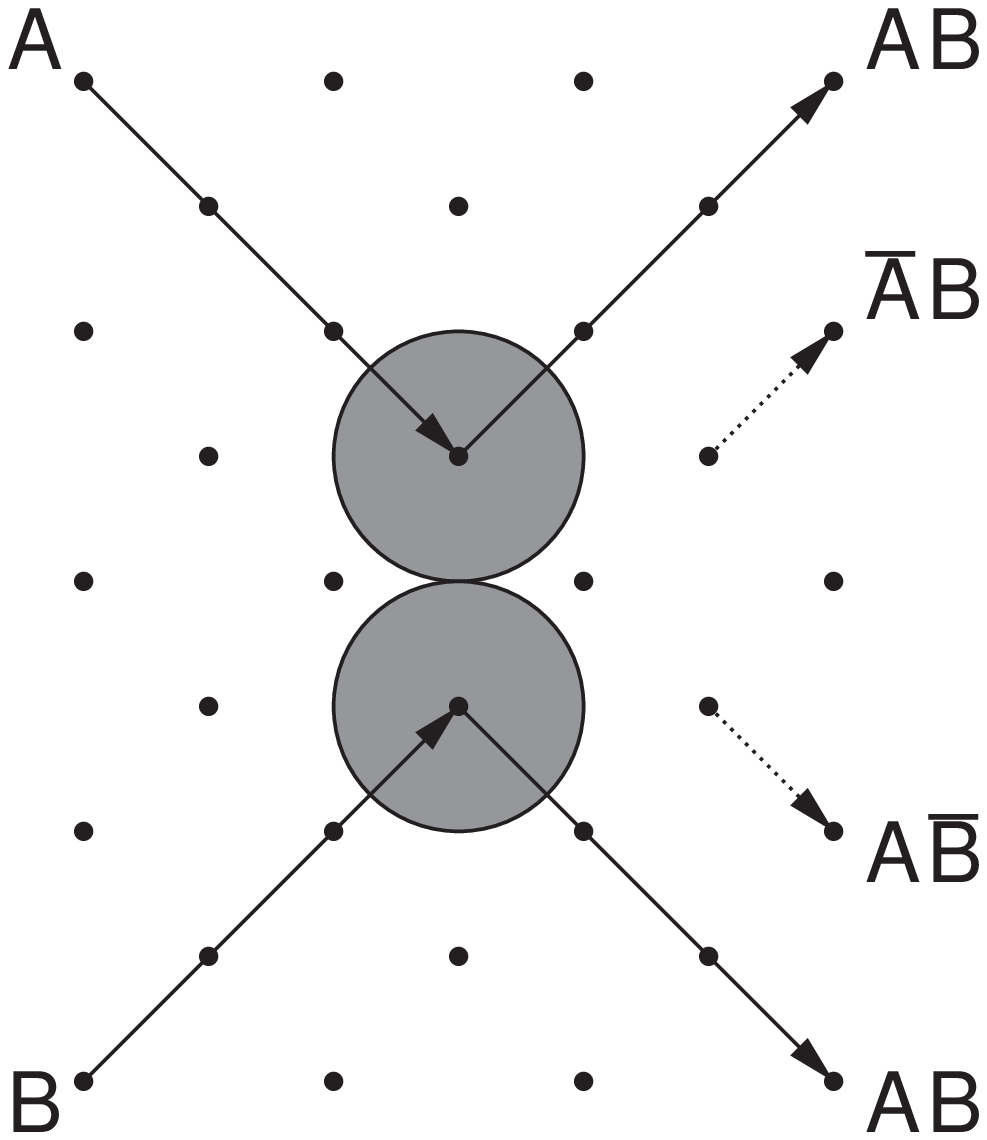,height=2in}}\hfill}%
{Fredkin's Billiard Ball Model.  (a)~Balls
heading toward a collision.  (b)~Paths taken in collision are
displaced from straight paths.}

In Figure~\ref{fig.fred-coll}a we show two finite-diameter billiard
balls heading toward a collision, both moving at the same speed.
Their centers are initially located at integer coordinates on a
Cartesian lattice---we will refer to these as {\em lattice points}.
At regular intervals, the balls will be found at consecutive lattice
points, until they collide.  In Figure~\ref{fig.fred-coll}b we show a
collision.  The outer paths show the actual course that the balls take
after a collision; the inner paths illustrate where each of the two
balls would have gone if the other one wasn't there to collide with
it.  Thus we see that a locus at which a collision might or might not
happen performs logic: if the presence of a ball at a lattice point at
an integer time is taken to represent a 1, and the absence a 0, then
we get 1's coming out on the outer paths only if balls at $A$ AND $B$
came in at the same time.  The other output paths correspond to other
logical functions of the inputs.  It is easy to verify that this
collision is a universal and invertible logic element (just reverse
all velocities to run BBM circuits backwards).  We also allow fixed
mirrors in our model to help route ball-signals around the
system---these are carefully placed so that the centers of balls are
still always found at lattice points at integer times.

In order to make a simple CA model of the BBM, we will represent
finite diameter balls in our CA by spatially separated {\em pairs} of
particles, one following the other---the leading edge of the ball
followed by the trailing edge.  When such a ball collides with
something, the front-edge particle collides first, and then the
influence is communicated to the trailing edge.  This kind of ``no
rigid bodies'' approach to collisions is more consonant with the
locality of interaction that we are trying to capture in CA's than a
larger-neighborhood model in which distant parts of a fixed-size ball
can see what's going on right away.

\figfig{bbmca}{ \hfill \hbox{%
\epsfig{figure=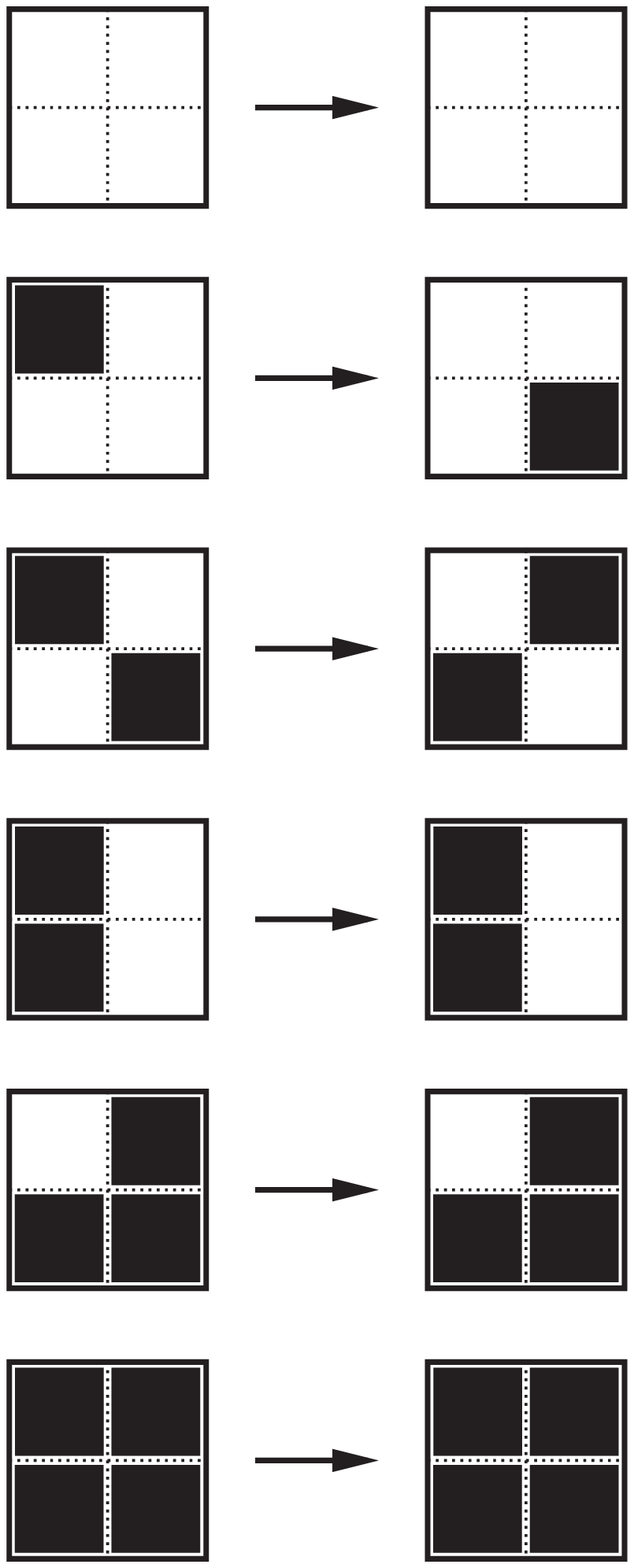,height=2in}}\hskip 1in \fbox{%
\epsfig{figure=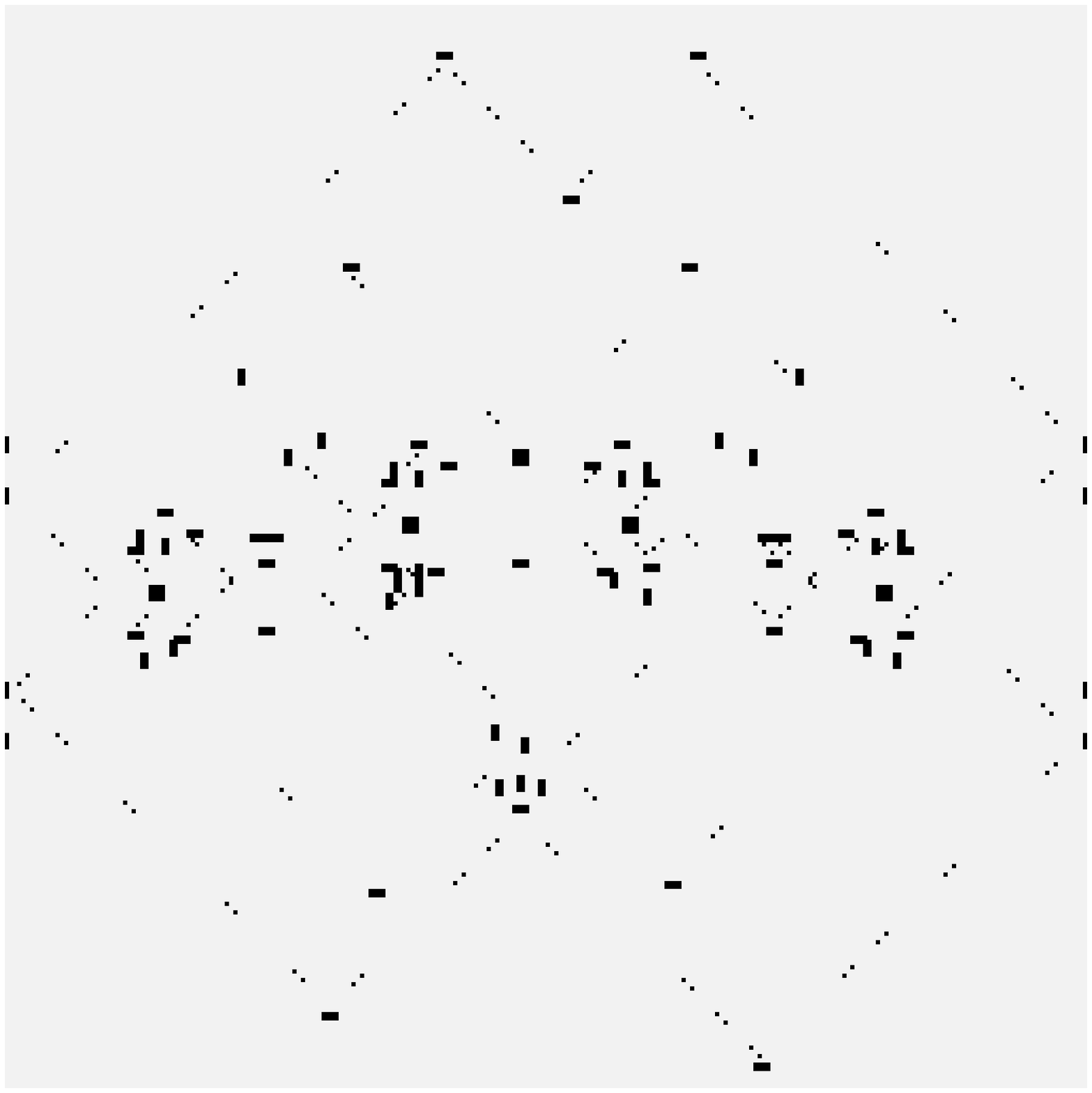,height=2in}}\hfill}%
{A Billiard Ball Model CA.  (a)~The BBMCA
rule.  (b)~A BBMCA circuit.}

Figure~\ref{fig.bbmca}a shows the BBMCA rule.  Like the Critters rule,
this rule is rotationally symmetric and so, again, only one case out
of every rotationally equivalent set is shown.  Note that the rule
conserves 1's (particles), and that only two cases change.  This is
the complete rule that is applied alternately to the even and odd
2$\times$2 blockings.  Note that, much like the Ising CA, this rule is
its own inverse: if we simply apply the update rule to the same
blocking twice in a row, the net effect is no change.\footnote{The Ising CA is
actually very closely related.  It can be put into the same 2$\times$2
block-partitioned format if we model bonds instead of
sites\cite{cambook}.}

\figfig{bbmca-coll}{ \hfill \hbox{%
\epsfig{figure=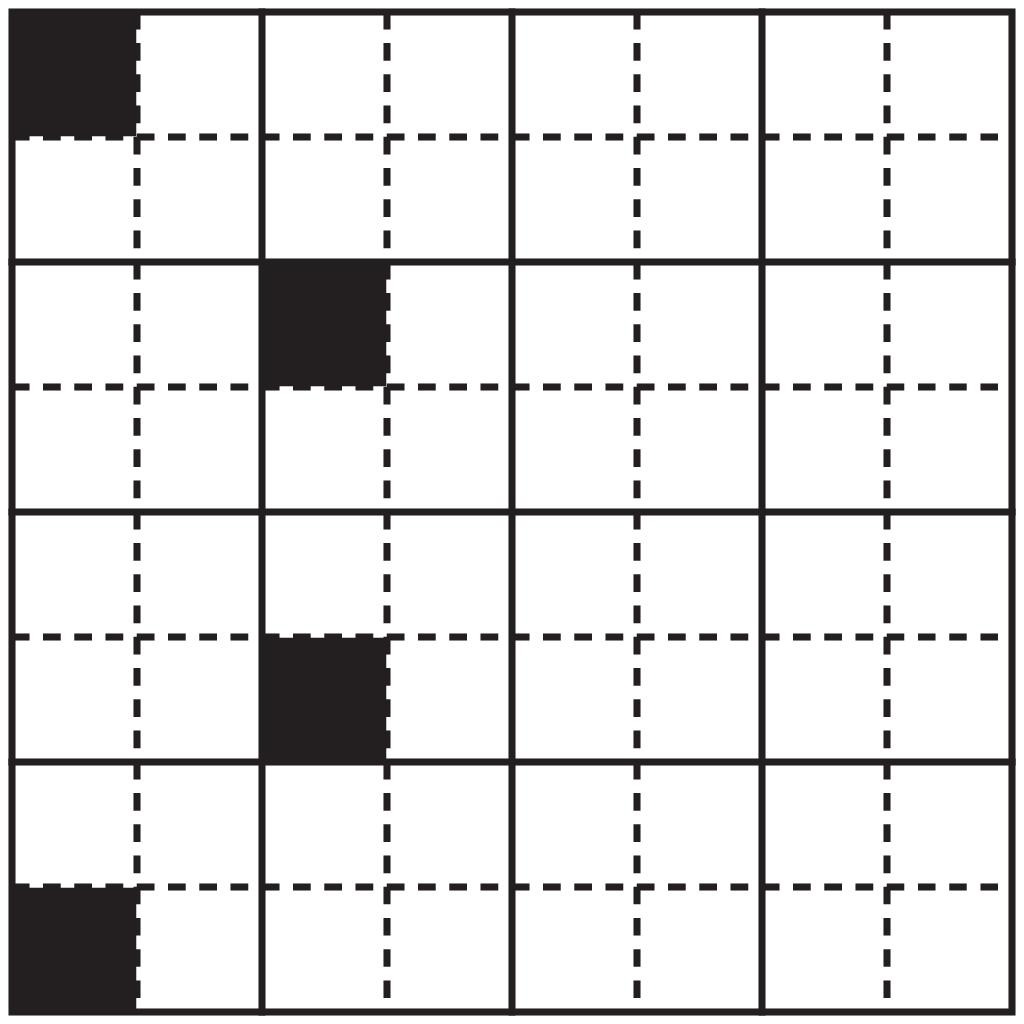,height=.7in}}\hskip .1in \hbox{%
\epsfig{figure=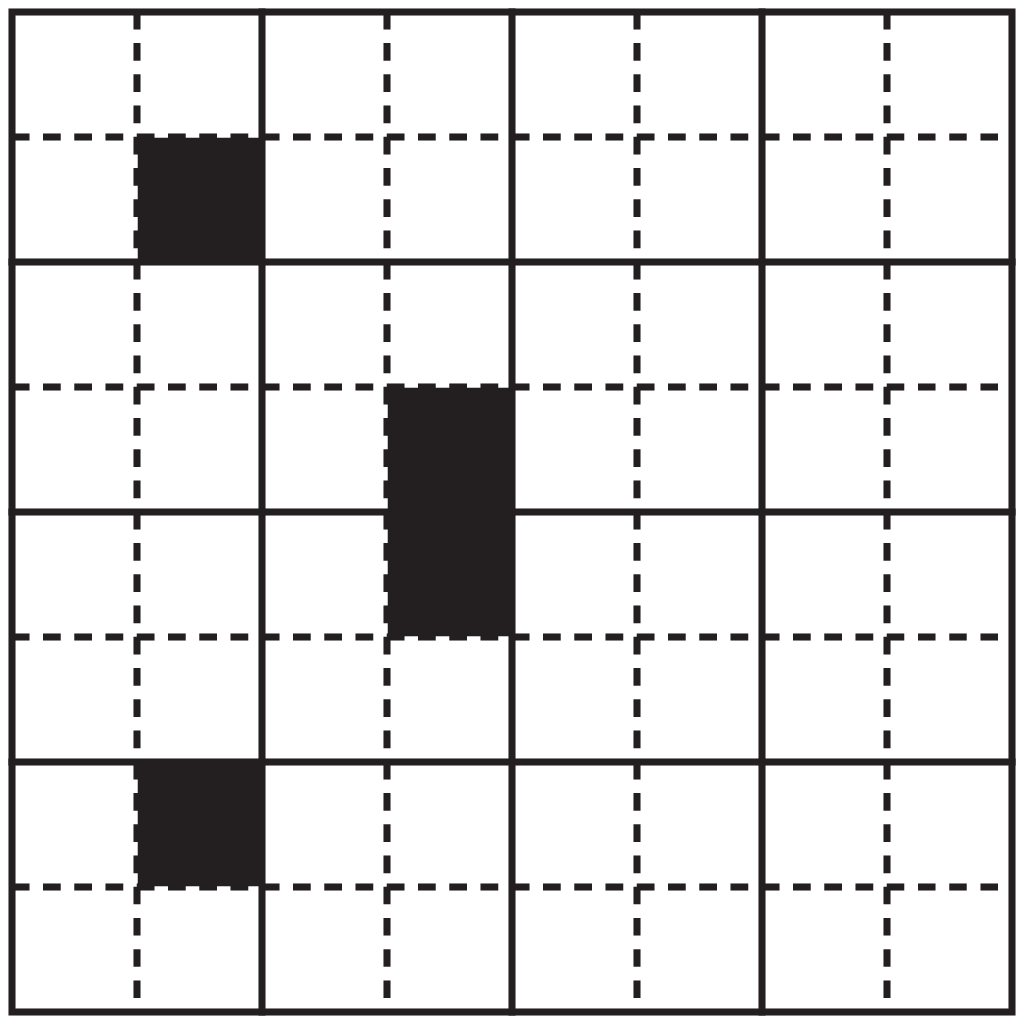,height=.7in}}\hskip .1in \hbox{%
\epsfig{figure=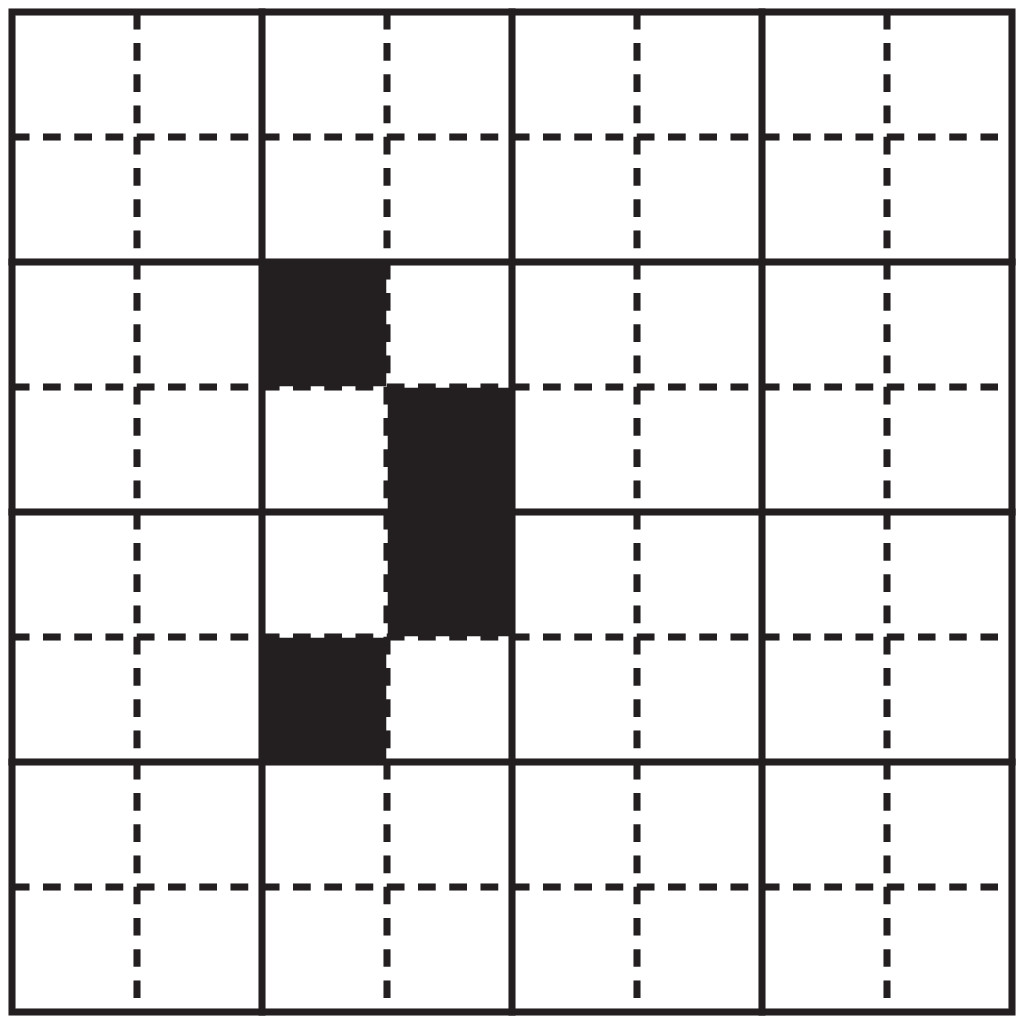,height=.7in}}\hskip .1in \hbox{%
\epsfig{figure=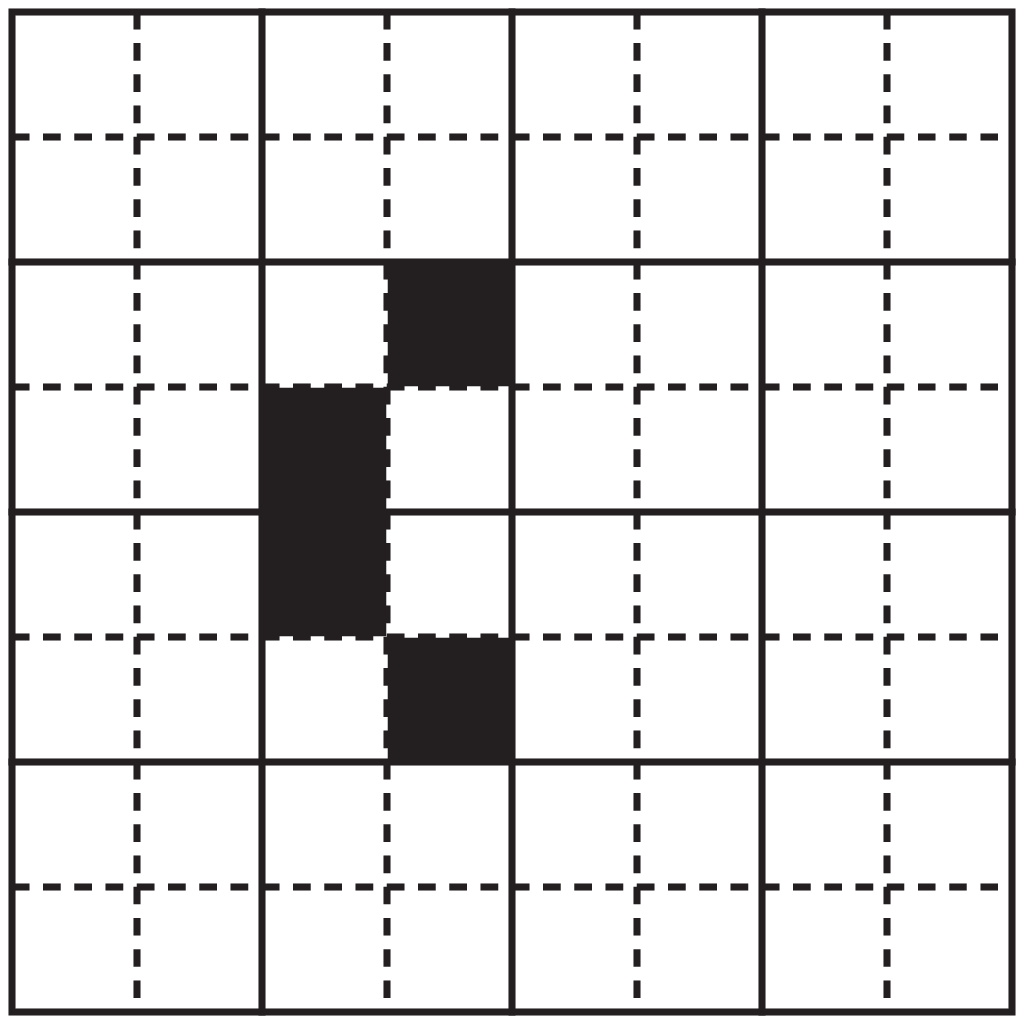,height=.7in}}\hskip .1in \hbox{%
\epsfig{figure=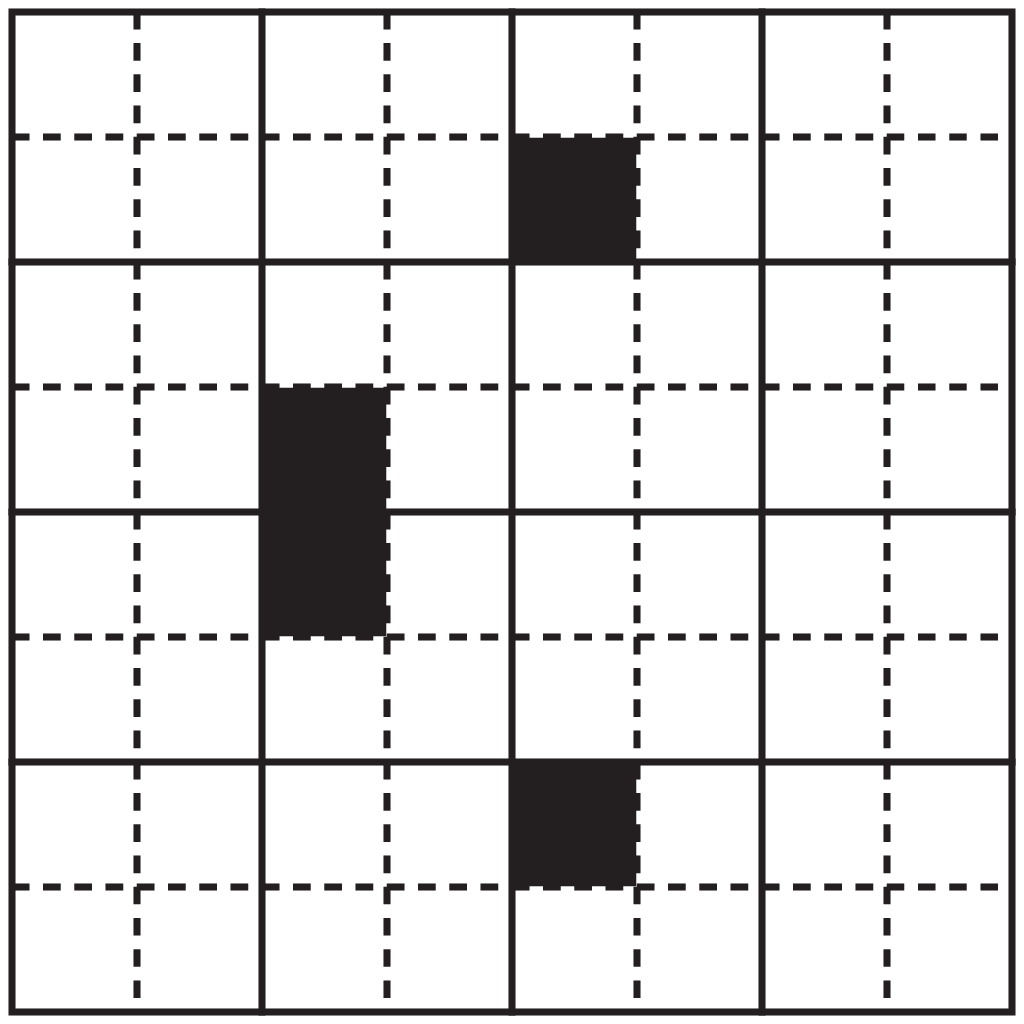,height=.7in}}\hskip .1in \hbox{%
\epsfig{figure=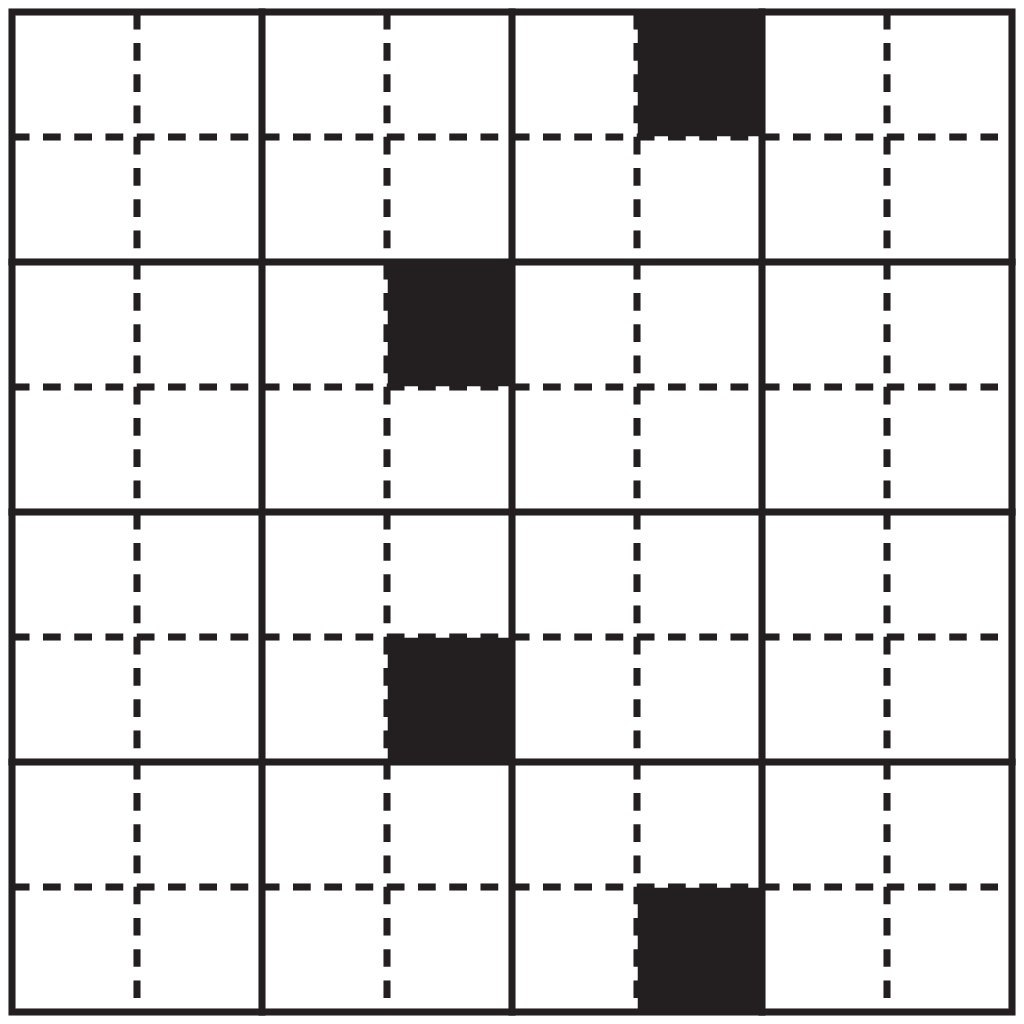,height=.7in}}\hfill}%
{A BBMCA collision.  We show succesive
snapshots of a small area where a collision is happening.  In the
first image, the solid-blocks are about to be updated.  The blocking
alternates in successive images.}

Figure~\ref{fig.bbmca-coll} shows a BBMCA collision between two
minimum-size balls---the gap between the two particles that make up
the front and back of the ball can be any odd number of empty sites.
Until the balls get close together the particles that form them all
propagate independently: a single 1 in one corner of a block moves to
the opposite corner.  When we change the blocking, the particle again
finds itself alone in a block in the same corner it started in, and
again moves in the same direction.  When two leading-edge particles
find themselves in the same block, the collision begins.  These
particles are stuck for one step---this case doesn't change.
Meanwhile the trailing edge particles catch up, each colliding head-on
with a leading-edge particle which was about to head back to meet it
(if the gap had been wider).  New particles come out at right angles
to the original directions, due to the ``two-on-a-diagonal'' case of
the rule, which switches diagonal.  Now one of the particles from each
head-on collision becomes the new leading edge particle; these are
done with the collision and head away from the collision locus, once
again propagating independently of the trailing particles.  Meanwhile
the two new trailing-edge particles are headed toward each other.
They collide and are stuck for one step before reflecting back the way
they came, each following along the path already taken by a leading
edge particle.  Each two-particle ball has been displaced from its
original path.  If the other two-particle ball hadn't been there, it
would have gone straight.

Mirrors are built by placing square patterns of four particles
straddling two adjacent blocks of the partition.  It is easy to verify
that such squares don't change under this rule, even if you put them
right next to each other.  Single particles just bounce back from such
mirrors.  The collision of a two-particle ball with such a mirror
looks just like the collision of two balls that we have already seen;
we just replace one of the balls with a mirror whose surface lies
along the axis of symmetry of the two-ball collision.  The remaining
ball can't tell the difference.  For more details about the BBMCA, see
\cite{cambook,margolus-thesis}.

\figfig{crit-universal}{ \hfill \fbox{%
\epsfig{figure=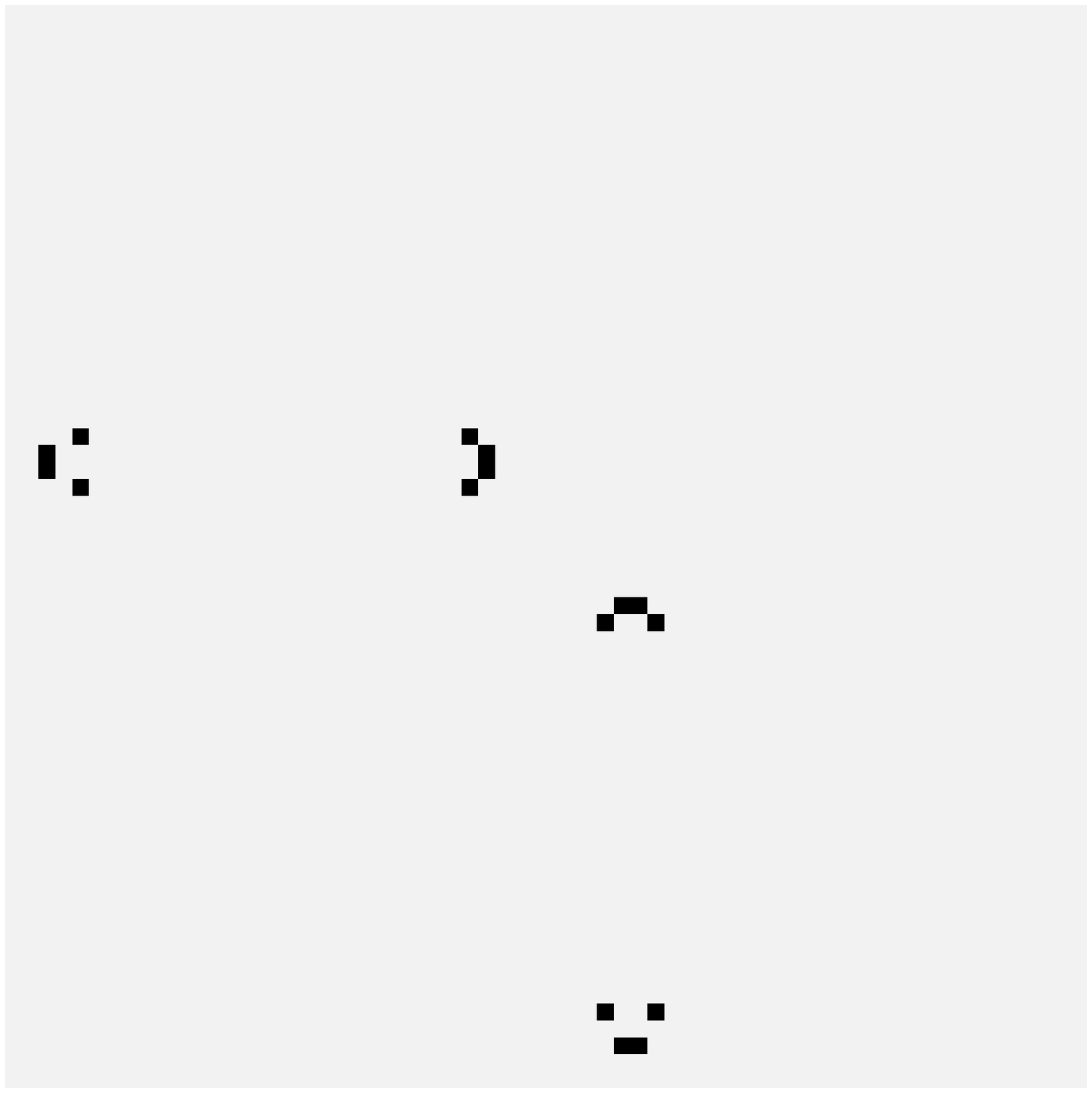,height=.7in}}\hskip .1in \fbox{%
\epsfig{figure=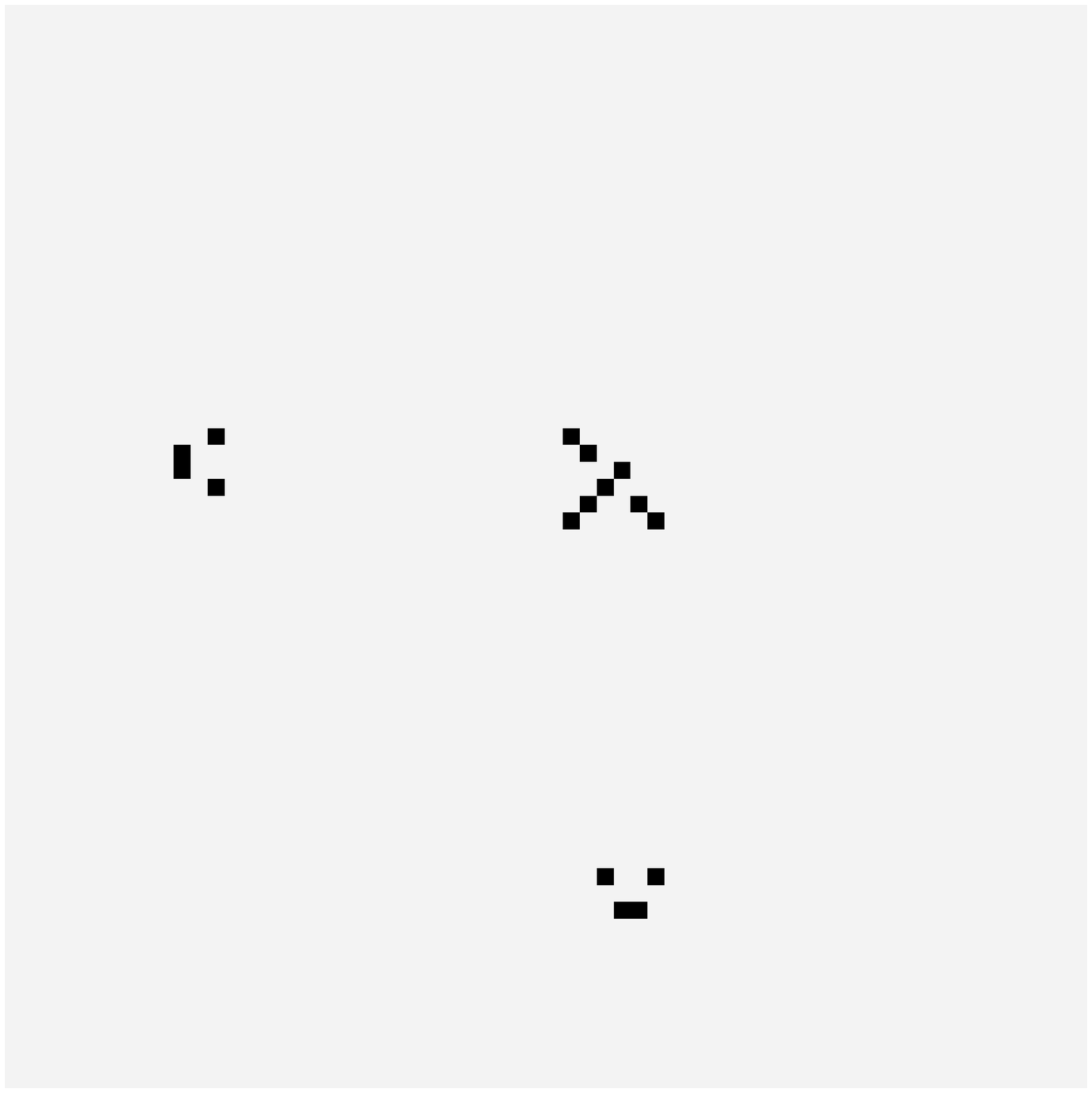,height=.7in}}\hskip .1in \fbox{%
\epsfig{figure=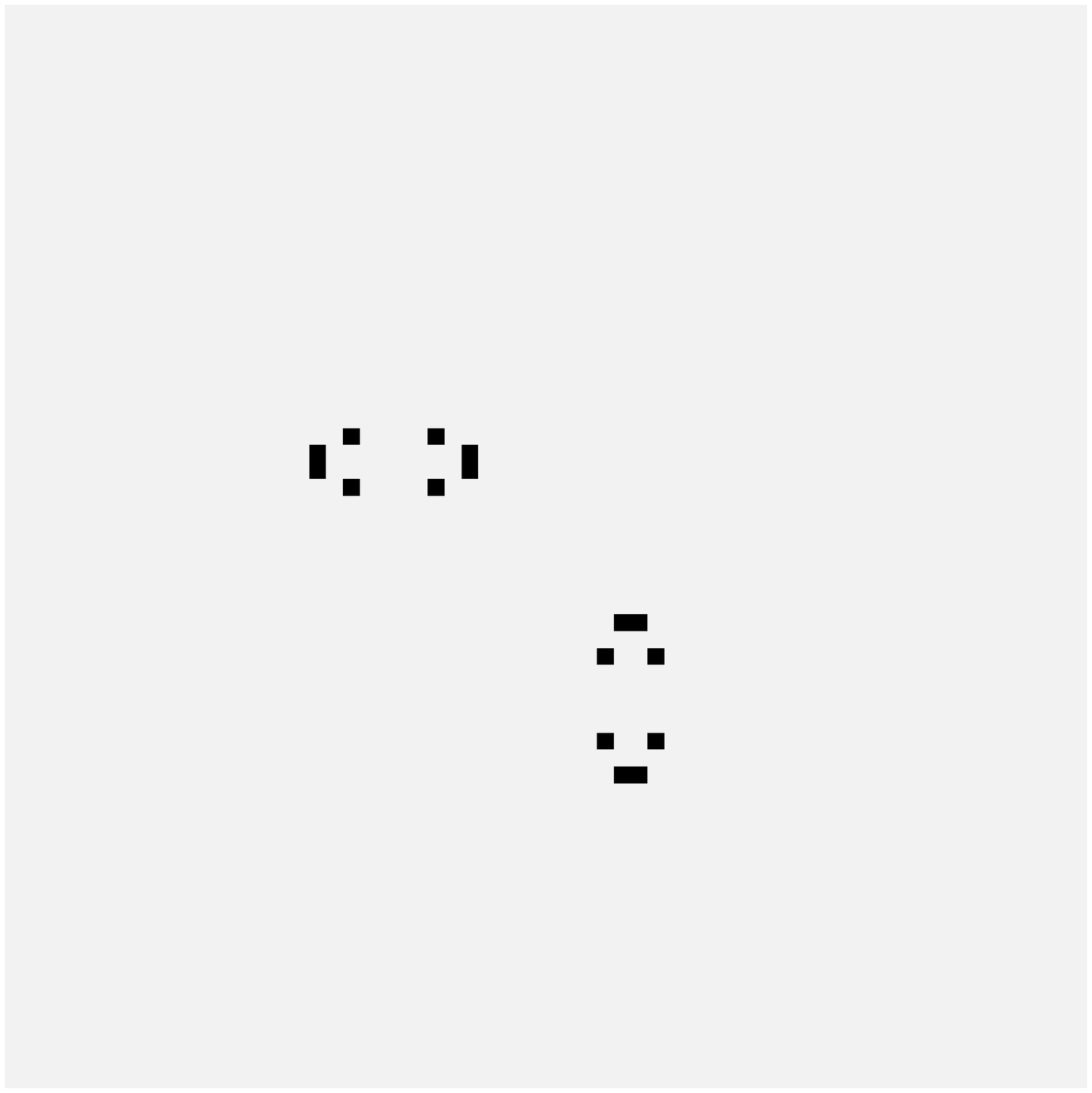,height=.7in}}\hskip .1in \fbox{%
\epsfig{figure=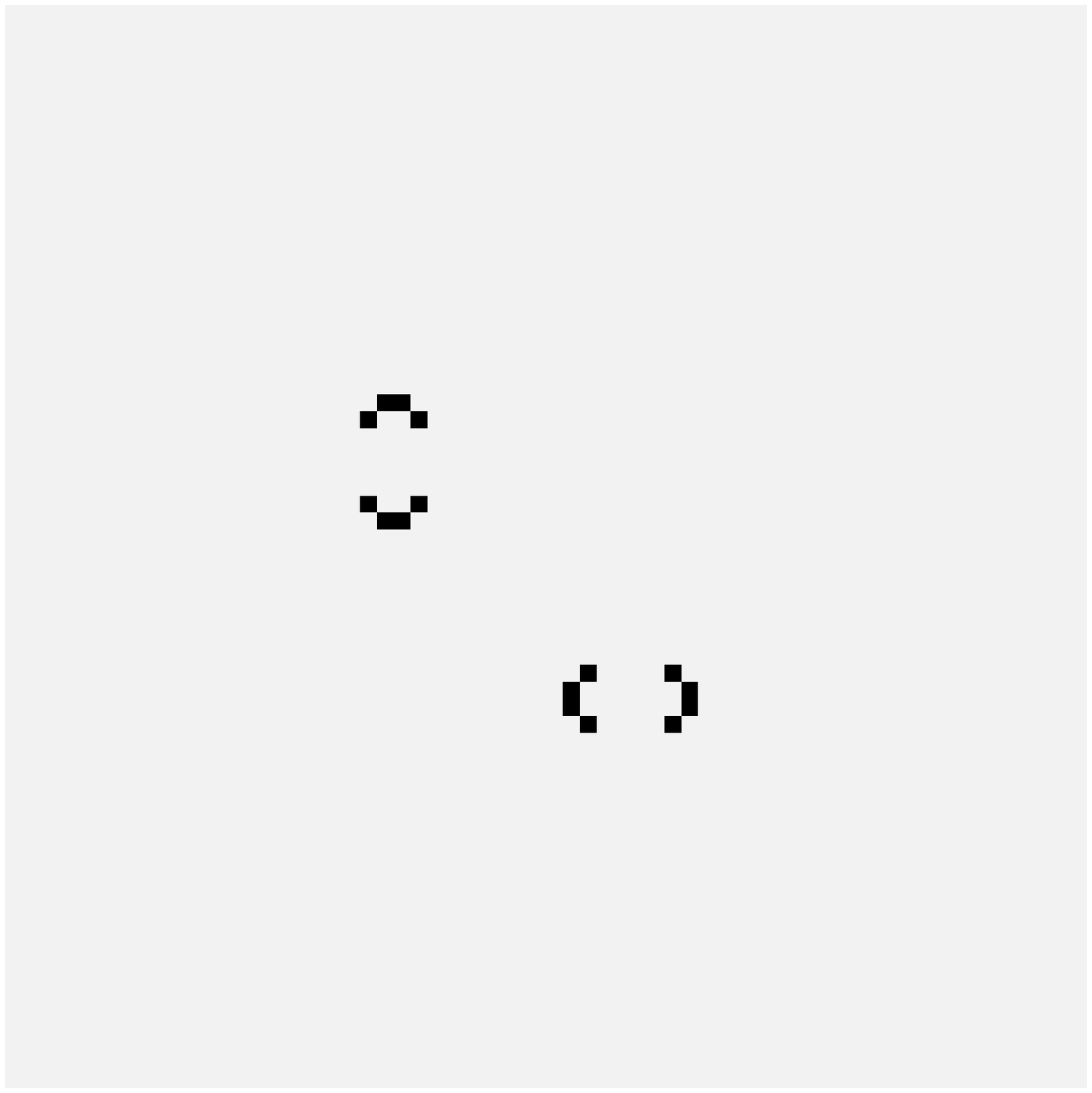,height=.7in}}\hskip .1in \fbox{%
\epsfig{figure=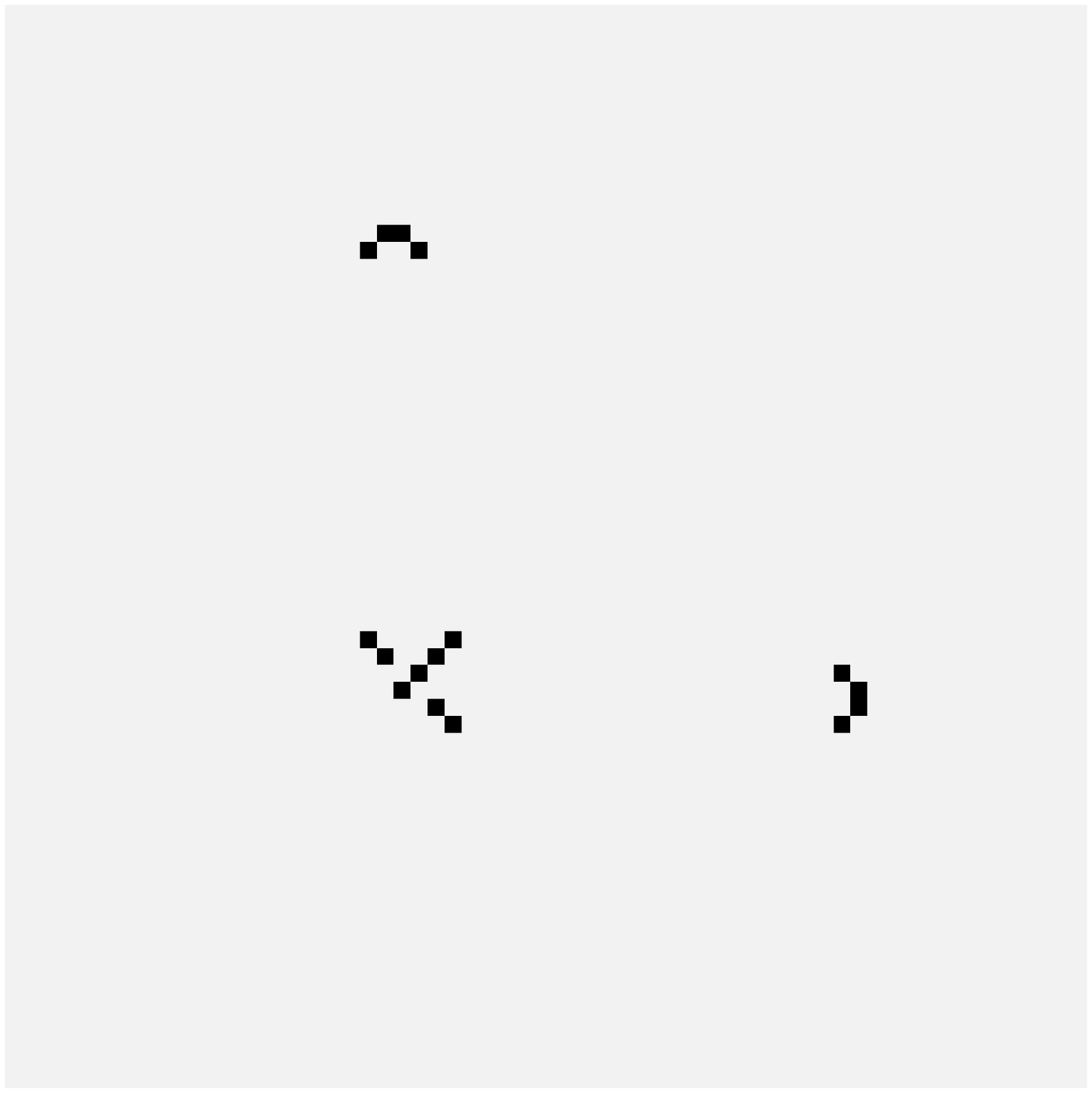,height=.7in}}\hskip .1in \fbox{%
\epsfig{figure=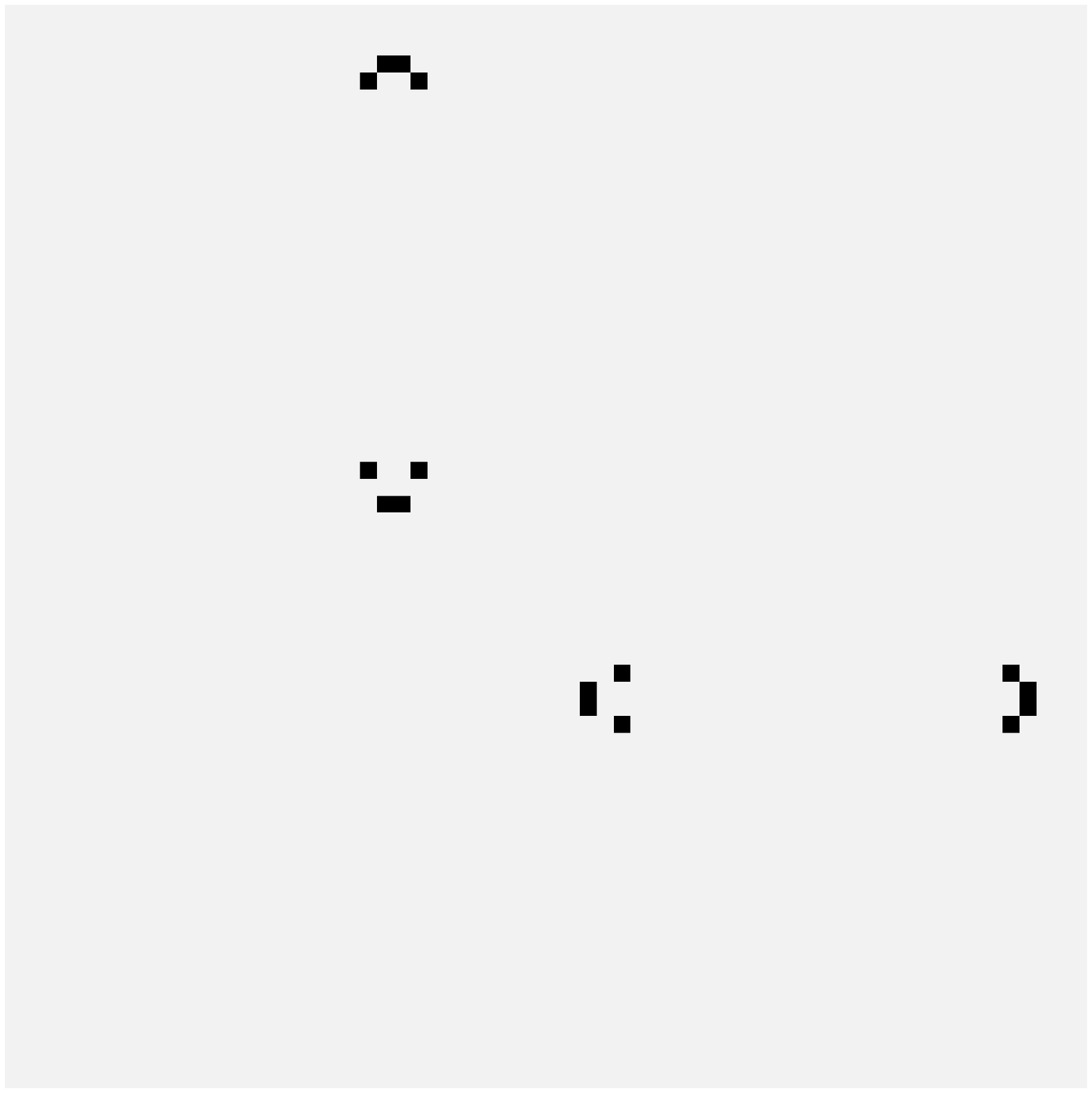,height=.7in}}\hfill}%
{A BBMCA-style collision of pairs of gliders in the Critters CA.  The
images shown are not consecutive states of the lattice, but are
instead spaced in time to correspond (with a $45^\circ$ rotation) to
the images in the previous figure.}

Figure~\ref{fig.bbmca}b shows a BBMCA circuit, computing a permutation
sequence.  Because of their long cycle times, invertible circuits tend
to make good pseudo-random number generators.  In fact, a perfect
random number generator would go through all of its internal states
before cycling, and so it would be perfectly invertible.  It is also
interesting to use the BBMCA to construct circuits that are more
directly analogous to thermodynamic systems, since the constraint of
invertibility means that {\em it is impossible to design a BBMCA
circuit that, acting on unpredictable statistical inputs that it
receives, can reduce the entropy of those data}---for the reasons
discussed in the previous
section\cite{margolus-thesis,bennett-thermo,bennett-demons}.  The
BBMCA is simple enough that it 
provides a good theoretical model to use for other inquiries about
connections between physics and computation.  For example, one can use
its dynamics as the basis of quantum spin models\cite{margolus-pqc}.


Using Figure~\ref{fig.crit-universal}, we sketch a simple
demonstration that the Critters rule of the previous section is
universal---we show that pairs of Critters-gliders suitably arranged
can act just like the ``balls'' in the BBMCA, which is universal.
Figure~\ref{fig.crit-universal} shows a collision that is equivalent
to that shown in Figure~\ref{fig.bbmca-coll}.  We don't show every
step of the Critters time-evolution; instead we show the pairs of
gliders at points corresponding to those in the collision of
Figure~\ref{fig.bbmca-coll}.  Mirrors can be implemented by two single
Critters particles, one representing each end of the mirror.

\section{Discrete molecular dynamics}\label{sec.md}

Having constructed a CA version of billiard ball dynamics, it seems
natural to try to construct CA's that act more like real
gases\cite{toffoli-pde,margolus-bbm}.  With enough particle directions
and particle speeds, our discrete Molecular Dynamics (MD) should
approximate a real gas.

The BBMCA has just four particle directions and a single particle
speed.  We will make our first MD model by modifying the BBMCA rule.
For simplicity, we won't worry about modeling finite-diameter balls:
single 1's will be our gas molecules.  The simplest such rule would
be, ``During each 2$\times$2 block update, each molecule ignores
whatever else is in the block and simply moves to the opposite corner
of its block.''  Then, when we switch partitions, that molecule would
again be back in the same kind of corner that it started in and so it
would again move in the same direction---moving exactly like an
isolated particle in the BBMCA.  This simple rule gives us a
non-interacting gas, with four directions and one speed.

\figfig{hpp}{ \hfill \hbox{%
\epsfig{figure=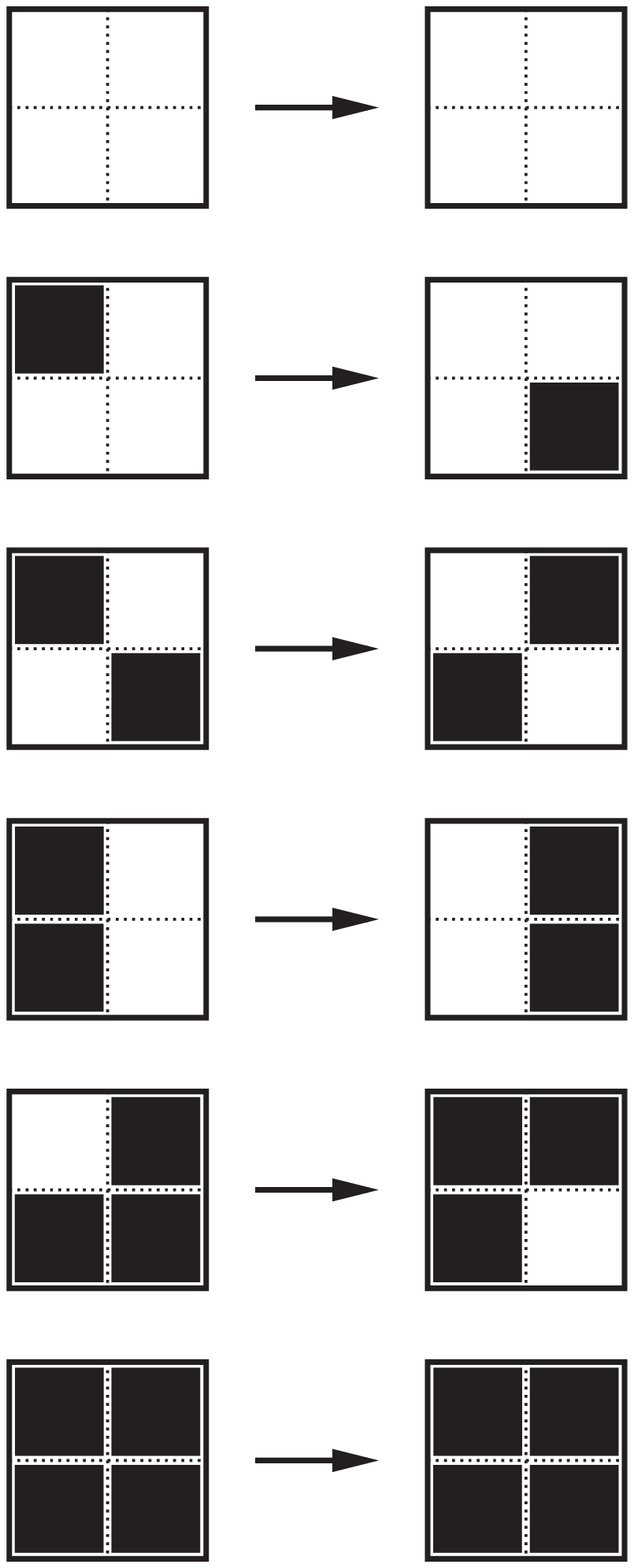,height=1.85in}}\hskip .35in \hbox{%
\epsfig{figure=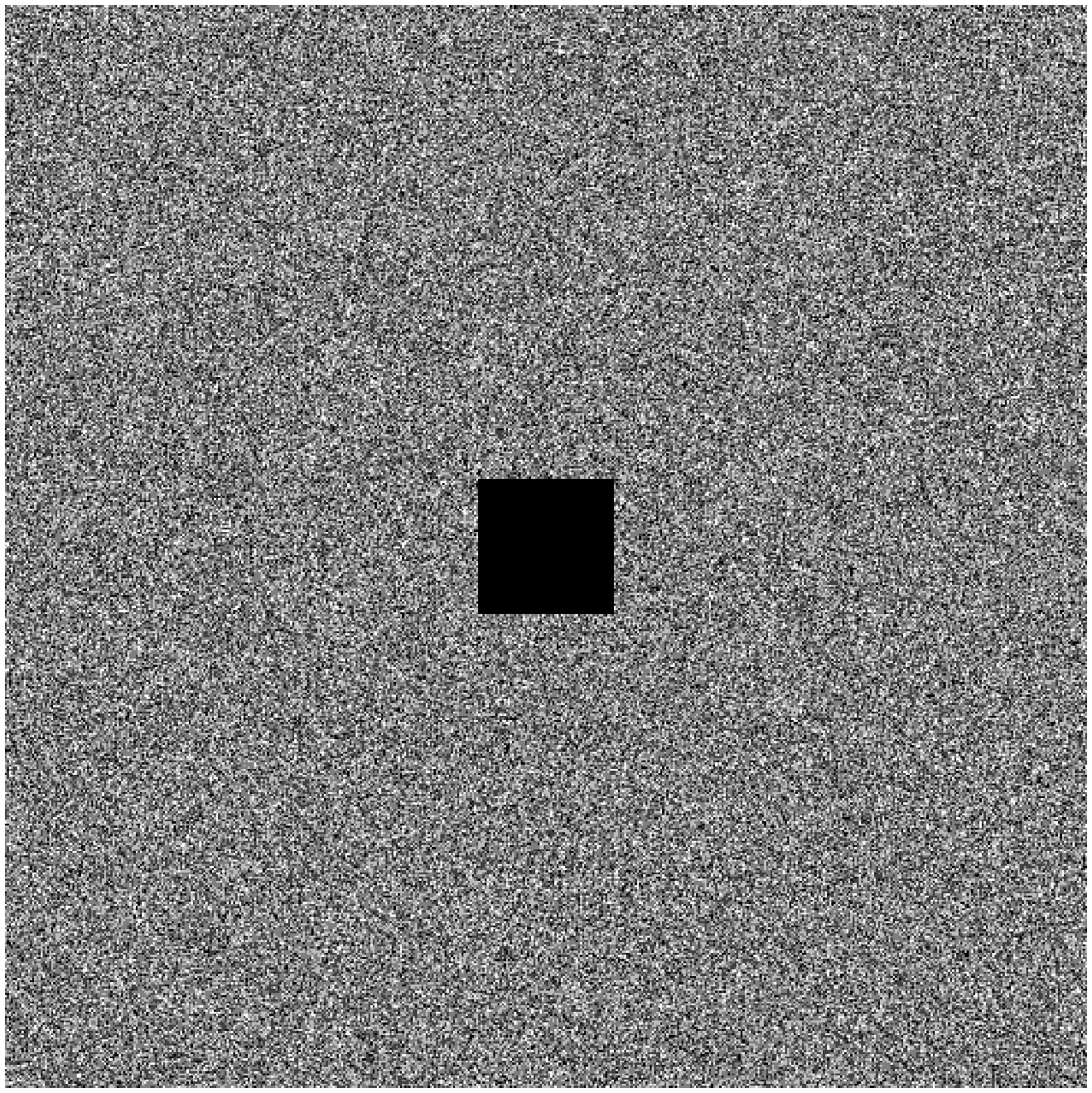,height=1.85in}}\hskip .2in \hbox{%
\epsfig{figure=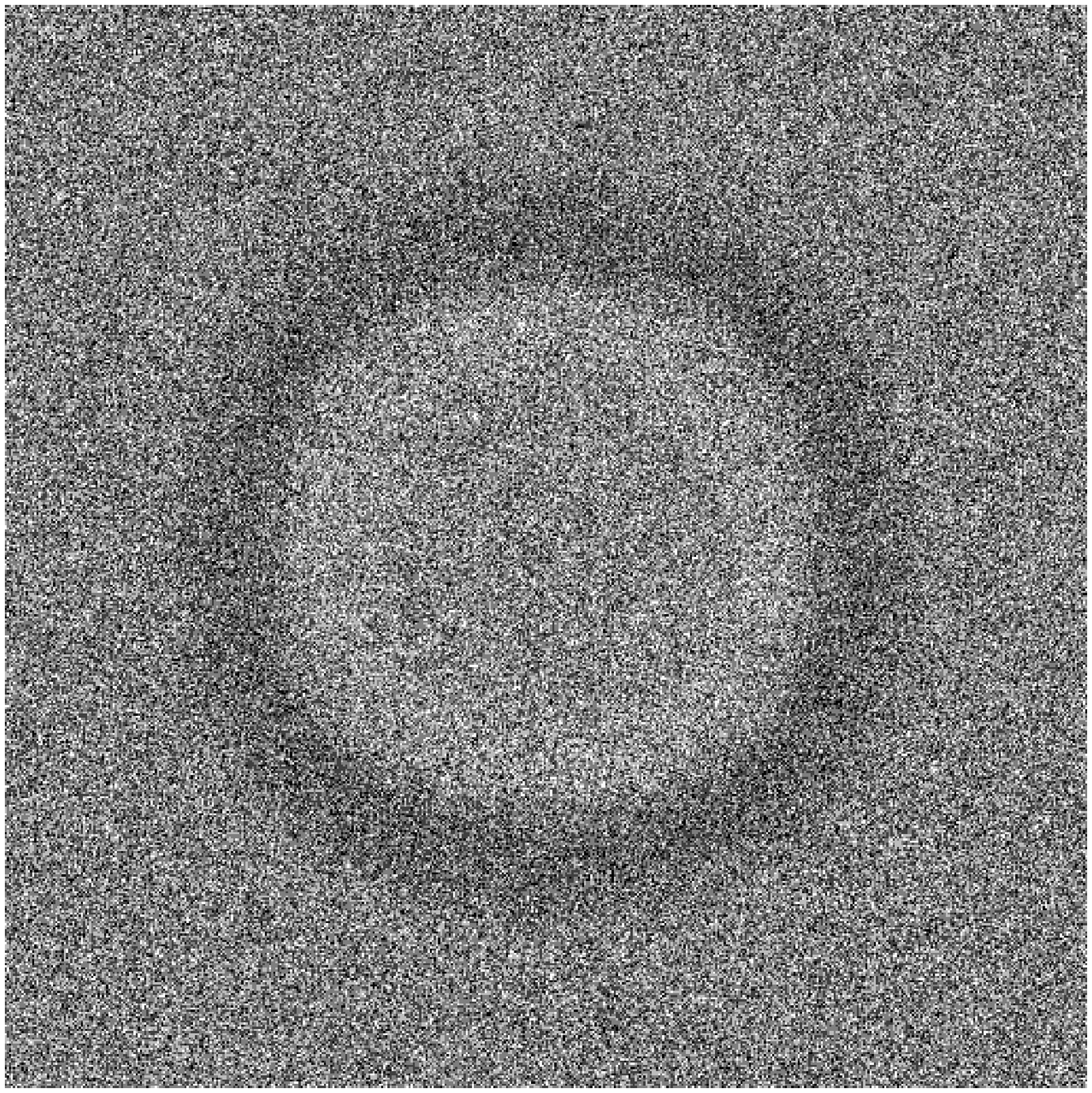,height=1.85in}}\hfill}%
{A simple four-direction lattice gas.  (a)~A
momentum conserving invertible rule.  (b)~A 512$\times$512 lattice
filled randomly with particles, with a square block of ones's in the
center. (c)~A round pressure wave spreads out from the center.}

We would like to add a {\em momentum conserving} collision to this
non-interacting gas rule.  We can begin by defining what we mean by
momentum.  If we imagine that our discrete lattice dynamics is simply
a succession of snapshots of a continuum system, as it was in the case
of the BBM, then we automatically inherit definitions of energy and
momentum from the continuous system.  To add a momentum conserving
collision to our simple four-direction gas, we should have two
molecules that collide head-on come out at right angles.
Figure~\ref{fig.hpp}a shows the non-interacting gas rule with one case
modified to add this collision: when exactly two molecules appear on
one diagonal of a block, they come out on the other diagonal.

We would not expect such a simple model to behave very much like a
real gas.  In Figure~\ref{fig.hpp}b, we show a 512$\times$512 2D space
filled with a random pattern of 1's and 0's, with a square block in
the center of the pattern that is all 1's.  Figure~\ref{fig.hpp}c
shows this system after about 200 updates of the space: we see a round
pressure wave.  We were amazed when we first ran this simulation in
the early 1980's\cite{toffoli-pde}.  How could we get such continuous
looking behavior from such a simple model?  This is the point at which
we began to think that perhaps CA MD might be immediately practical
for fluid modeling\cite{super,cambook}.  Discrete lattice models are
well adapted to meshes of locally interconnected digital hardware,
which can be very large and very fast if the models are simple.  It
turns out, though, that this particular model is too simple to
simulate fluid flow---though it is useful for other purposes.  This
four-direction {\em lattice gas automaton} (LGA) is now commonly known
as the HPP gas after its originators\cite{hpp}, who analyzed it about
a decade before we rediscovered it.  Their analysis showed that this
four-velocity model doesn't give rise to normal isotropic
hydrodynamics.

Notice that the HPP gas is perfectly invertible---like the BBMCA, its
rule is its own inverse.  Thus we can run our pressure wave backwards,
getting back exactly to the square block we started from.  This doesn't
contradict what we said earlier about entropy in invertible CA's,
since a messy state can always be cleaned up if you undo {\em exactly}
the sequence of actions that produced the mess.  Normal forward time evolution
doesn't do this.

\figfig{refract}{ \hfill \hbox{%
\epsfig{figure=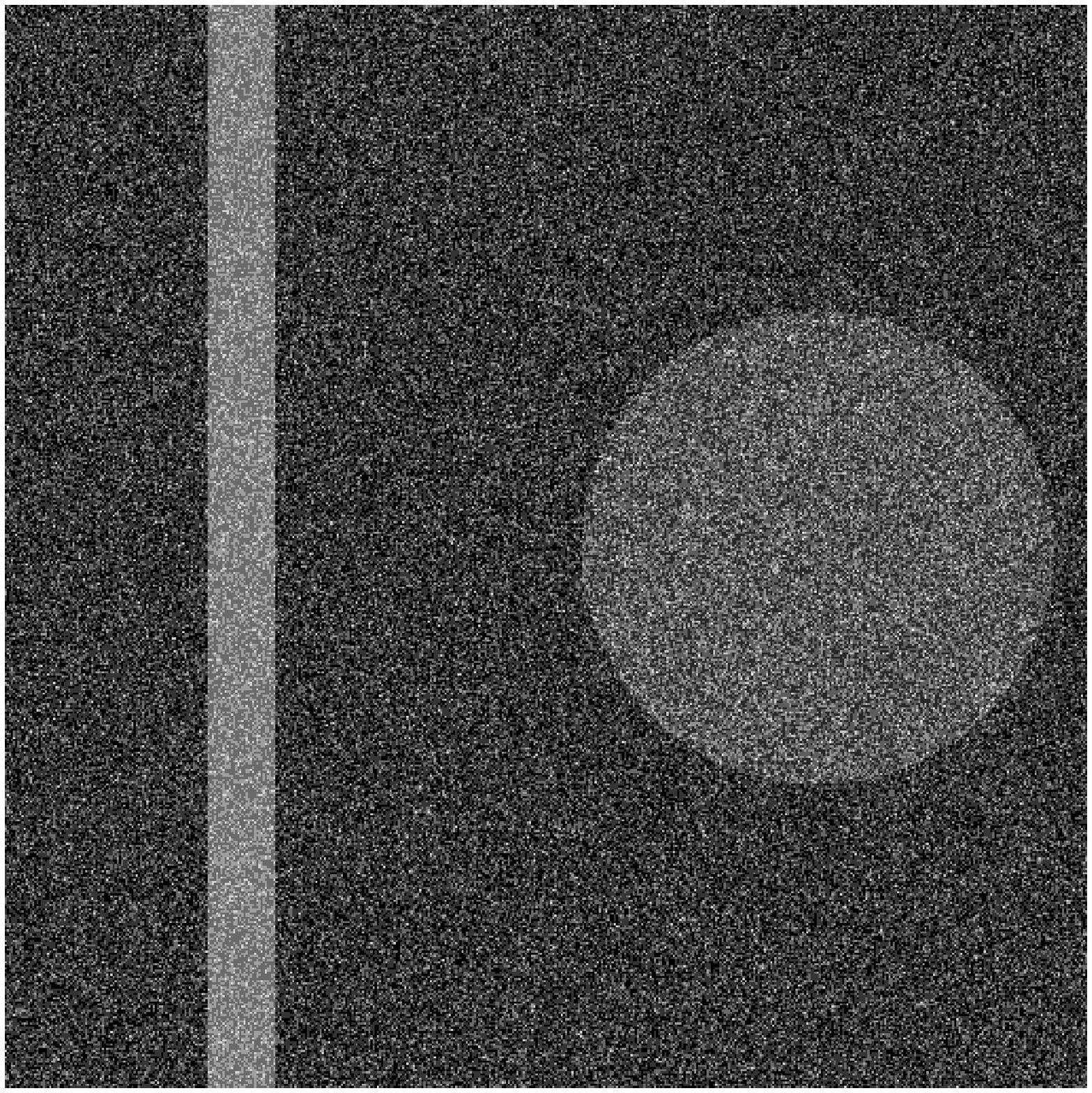,height=1.5in}}\hskip .2in \hbox{%
\epsfig{figure=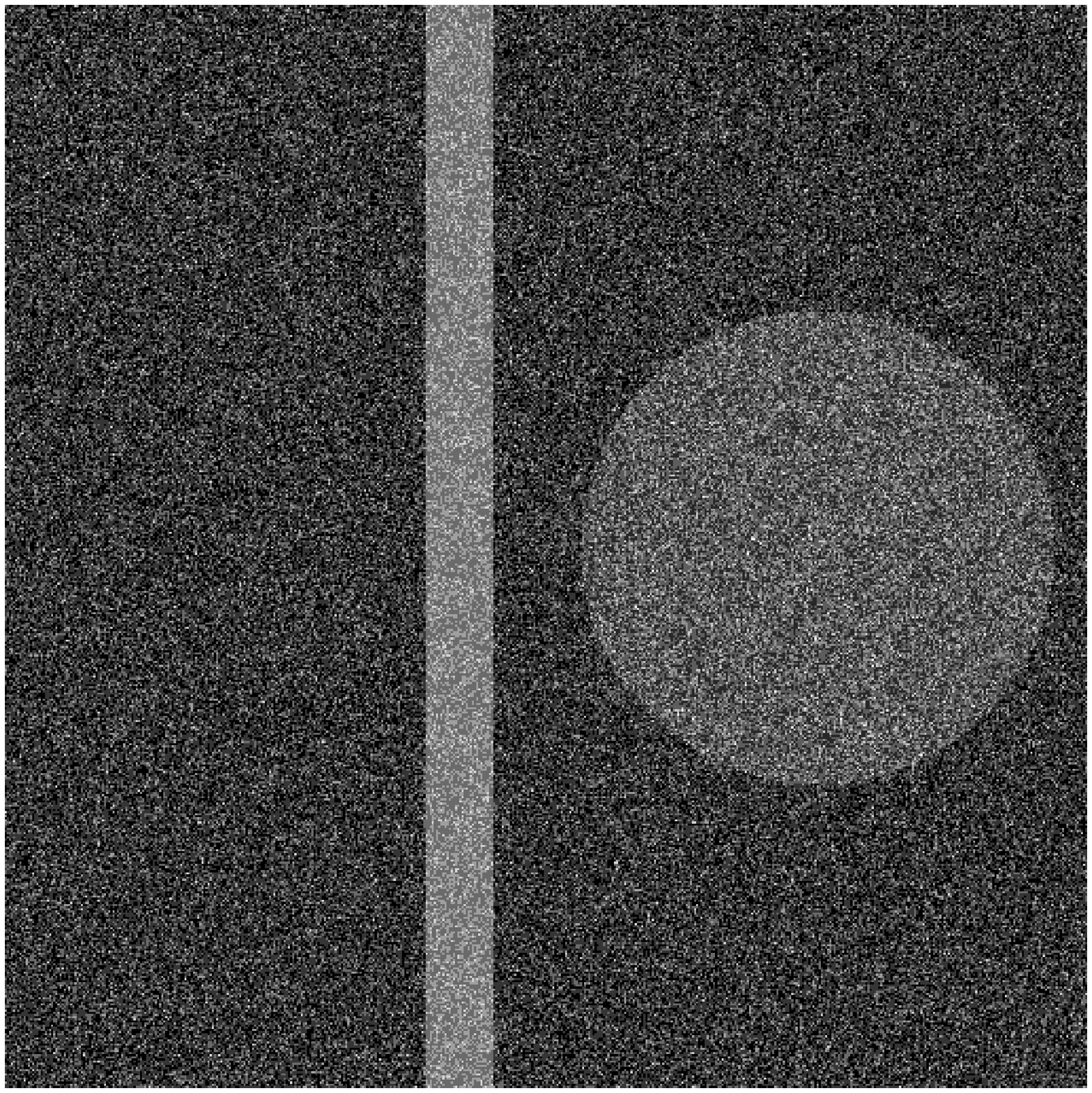,height=1.5in}}\hskip .2in \hbox{%
\epsfig{figure=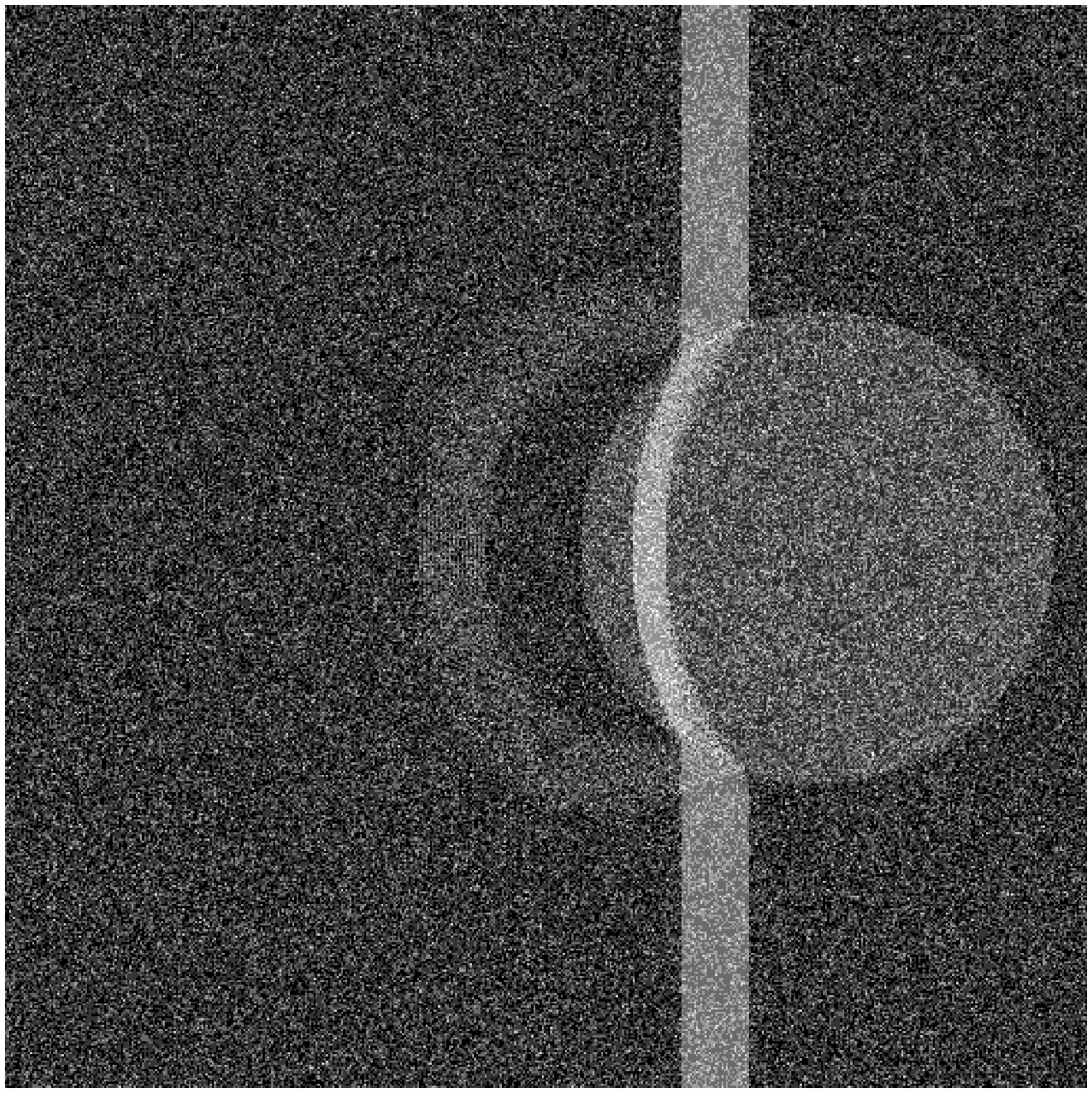,height=1.5in}}\hfill}%
{A slightly modified gas CA.  The vertical bar
is a density wave traveling to the right, the circle is a region in
which waves are slower.  We see the wave reflect and refract.}

Once we are able to model one macroscopic phenomenon, it is often
obvious how to model other phenomena.  Starting from a model with
sound waves, we can make a model with reflection and refraction of
such waves.  In Figure~\ref{fig.refract} we show a simulation using a
2-bit variant of the HPP CA.  Here we have added a bit to each site,
and used it to mark a circular region of the space: one bit at each
site is a {\em gas} bit, and the other bit is a {\em mark} bit.  We
now alternate the rule with time so that, for one complete
even-time/odd-time update of the lattice we apply the HPP rule to the
gas bits at all sites; then we do a complete even/odd update only of
unmarked blocks, with all gas particles in blocks containing non-zero
mark-bits left unchanged.  This gives us two connected HPP systems,
one running half as fast as the other.  In particular, waves travel
half as fast in the marked region, and so we get refraction of the
wave that is incident from the left.  Notice that the dynamics is
still perfectly invertible, and that we can make our ``lens'' any
shape we desire---it is a general feature of CA MD that we can simply
``draw'' whatever shaped obstacles and potentials we need in a
simulation\cite{smith-topol}.  Related LGA techniques have been used
for complex antenna simulations\cite{simons}.

We can model many other phenomena by coupling 2$\times$2 block rules.
We can, for example, use the HPP gas or a finite-impact-parameter
variant of it (the TM gas\cite{super,cambook}, which has better
momentum-mixing behavior) as a source of pseudo-randomness in a
diffusion model.  We start by again putting two bits at each lattice
site---one bit will belong to the {\em diffusing system}, while the
other belongs to the {\em randomizing system}.  Let the four bits of
the randomizing system in each block simply follow the HPP dynamics
described above.  Let the four bits of the diffusing system be rotated
$90^\circ$ clockwise or counterclockwise, depending on the parity of
the number of 1's in the four ``random'' bits.  This results in a
perfectly invertible diffusion in which no more than one diffusing
particle can ever land at the same site\cite{cambook}.  Using this
approach with enough bits at each site we can, for example, model the
diffusion and interaction of different chemical species.

The HPP CA was originally presented in a different format than we have
used above\cite{hpp}.  Since this other format is in many cases very
natural for discussing MD models, we will describe it here, and then
relate it to a partitioned description.  We start by putting four bits
of state at each site of a square lattice.  We will call the bits at
the $i^{\rm th}$ site $N_i$, $S_i$, $E_i$ and $W_i$.  The dynamics
consists of alternating two steps: (1) {\em move} the data, and then
(2) let the groups of bits that land at each site {\em interact}
separately.  The first step moves the data in sheets: if we think of
the directions on our lattice as being North, South, East and West,
then all of the $N$ bits are moved one position North, all the $S$
bits one position South, etc.  Our interaction rule at each site
combines the bits that came from different directions and sends them
out in new directions.  A state consisting of two 1's (particles) that
came in from opposite directions and two 0's from the other
directions, is changed into a state in which the 1's and 0's are
interchanged---the particles come out at right angles to their
original directions.  In all other cases, particles come out in the
same directions they came in.

We can think of this as a particular kind of partitioning rule, where
the four bits at each site are the groups, and we use the
data-movement step to rearrange the bits into new groups.  Although in
some ways this {\em site-partitioned} description of the HPP gas is
simpler, it also suffers from a slight defect.  If we imagine, as we
did in our discussion of the Ising model, that our lattice is a giant
black and white checkerboard, then we notice that in one data movement
step all of the bits that land at black squares came from white
squares, and vice versa.  No data that is currently on a black square
will ever interact with data that is currently on a white square: we
have two completely independent subsystems.  The 2$\times$2 block
version of the HPP rule is isomorphic to just one of these subsystems,
and so lets us avoid simulating two non-interacting systems.  Of
course we can also avoid this problem in the site-partitioned version
with more complicated time-dependent shifts: we can always reexpress
any partitioned CA as a site-partitioned CA.  This fact has been
important in the design of our latest CA machines.

The HPP lattice gas produces a nice round-looking sound wave, but
doesn't reproduce 2D hydrodynamics in the large-scale limit.  We can
clearly make a CA model with more speeds and directions by having
molecules travel several lattice positions horizontally and/or
vertically at each step---just add more particles at each site, and
shift the different momentum fields appropriately during the movement
step.  With enough speeds and directions, it seems obvious that we can
get the right macroscopic limit---this should be very much like a
hard-sphere gas, which acts like a fluid.  The fact that so many
different fluids obey the same hydrodynamic equations also suggests
that the details of the dynamics can't matter very much, just the
constraints such as momentum and particle conservation.

\figfig{fhp2}{ \hfill \hbox{%
\epsfig{figure=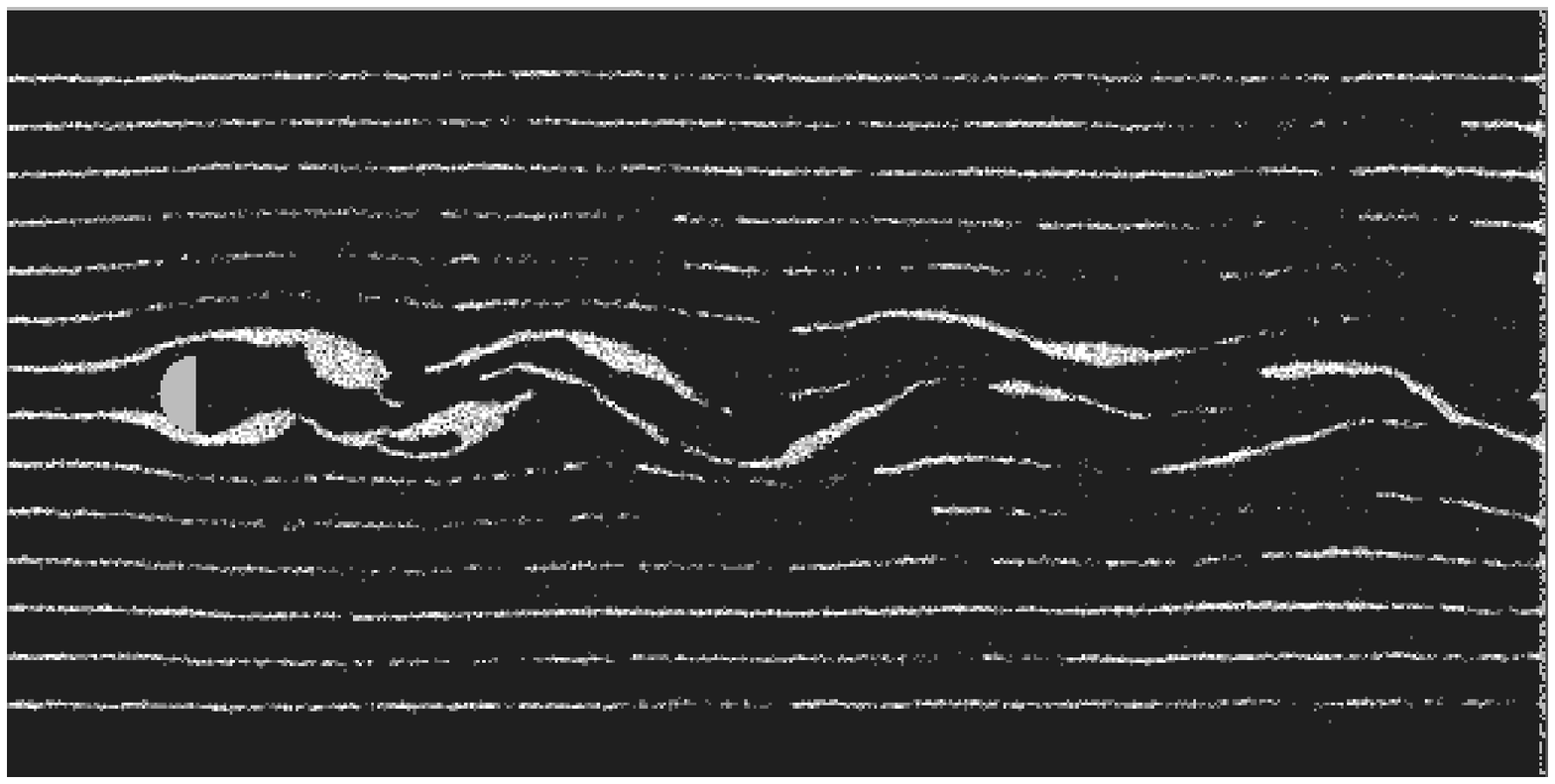,height=2in}}\hfill}%
{Flow past a half-cylinder using a six-direction
lattice gas on a triangular lattice.  We simulate ``smoke'' streamers
to visualize the flow.}

So how simple a model can work?  It was found\cite{fhp} that we can
recover macroscopic 2D hydrodynamics from a model that is only
slightly more complicated than the HPP gas.  A single-speed model with
six particles per site, moving in six directions on a triangular
lattice, will do.  If all zero-net-momentum collisions cause the
molecules at the collision site to scatter into a rotated
configuration, and otherwise particles go straight, then in the low
speed limit we recover isotropic macroscopic fluid dynamics.
Figure~\ref{fig.fhp2} shows a simulation of a slightly more
complicated six-direction lattice gas\cite{filga}.  The simulation
shown is 2K$\times$1K, and we see vortex shedding in flow past a
half-cylinder.  The white streamers are actually a second gas (more
bits per site!), inserted into the hydrodynamic gas as a kind of smoke
used to visualize flows in this CA wind tunnel.  This is an invertible
CA rule, except at the boundaries which are irreversibly being forced
(additional bits per site mark the boundaries).  Simple single-speed
CA's have also been used to simulate 3D
hydrodynamics\cite{doolen,rothman-cam}.

When it was discovered that lattice gases could simulate hydrodynamic
behavior, there was a great deal of excitement in some circles and
skepticism in others.  The exciting prospect was that by simplifying
MD simulations to the point where only the essence of hydrodynamic
behavior remained, one could extend the scale of these simulations to
the point where interesting hydrodynamics could be done directly with
an MD method.  This spawned an entire new field of
research\cite{doolen,doolen91,torino,nice,waterloo,princeton,boston-lga}.
This optimistic scenario has not yet been realized.  One problem is
that simple single speed models aren't well suited for simulating
high-speed flows.  As in a photon gas, the sound speed in a
single-speed LGA is almost the same as the maximum particle speed,
making supersonic flows impossible to simulate.  You need to add more
particle speeds to fix this.  The biggest problem, though, is that you
need truly enormous CA systems to get the resolution needed for
hydrodynamic simulations with high Reynold's numbers\cite{yakhot}.




\figfig{chem-xtal}{ \hfill \hbox{%
\epsfig{figure=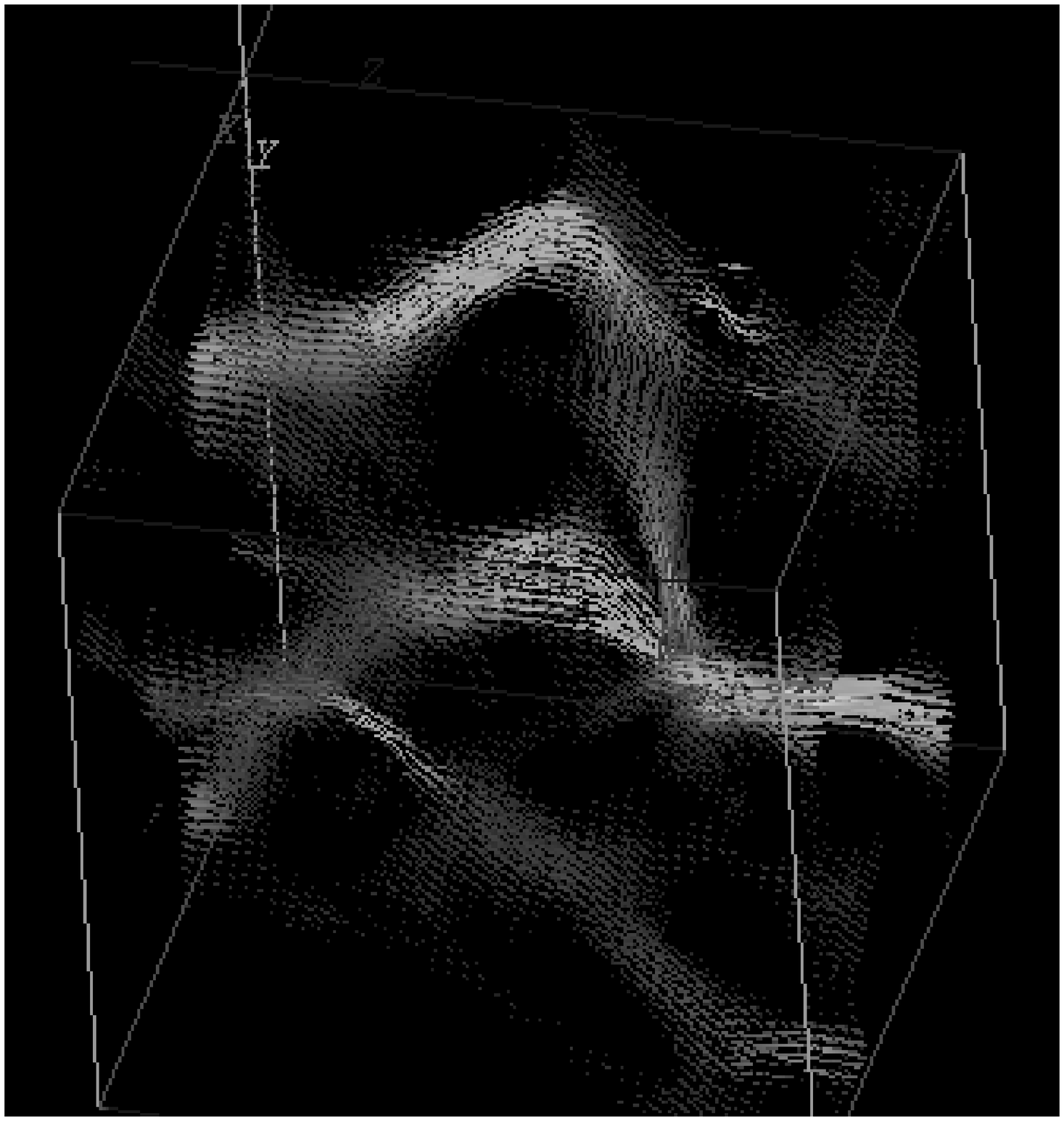,height=1.5in}}\hskip .2in \fbox{%
\epsfig{figure=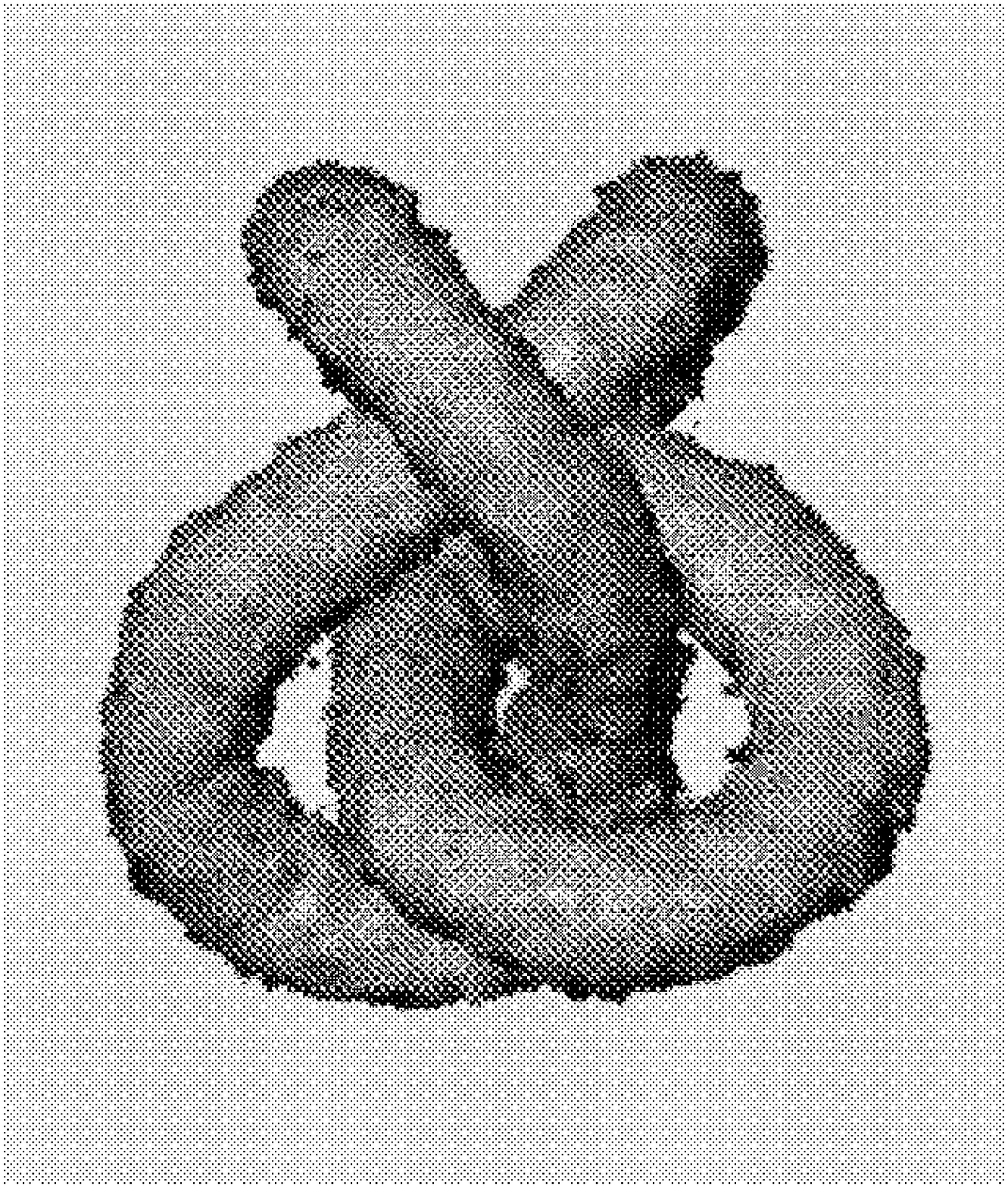,height=1.5in}}\hskip .2in \hbox{%
\epsfig{figure=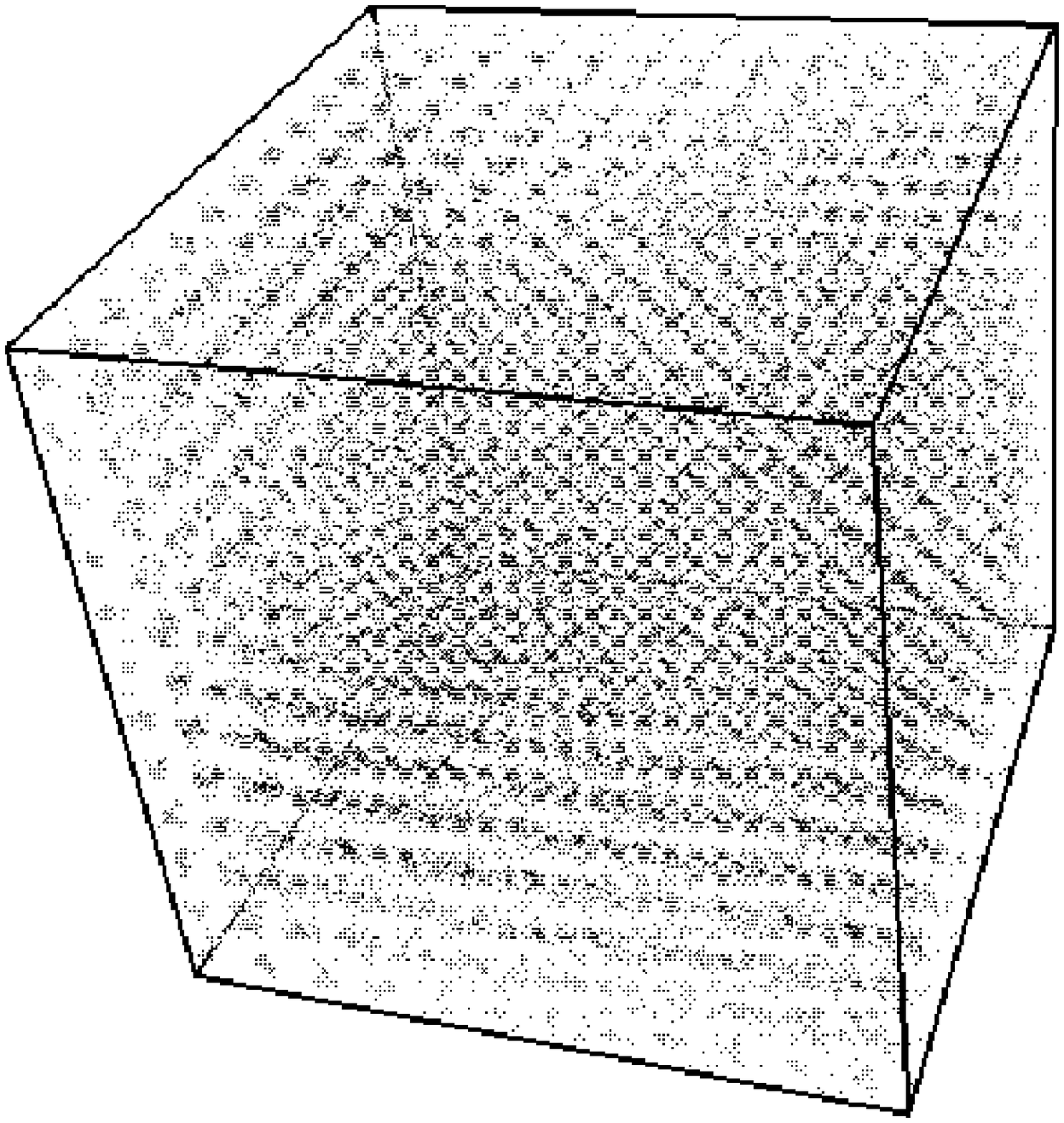,height=1.5in}}\hfill}%
{Some CA MD simulations.  (a)~Flow
through a porous medium.  (b)~A topologically complicated structure
within a chemical reaction simulation.  (c)~Crystallization
using irreversible discrete forces.}

For the near term, for those interested in practical modeling, it
makes sense to avoid high-Reynold's numbers and fast fluid flows, and
to use MD CA models to simulate other kinds of systems that are hard
to simulate by more conventional numerical techniques.  Suitable
candidates would include systems for which the best current simulation
techniques are in fact some form of molecular dynamics, as well as
systems for which there are at present no good simulation techniques
because traditional MD cannot reach the hydrodynamic regime.  An
example would be systems with very complicated flows.
Figure~\ref{fig.chem-xtal}a shows a simulated flow through a piece of
sandstone.  The shape of the sandstone was obtained from MRI imaging
of an actual rock, taking advantage of the ability of CA MD
simulations to handle arbitrarily shaped obstacles.  Shading in the
figure indicates flow velocity.  Simulations were compared against
experiments on the same rock that was imaged, and agreement was
excellent\cite{rothman-book,rothman-cam}.  More complicated flows,
involving immiscible liquids, have been simulated with this same
technique.

CA models of complex systems can be built up by combining simpler
models.  We simply pile up as many bits as we need at each lattice
site, representing as many fluids, random variables, heat baths, and
other fields as we desire.  Then we update them in some repeated
sequence, including extra steps that make different groups of
subsystems interact, much as we did in our diffusion and our
refraction examples.  For practical purposes we will often dispense
with invertibility, and be satisfied with irreversible rules coupled
to pseudo random subsystems.  Figure~\ref{fig.chem-xtal}b shows an
example of a 3D chemical reaction simulation of this sort, which
simulates the FitzHugh-Nagumo reaction-diffusion
dynamics\cite{kapral-pre}.  The knot and its surroundings are composed
of two different chemical phases.  The connectivity of the knot in
conjunction with domain repulsion keeps the knot from shrinking away.
Many kinds of multiphase fluids, microemulsions, and other complex
fluids have been simulated using related
techniques\cite{reaction-diffusion,microemulsion,rothman-book}.

We can easily add discrete forces by having particles at a discrete
set of vector separations interact.  If two such particles are heading
away from each other we can point them toward each other and otherwise
leave them unchanged---this results in an attraction.  This kind of
rule isn't invertible, but it is energy and momentum conserving.
Figure~\ref{fig.chem-xtal}c shows a 3D crystallization simulation
using a potential built up out of such
interactions\cite{yepez-crystal}.  This is not currently a very
practical way to simulate crystals, but this kind of technique is
generally useful\cite{apert,yepez-blobs}.  For example, the ``smoke''
in Figure~\ref{fig.fhp2} has a weak cohesive force of this kind, which
makes the smoke streams thinner.

There are many other ways to build new CA MD models.  We often appeal
to microdynamical analogy, or to simulating ``snapshots'' of a
hypothetical continuous dynamics.  We can take aspects of existing
atomistic models and model them statistically at a higher level of
aggregation using exact integer counts and conservations, to avoid any
possibility of numerical instability\cite{filga}.  We can combine CA's
with more traditional numerical mesh techniques, using discrete
particles to handle difficult interface regions\cite{interface}.  We
can adapt various energy-based techniques from statistical
mechanics\cite{gunn,microemulsion}.  We can also build useful models in a less
systematic manner, justifying their use by simulation and careful
measurements\cite{rothman-book}.  Combining well understood CA MD
components to build up simulations of more complex systems is a kind
of iterative programming exercise that involves testing components in
various combinations, and adjusting interactions.

Although there is already a role for CA MD models even on conventional
computers, there is a serious mismatch on such machines between
hardware and algorithms.  If we are going to design MD simulations to
fit into a CA format, we should take advantage of the uniformity and
locality of CA systems, which are ideally suited to efficient and
large-scale hardware realization.

\section{Crystalline computers}

Computer algorithms and computer hardware evolve together.  What we
mean by a good algorithm is that we have found some sort of efficient
mapping between the computation and the hardware.  For example, CA and
other lattice algorithms are sometimes ``efficiently'' coded on
conventional machines by mapping the parallelism and uniformity of the
lattice model onto the limited parallelism and uniformity of word-wide
logical operations---so-called ``multi-spin coding.''  This is a
rather grotesque physical realization for models that directly mimic
the structure and locality of physics: we first build a computer that
hides the underlying spatial structure of nature, and then we try to
find the best way to contort our spatial computation to fit into that
mold!

Ultimately all physical computations have to fit into a spatial mold,
and so our most efficient computers and algorithms will eventually
have to evolve toward the underlying spatial ``hardware'' of
nature\cite{toffoli-nature}.  Because physical information can travel
at only a finite 
velocity, portions of our computation that need to communicate quickly
must be physically located close together.  Computer architects can
only hide this fact-of-life from us for so long.  At some point, if we
want our computations to run faster, our algorithms must take on the
responsibility of dealing with this constraint.

Computer engineers are not unaware of this spatial constraint.
Various locally-interconnected parallel computers have been built and
studied\cite{ftl}.  Mesh architectures are organized like a kind of
CA, but usually with a rather powerful computer with a large memory at
each lattice site.  Unlike CA's, they normally don't have the same
operation occurring everywhere in the lattice at the same time.  {\em
SIMD} or {\em data parallel} mesh machines are more CA-like, since
they typically have a simpler processor at each site, and they do
normally have the operation of all processors synchronized in perfect
lockstep.


Another important spatial computing device is the {\em gate array}.
These regular arrays of logic elements are very much like a universal
CA.  Initially, we build these chips with arrays of logic elements,
but we leave out the wiring.  Later, when we need a chip with a
specific functionality, we can quickly ``program'' the gate array by
simply adding wires to connect together these elements into an
appropriate logic circuit.  FPGA's (field programmable gate arrays)
make programming the interconnections even easier.  What should be
connected to what is specified by some bits that we communicate
directly to the chip: this rapid transition from bits to circuitry
eliminates much of the distinction that is normally made between
hardware and software\cite{fccm}.

As general purpose computing devices, none of these CA-like machines
are significant mainstream technologies---the evolutionary forces that
are pushing us in the CA direction haven't yet pushed hard enough.
This was even more true when Tom Toffoli and I first started playing
with CA's together, almost two decades ago.  There were no machines
available to us then that would let us run and display CA's quickly
enough that we could experience them as dynamical worlds.  We became
increasingly frustrated as we tried to explore the new and exciting
realm of invertible CA's on available graphical workstations: each
successive view of our space took minutes to compute.

\figfig{cams}{ \hfill \fbox{%
\epsfig{figure=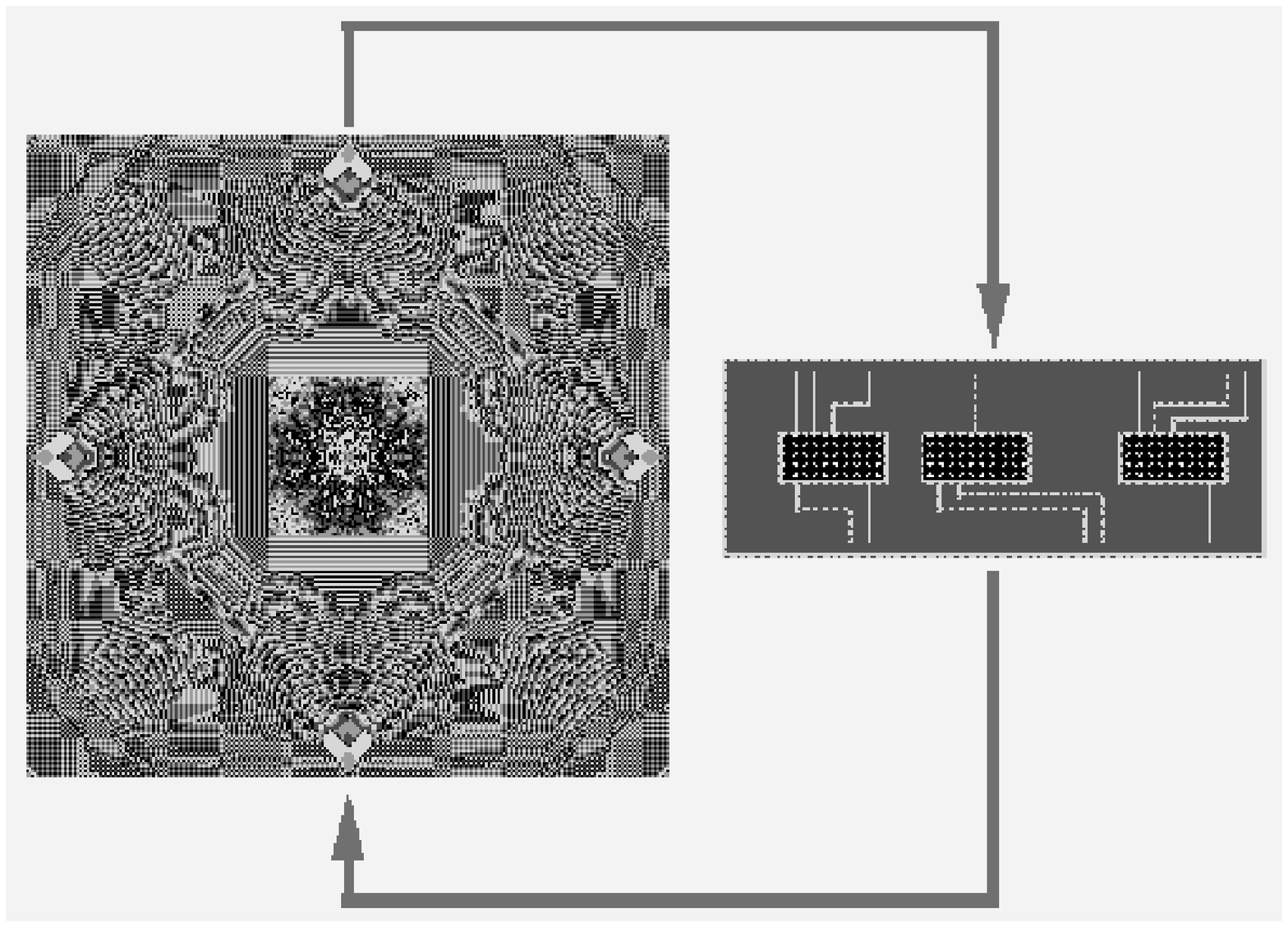,height=1.9in}}\hskip .4in \fbox{%
\scalebox{.864}{\includegraphics*[1.65in,7.15in][3.85in,9.35in]{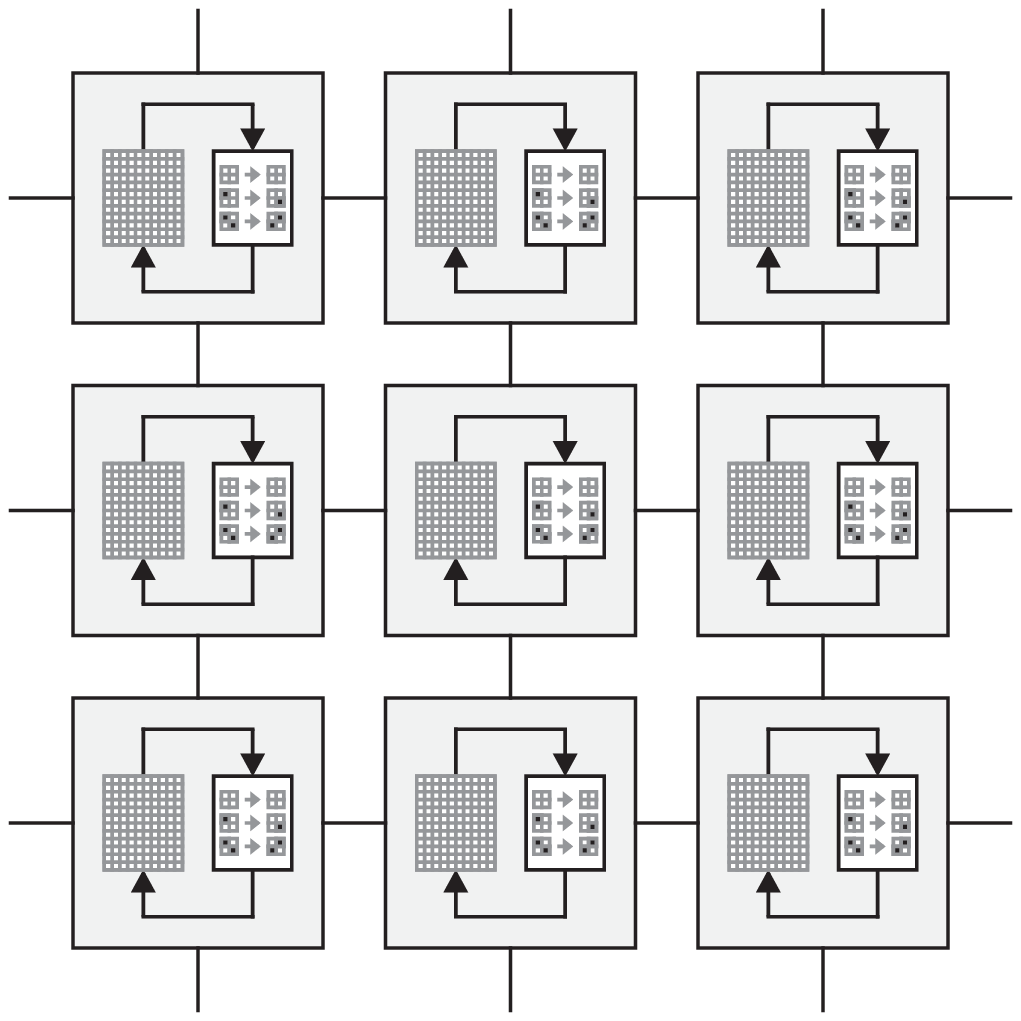}}}\hfill}%
{CA machines.  (a)~Our earliest CA machines scanned their memory
like a framebuffer, applying discrete logic to a sequence of
neighborhood windows. (b)~CAM-8 uses a 3D mesh array of SIMD
processors.  Each processor handles part of the lattice, alternating
data movement with sequential lookup-table updating of the groups of
bits that land at each lattice site.} 

Tom designed and built our first CA simulation hardware.  This CA
machine was a glorified frame-buffer that kept the CA space in memory
chips.  As it scanned a 2D array of pixels out of the memory, instead
of just showing them on a screen, it first applied some neighborhood
logic to the data, then sent each resulting pixel to the screen while
also putting it back into memory.  At first the neighborhood logic was
on a little prototyping board (this is shown schematically in
Figure~\ref{fig.cams}a).  I learned about digital circuitry by
building site-update rules out of TTL chips---each small logic circuit
I built became the dynamical law for a different video-rate universe!
Eventually, we switched to lookup tables, which largely replaced the
prototype circuits.  The first few generations of machines, however,
all had wires that you could move around to change things---whenever
you wanted to demonstrate something, you would invariably find that
someone else had rewired your machine\cite{cam5}.

We went through several generations of hardware, playing with models
that no one else had ever seen\cite{cambook}.  At a certain point we
rediscovered the HPP lattice gas, and our simulations rekindled
interest in this kind of model.  At this point, our machines became
inadequate.  They had been designed for interacting and experimenting
with small 2D systems that you could watch on a monitor.  Real CA MD
was going to require large-scale 3D machines designed for serious
scientific simulation, with provisions for extensive data analysis and
visualization.  A new dedicated CA MD machine could be about 1000
times as cost effective as existing supercomputers for this task, and
would provide the interactivity and flexibility of a personal CA
workstation.


I designed the new architecture (CAM-8) based on the experience that
Tom and I had with our earlier machines\cite{margolus-waterloo}.  As
shown in Figure~\ref{fig.cams}b, this machine uses a 3D mesh array of
SIMD processors running in perfect lockstep.  The lattice being
simulated (which can be $n$ dimensional) is divided up evenly among
the processors, each processor handling an equal {\em sector} of the
overall lattice.  As in our earlier machines, the state of the lattice
is kept in ordinary memory chips, and the updating is done by lookup
tables---also memory chips.  Data cycles around and around between
these two sets of memory chips.

Unlike our previous machines, which provided a fixed set of
traditional CA neighborhoods, the only neighborhood format supported
by CAM-8 is {\em site-partitioning} (discussed in
Section~\ref{sec.md}).  Any {\em bit field} (i.e., set of
corresponding bits, one from every site) can be shifted uniformly
across the lattice in any direction.  Whatever data land at a given
lattice site are updated together as a group.  This is a particularly
convenient neighborhood format from a modeling point of view, since
any CA dynamics on any neighborhood can be accomplished by performing
an appropriate sequence of data shifts and site updates, each acting
on a limited number of site bits at a time.  Also, as we've seen,
partitioning is a particularly good format for constructing models
that incorporate desired physical properties.

Site partitioning is also a very convenient neighborhood format from a
hardware standpoint.  Since the pattern of data movement is very
simple and regular, mesh communication between processors is also very
simple.  Since the updating is done on each site independently, it
doesn't matter what order the sites are updated in, or how many
different processors are involved in the updating.  All of this can be
organized in the manner most efficient for the hardware.

CAM-8 machines were built and performed as expected.  All of the
simulations depicted in this chapter were done on CAM-8, except for
that of Figure~\ref{fig.chem-xtal}a (which is similar to
\cite{rothman-cam}).  CAM-8 has not, unfortunately, had the impact
that we hoped it might.  First of all, during the time that we were
building the machine, it was found that lattice gases weren't as well
suited for high Reynold's number hydrodynamic flow simulations as
people had hoped.  In addition, in the absence of any good CA
machines, interest in lattice dynamics calculations had shifted to
techniques that make better use of the floating point capabilities of
conventional computers.  Also, most researchers interested in
developing practical applications already had good access to
conventional supercomputers, which were 1000 times less cost effective
than CAM-8, but had familiar and high-quality software and system
support.  Finally, the evolutionary forces favoring CA-like machines
were temporarily on the wane at the time when CAM-8 was completed, as
multiprocessor funding dried up and fine-grained parallel computing
companies folded.  We didn't build any versions of our indefinitely
scalable CAM-8 machine that were large enough to make previously
unreachable realms of CA modeling accessible---as our early machines
first opened the CA world to us.

In the near term, prospects again look good for CA machines.  Although
our small personal-computer-scale CAM-8 machines are still about as
good as any supercomputer for LGA computations, advances in technology
make radical improvements possible.  By putting logic directly on DRAM
memory chips, which is now routinely done, and by exploiting the
enormous memory bandwidth that can be made available on-chip, it is
possible today to make a SIMD machine similar to the CAM-8 that is
over 10,000 times faster {\em per memory chip} than the current CAM-8
hardware\cite{margolus-hsc}.  Putting together arrays of such chips,
qualitatively new physical simulations will become possible.  Other
SIMD applications such as logic simulation, image processing and 3D
bit-map manipulation/rendering will also run at about a trillion
bit-operations per second per chip.  Whether we manage to make our
next dream machine, the time is ripe for commercial SIMD-based CA
machines.

What of the more distant future?  In the preceding sections, we have
emphasized invertible CA's.  Aside from their intrinsic interest, they
have the virtue that they mimic the microscopic invertibility of
physical dynamics.  From a macroscopic point of view, this means that
these CA's can in principle be implemented using frictionless
reversible mechanisms---they don't depend on dissipative
thermodynamically irreversible processes in order to
operate\cite{frank-thesis,vieri-thesis,younis-knight,younis-thesis}.  Thus 3D
machines based on invertible CA's won't have the same problem getting
rid of dissipated heat that irreversible machines
do\cite{bennett-thermo,bennett-fund,feynman-comp}.  From a more
microscopic point of view, we can see the match between invertible
computation and invertible quantum physics as making possible direct
use of quantum scale elements and processes to do our computations.
We can make use of discrete properties of these quantum elements to
represent our digital quantities and perform our digital computations.

Thus in the distant future I expect that our most powerful large-scale
{\em general purpose} computers will be built out of macroscopic
crystalline arrays of identical invertible computing elements.  We
would make such large arrays out of identical elements because they
will then be easier to control, to design, to build and to test.
These will be the distant descendants of todays SIMD and FPGA
computing devices: when we need to perform inhomogeneous computations,
we will put the irregularities into the program, not the hardware.
The problem of arranging the pieces of a computation in space will be
part of the programming effort: architectural ideas that are used
today in physical hardware may reappear as data structures within this
new digital medium.  With molecular scale computing elements, a small
chunk of this {\em computronium}\cite{margolus-fund} would have more
memory and processing power than all of the computers in the world
today combined, and high Reynold's number CA MD calculations of fluid
flow would be practical on such machines.

Note that I don't expect our highest performance general purpose
computers to be quantum spin computers of the sort discussed in
Section~\ref{sec.intro}.  In such a machine, the whole computer
operates on a superposition of distinct computations simultaneously.
This kind of {\em quantum parallelism} is very delicate, and the
overhead associated with the difficult task of maintaining a
superposition of computations over a large spatial scale will be such
that it will only be worth doing in very specialized situations---if
it is possible at all\cite{preskill}.  This won't be something that we
will do in our general purpose computers.

\section{What makes a CA world interesting?}

Future CA machines will make extensive large-scale CA simulations
possible---we will be able to study the macroscopic properties of CA
worlds that we design.  Aside from issues of size and speed, there
doesn't seem to be any obvious reason why exact classical information
models cannot achieve as high a level of rich macroscopic complexity
as we see in our universe.  This is a very different modeling
challenge than trying to simulate quantum mechanics with CA's.  We
would like to simulate an interesting macroscopic world which is built
out of classical information.  It is instructive to try to see what
the difficulties might be.

The most important thing in our universe that makes it interesting to
us is of course {\em us}.  Or more generally, the existence of complex
organisms.  Thus let's begin by seeing what it might take to simulate
a world in which Darwinian evolution is {\em likely} to take place.
Since no one has yet made a CA that does this, our discussion will be
quite speculative.

One of the most successful computer models of evolution is Tom Ray's
Tierra\cite{ray1,ray2}, which was designed to capture---in an exact
digital model---an essential set of physical constraints abstracted
{from} biology.  His model did not include spatial locality or
invertibility, but we could try to add these features.


Modeling evolution in a robust spatial fashion may, however, entail
incorporating some physical properties into our CA systems that are
not so obvious\cite{bridge}.  For example, in nature we have the
property that we can take complicated objects and set them in motion.
This property seems to be essential for robust evolution: it is hard
to imagine the evolution of complex organisms if simpler pieces of
them can't move toward each other!  No known universal CA has this
property (but see \cite{chopard-book,apert,yepez-blobs}).  There is
nothing in the Life CA, for example, that corresponds to a glider-gun
in motion.

The general property of physics that allows us to speak about an
object at rest and then identify {\em the same object} in motion is
{\em relativistic invariance}.  The fact that the laws of physics look
the same in different relativistic frames means that we can have the
same complex macroscopic objects in all states of motion: an
organism's chemistry doesn't stop working if it moves, or if the place
it lives moves!  In a spatial CA dynamics, some sort of spatial
macroscopic motion invariance would clearly make evolution more
likely.  Since our CA's have a maximum speed at which information can
travel---a finite speed of light---relativistic invariance is a
possible candidate.  Full relativistic invariance may be more than we
need, but it is interesting to ask, ``Can we incorporate relativistic
invariance into a universal CA model?''

We have already seen that we can have macroscopic rotational
invariance in our lattice gas models, and we know that numerical mesh
calculations of relativistic dynamics are possible.  Thus achieving a
good approximation of relativistic invariance in the macroscopic limit
for an exact CA model seems possible\cite{hrgovcic}.  Such a system
would, at least in the limit, have the conservations associated with
the continuous Lorentz group of symmetries.  Although it is not
possible to put a continuous symmetry directly into a discrete CA
rule, it is certainly possible to put these conservations into the
rule, along with a discrete version of the symmetries---just as we did
in our lattice gas models.\footnote{In continuum physics, continuous
symmetries are regarded as fundamental and conservations arise as a
consequence of symmetry.  Fredkin has pointed out that in discrete
systems, it must be the conservation that is fundamental.}

Thus we might imagine our relativistically invariant CA to be a
refinement of lattice gases---we would also like to make it invertible
for the reasons discussed in Section~\ref{sec.rev}.  But we also
demand that this CA incorporate computation universality.  This may
not be easy: since a relativistically invariant system must have
momentum conservation, we will need to worry about how to hold complex
interacting structures together.  Thus we may need to incorporate some
kind of relativistically invariant model of forces into our system.

Simulating forces in an exact and invertible manner is not so easy,
particularly if we want the forces to last for a long
time\cite{yepez-thesis}.  Models in which forces are communicated by
having all force-sources continuously broadcast field-particles have
the problem that the space soon fills up with these
field-particles---which cannot be erased because of local
invertibility---and then the force can no longer be communicated.
Directly representing field gradients works better, but making this
work in a relativistic context may be hard.

At this point, we might also begin to question our basic CA
assumptions.  We introduced crystalline CA's to try to emulate the
spatial locality of physics in our informational models, but we are
now discussing modeling phenomena in a realm of physics in which
modern theories talk about extra dimensions and variable topology.
Perhaps whatever is essential fits nicely into a simple crystalline
framework, but perhaps we need to consider alternatives.  We could
easily be led to informational models in which the space and time of
the microscopic computation becomes rather divorced from the space and
time of the macroscopic phenomena.

\smallskip

We started this section with the (seemingly) modest goal of using a CA
to try to capture aspects of physics necessary for a robust evolution
of interesting complexity, and we have been led to discuss
incorporating larger and larger chunks of physics into our model.
Perhaps our vision is too limited, and there are radically different
ways in which we can have robust evolution in a spatial CA model.  Or
perhaps we can imitate nature, but cheat in various ways.  We may not
need full relativistic invariance.  We may not need exact
invertibility.  On the other hand, it is also perfectly possible that
we can't cheat very much and still get a system that's nearly as
interesting as our world.

\section{Conclusion}

I was in high school when I first encountered cellular automata---I
read an article about Conway's ``Game of Life.''  At that time I was
intensely interested in both Physics and Computers, and this game
seemed to combine the two of them.  I immediately wondered if our
universe might be a giant cellular automaton.

The feeling that physics and computation are intimately linked has
remained with me over the years.  Trying to understand the
difficulties of modeling nature using information has provided an
interesting viewpoint from which to learn about physics, and also to
learn about the ultimate possibilities of computer hardware.  I have
learned that many properties of macroscopic physics can be mirrored in
simple informational models.  I have learned that quantum mechanics
makes both the amount and the rate-of-change of information in
physical systems finite---all physical systems do finite information
processing\cite{margolus-speed}.  I have learned that the
non-separability of quantum systems makes it hard to model them
efficiently using classical information---it is much easier to
construct quantum spin computer models\cite{lloyd-sim}.  I have
learned that physics-like CA models can be the best possible
algorithms when the computer hardware is also adapted to the
constraints and structure of physical dynamics.  I have learned that
developing computer hardware that promotes this viewpoint can consume
an enormous amount of time!

Since classical information is much easier to understand than quantum
information I have mostly studied classical CA models.  In these
systems, a macroscopic dynamical world arises from {\em classical}
combinatorics.  Continuous classical-physics behavior can emerge in
the large-scale limit.  We can try to model and understand (and
perhaps teach our students about) all sorts of physical phenomena
without getting involved in quantum complications.  We can also try to
clarify our understanding of the fundamental quantities, concepts and
principles of classical mechanics {\em and of classical computation}
by studying such systems\cite{fredkin-comp}.  The principle of
stationary action, for example, must arise in such systems solely from
combinatorics---there is no underlying quantum substratum in a CA
model.  Conversely, we should remember that information (in the guise
of entropy) was an important concept in physics long before it was
discovered by computer scientists.  Just as Ising-like systems have
provided intuitive classical models that have helped clarify issues in
statistical mechanics, CA's could play a similar role in dynamics.

Since we have focused so much on discrete classical CA models of
physics, it might be appropriate to comment briefly on their
relationship to discrete quantum models---Feynman's quantum spin
computer of Section~\ref{sec.intro}.  Exactly the same kinds of
grouping and sublattice techniques that we have used to construct
invertible CA's also allow us to construct quantum
CA's---QCA's\cite{margolus-qc}.  We simply replace invertible
transformations on groups of bits with unitary transformations on
groups of spins.  Just as it is an interesting problem to try to
recover classical physics from ordinary CA's, it is also interesting
to try to find QCA's that recover the dynamics of known quantum field
theories in the macroscopic
limit\cite{boghosian-qc,meyer-qlga,meyer-boston,yepez-thesis}.
Following our Ising CA example, it might be instructive to investigate
classical CA's that are closely related to such QCA's.

Although people have often studied CA's as abstract mathematical
systems completely divorced from nature, ultimately it is their
connections to physics that make them so interesting.  We can use them
to try to understand our world better, to try to do computations
better---or we can simply delight in the creation of our own toy
universes.  As we sit in front of our computer screens, watching to
see what happens next, we never really know what new tricks our CA's
may come up with.  It is really an exploration of new worlds---live
television from other universes.  Working with CA's, anyone can
experience the joy of building simple models and the thrill of
discovering something new about the dynamics of information.  We can
all be theoretical physicists.

\section{Acknowledgments}

Much of what I know of this subject I learned from Edward Fredkin,
Tommaso Toffoli and Charles Bennett.  They were my close collaborators
in the MIT Information Mechanics Group and my dear friends.  They
provided me with the support and encouragement I needed to begin work
in this field.  It was a privilege to be a part of their group.

Richard Feynman has been a constant source of inspiration and insight
in my physics career.  His {\em Lectures on Physics}\cite{feynman-lec}
sustained me through my undergraduate years, and I spent some
wonderful months as a graduate student visiting with him at CalTech
and giving lectures on CA's and on reversible computing in his course
on computation\cite{feynman-comp}.  I have many fond memories of his
warmth, charm, humor and {\em joie de vivre}.

I learned much from conversations with Mark Smith, Mike Biafore,
Gerard Vichniac and Hrvoje Hrgov\u{c}i\'{c}---colleagues in the MIT IM
Group.  I have also learned much from my colleagues in the lattice gas
field who have helped me in my research, particularly Jeff Yepez, who
has been a close collaborator on CA research with the CAM-8 machine.
The construction of CAM-8 itself was a large effort that involved many
people---including Tom Toffoli, Tom Durgavich, Doug Faust, Ken
Streeter and Mike Biafore---each of whom contributed in a vital way.
My current 
physics/computation related research has been made possible by the
support, collaboration and encouragement of Tom Knight.

This manuscript has benefitted greatly from discussions with and
comments by Raissa D'Souza, Lov Grover, David Meyer, and Ilona Lappo.
Thanks to Dan Rothman for providing Figure~\ref{fig.chem-xtal}a, Ray
Kapral for providing Figure~\ref{fig.chem-xtal}b, and Jeff Yepez for
providing Figure~\ref{fig.chem-xtal}c.  Figure~\ref{fig.other-ising}b
comes from a simulation written by Tom Toffoli and
Figure~\ref{fig.same} comes from an investigation that I'm working on
jointly with Raissa D'Souza and Mark Smith.

Support for this work comes from DARPA contract DABT63-95-C-0130, the
MIT AI Lab's Reversible Computing Project, and Boston University's
Center for Computational Science.

\section*{Notes on the references}

Much of the material in this chapter is
discussed at greater length in \cite{margolus-thesis} and
\cite{cambook}.  These documents were both strongly influenced by
ideas and suggestions from Edward Fredkin, some of which are also
discussed in \cite{fredkin-bbm,fredkin-dm,fredkin-comp,fredkin-nature,fredkin-cosmogony,fredkin-perspective}.  Some
recent books on CA modeling of physics are \cite{rothman-book} and
\cite{chopard-book}.  Many of the early lattice-gas and quantum
computing papers are reproduced in \cite{doolen} and \cite{lloyd}
respectively.  Recent related papers can be found online in the {\tt
comp-gas} and {\tt quant-ph} archives at {\tt http://xxx.lanl.gov},
and cross-listed there from other archives at LANL such as {\tt
chao-dyn}.  Pointers to papers in these archives are given in some of
the references below.


\begin{thebibliography}{199}

        \bibitem{lloyd-sim} D.~S. Abrams and S. Lloyd,
``Simulation of many-body fermi systems on a universal quantum
computer,'' {\em Phys.\ Rev.\ Lett.\ \bf 79}, 2586--2589 (1997) and
{\tt quant-ph/9703054}.
        \bibitem{rothman-cam} C. Adler, B.  Boghosian, 
E. Flekkoy, N. Margolus and D. Rothman, 
``Simulating three-dimensional hydrodynamics on a
cellular-automata machine,'' in \cite[p.\ 105--128]{princeton} and
{\tt chao-dyn/9508001}.
        \bibitem{apert} C. Appert, V. Pot and S. Zaleski,
``Liquid-gas models on 2D and 3D lattices,'' in \cite[p.\
1--12]{waterloo}.
        \bibitem{bennett-thermo} C.~H. Bennett, ``The thermodynamics
of computation---a review,'' in \cite[p.\ 905--940]{endicott}.
        \bibitem{bennett-fund} C.~H. Bennett and R. Landauer,
``The fundamental physical limits of computation,'' {\em Scientific
American \bf 253:}1,  38--46 (1985).
        \bibitem{bennett-demons} C.~H. Bennett, ``Demons,
engines, and the second law,'' {\em Scientific American \bf 257:}5, 88-96
(1987).
        \bibitem{life} E. Berlekamp, J. Conway and
R. Guy, {\em Winning Ways For Your Mathematical Plays, Volume 2}
(Academic Press, 1982).
        \bibitem{biafore} M. Biafore, ``Few-body cellular
automata,'' MIT Ph.D. Thesis (1993), available as MIT/LCS/TR-597 (MIT
Laboratory for Computer Science, 1993).
        \bibitem{microemulsion} B.~M. Boghosian, P. Coveney 
and A.~N. Emerton, ``A lattice-gas model of
microemulsions,'' {\em Proc.\ Roy.\ Soc.\ Lond.\ A \bf 452}, 1221--1250 (1996)
and {\tt comp-gas/9507001}.
        \bibitem{boston-lga} B.~M. Boghosian, F.~J. Alexander 
and P.~V. Coveney (eds.), {\em Discrete Models of
Complex Fluid Dynamics}, special issue of {\em Int.\ J.\ Mod.\ Phys.\ C  \bf
8}:4 (1997).
        \bibitem{boghosian-qc} B.~M. Boghosian and W. Taylor IV, 
``Quantum lattice-gas models for the many-body
Schr\"odinger equation,'' in \cite{boston-lga} and {\tt
quant-ph/9701016}.
        \bibitem{filga} B.~M. Boghosian, J. Yepez, F.~J.  Alexander 
and N. Margolus, ``Integer Lattice Gases,'' {\em
Phys.\ Rev.\ E.} {\bf 55}:4, 4137--4147 and {\tt comp-gas/9602001}.
	\bibitem{nice} J.~P. Boon (ed.), {\em Advanced
Research Workshop on Lattice Gas Automata}, special issue of {\em
J. Stat.\ Phys.\ \bf 68}:3/4 (1992).
        \bibitem{reaction-diffusion} J.~P. Boon, D. Dab,
R. Kapral and A. Lawniczak, ``Lattice gas automata for reactive
systems,'' {\em Phys.\ Rep.\ \bf 273}, 55--147 (1996) and {\tt
comp-gas/9512001}.
        \bibitem{brooks} R. Brooks and P. Maes (eds.),
{\em Artificial Life IV Proceedings} (MIT Press, 1994).
        \bibitem{burks} A. Burks, (ed.), {\em Essays on Cellular
Automata} (Univ.\ Illinois Press, 1970).
        \bibitem{paris} H. Chate (ed.), {\em International
Workshop on Lattice Dynamics,} special issue of {\em Physica D \bf
103:}1/4 (1997).
        \bibitem{chopard-book} B. Chopard, B. and M. Droz,
{\em Cellular Automata Modeling of Physical Systems}, (Cambridge
University Press, 1998).
        \bibitem{creutz} M. Creutz, ``Deterministic ising
dynamics,'' {\em Annals of Physics} {\bf 167}, 62--76 (1986).
        \bibitem{los-alamos} D. Farmer, T. Toffoli and
S. Wolfram (eds.), {\em Cellular Automata}, North-Holland
(1984); reprinted from {\em Physica D \bf 10}:1/2 (1984).
        \bibitem{raissa} R. D'Souza  and N. Margolus,
``Reversible aggregation in a lattice gas model using coupled
diffusion fields,'' preprint available in {\tt cond-mat/9810258}.
	\bibitem{doolen} G. Doolen (ed.), {\em Lattice-Gas
Methods for Partial Differential Equations} (Addison-Wesley, 1990).
	\bibitem{doolen91} G. Doolen (ed.), {\em Lattice-Gas
Methods for PDE's: Theory, Applications and Hardware} (North-Holland,
1991); reprinted from {\em Physica D \bf 47:}1/2.
        \bibitem{feynman-lec} R.~P. Feynman, R.~B. Leighton 
and M. Sands, {\em The Feynman Lectures on Physics}
(Addison-Wesley, 1963).
        \bibitem{feynman-sim} R.~P. Feynman, ``Simulating
physics with computers,'' in \cite[p.\ 467--488]{endicott}.
        \bibitem{feynman-comp} R.~P. Feynman, {\em Feynman
Lectures on Computation,} edited by J. G. {\sc Hey} and R. W. {\sc
Allen} (Addison-Wesley, 1996).
        \bibitem{frank-thesis} M.~P. Frank, ``Reversibility for
efficient computing,'' MIT Ph.D. Thesis (1999).
        \bibitem{fhp} U. Frisch, B. Hasslacher and
Y. Pomeau, ``Lattice-gas automata for the navier-stokes
equation,'' {\em Phys.\ Rev.\ Lett.} {\bf 56}, 1505--1508 (1986).
        \bibitem{endicott} E. Fredkin, R. Landauer
and T. Toffoli (eds.), {\em Proceedings of the Physics of
Computation Conference}, in {\em Int.\ J. Theor.\ Phys.}, issues {\bf
21}:3/4, {\bf 21}:6/7, and {\bf 21}:12 (1982).
        \bibitem{fredkin-bbm} E. Fredkin and T. Toffoli, ``Conservative logic,'' in \cite[p.\ 219--253]{endicott}.
        \bibitem{fredkin-dm} E. Fredkin, ``Digital mechanics: an
informational process based on reversible universal CA,'' in \cite[p.\
254--270]{gutowitz}.
        \bibitem{fredkin-comp} E. Fredkin, ``A physicist's model
of computation,'' {\em Proceedings of the XXVIth Recontre de Moriond,}
283--297 (1991).
        \bibitem{fredkin-nature} E. Fredkin, ``Finite nature,''
{\em Proceedings of the XXVIIth Recontre de Moriond}, (1992).
        \bibitem{fredkin-cosmogony} E. Fredkin, ``A new
cosmogony,'' in \cite[p.\ 116--121]{physcomp92}.
        \bibitem{fredkin-perspective} E. Fredkin, ``The digital
perspective,'' in \cite[p.\ 120--121]{physcomp96}.
        \bibitem{lwod} D. Griffeath and C. Moore, ``Life
without death is p-complete,'' Santa Fe Institute working paper
97-05-044, to appear in {\em Complex Systems} (1998); also available
at {\tt http://psoup.math.wisc.edu}, a general source of information
on pattern formation in irreversible CA's.
        \bibitem{grover} L.~K. Grover, ``A fast quantum
mechanical algorithm for database search,'' {\em Proceedings, $28^{\rm
th}$ Annual ACM Symposium on the Theory of Computing (STOC)}, 212--218 (1996)
and {\tt quant-ph/9605043}; reproduced in \cite{lloyd}.
        \bibitem{gunn} J.~R. Gunn, C.~M. McCallum and
K.~A. Dawson, ``Dynamic lattice-model simulation,'' {\em Phys.\
Rev.\ E}, 3069--3080 (1993).
        \bibitem{gutowitz} H. Gutowitz, {\em Cellular Automata:
Theory and Experiment} (North Holland, 1990); reprinted from {\em
Physica D \bf 45}:1/3 (1990). 
        \bibitem{hanson} J.~E. Hanson and J. P. Crutchfield, 
``Computational mechanics of cellular automata: an
example,'' in \cite[p.\ 169--189]{paris}.
        \bibitem{hpp} J. Hardy, O. de Pazzis and Y. Pomeau, ``Molecular dynamics of a classical lattice gas:
transport properties and time correlation functions,'' {\sl Phys.\
Rev. A \bf 13}, 1949--1960 (1976).
        \bibitem{herrmann} H.~J. Herrmann, ``Fast algorithm
for the simulation of Ising models,'' {\em J. Stat.\ Phys.\ \bf 45},
145--151 (1986).
        \bibitem{hrgovcic} H.~J. Hrgov\u{c}i\'{c}, ``Discrete
representations of the n-dimensional wave-equation,'' {\em J. Phys.\ A
\bf 25:}5, 1329--1350 (1992).
        \bibitem{joe-thesis} H.~J. Hrgov\u{c}i\'{c}, ``Quantum
mechanics on spacetime lattices using path integrals in a Minkowski
metric,'' MIT Ph.D. Thesis (1992), available online at {\tt
http://www.im.lcs.mit.edu/poc/hrgovcic/thesis.ps.gz}
        \bibitem{huang} K. Huang, {\em Statistical Mechanics},
(John Wiley \& Sons, 1987).
        \bibitem{ising} E. Ising, ``Beitrag zur theorie des
ferromagnetismus,'' {\em Zeits.\ f\"ur Phys.\ \bf 31}, 253--258 (1925).
        \bibitem{kari-uncomputable} J. Kari, ``Reversibility
and surjectivity problems of cellular automata,'' {\em Journal of
Computer and System Sciences \bf 48:}1, 149--182 (1994).
        \bibitem{kari-partitioning} J. Kari, ``Representation
of reversible cellular automata with block permutations,'' {\em
Mathematical Systems Theory \bf 29:}1, 47--61 (1996).
        \bibitem{kauffman} S.~A. Kauffman, ``Requirements for
evolvability in complex systems: orderly dynamics and frozen
components,'' in \cite[p.\ 151--192]{zurek}.
        \bibitem{interface} T. Kawakatsu  and K. Kawasaki,
``Hybrid models for the dynamics of an immiscible binary mixture with
surfactant molecules,'' {\em Physica A \bf 167:}3, 690--735 (1990).
        \bibitem{langton} C. Langton, C. Taylor,
J.~D. Farmer and S. Rasmussen (eds.), {\em Artificial Life
II, Santa Fe Institute Studies in the Sciences of Complexity,
vol. XI} (Addison-Wesley, 1991).
	\bibitem{waterloo} A. Lawniczak
and R. Kapral (eds.),  {\em Pattern Formation and Lattice-Gas
Automata} (American Mathematical Society, 1996).
        \bibitem{ftl} F.~T. Leighton, {\em Introduction to
Parallel Algorithms and Architectures: Arrays, Trees, Hypercubes}
(Morgan Kaufman, 1991).
        \bibitem{lloyd} S. Lloyd (ed.), {\em Quantum
Computation} (John Wiley \& Sons, 1999).
        \bibitem{lo} H.~K. Lo, T. Spiller and S. Popescu (eds.), 
{\em Introduction to Quantum Computation and
Information} (World Scientific, 1998).
        \bibitem{kapral-pre} A. Malevanets and R. Kapral,
``Microscopic model for FitzHugh-Nagumo dynamics,'' {\em Phys.\ Rev.\
E \bf 55:}5, 5657-5670 (1997).
        \bibitem{meyer-qlga} D.~A. Meyer, ``From quantum
cellular automata to quantum lattice gases,'' {\em J. Stat.\ Phys.\
\bf 85}, 551--574 (1996) and in {\tt quant-ph/9604003}.
        \bibitem{meyer-boston} D.~A. Meyer, ``Quantum lattice
gases and their invariants,'' in \cite[p.\ 717--736]{boston-lga} and
{\tt quant-ph/9703027}.
	\bibitem{super} N. Margolus, T. Toffoli
and G. Vichniac, ``Cellular-automata supercomputers for
fluid dynamics modeling,'' {\em Phys.\ Rev.\ Lett.} {\bf 56}, 1694--1696 (1986).
        \bibitem{margolus-bbm} N. Margolus, ``Physics-like
models of computation,'' in \cite[p.\ 81--95]{los-alamos}; reproduced in \cite{lloyd}.
        \bibitem{margolus-qc} N. Margolus, ``Quantum
computation,'' {\sl New Techniques and Ideas in Quantum Measurement
Theory} (Daniel {\sc Greenberger} ed.), 487--497 (New York Academy of Sciences,
1986); reproduced in \cite{lloyd}.
	\bibitem{margolus-thesis} N. Margolus, ``Physics and
computation'' MIT Ph.D. Thesis (1987).  Reprinted as {\em Tech.\
Rep.\ MIT/LCS/TR-415}, MIT Lab.\ for Computer Science, Cambridge MA
02139 (1988). 
        \bibitem{margolus-pqc} N. Margolus, ``Parallel quantum
computation,'' in \cite[p.\ 273--287]{zurek}; reproduced in \cite{lloyd}.
        \bibitem{margolus-fund} N. Margolus, ``Fundamental
physical constraints on the computational process,'' {\sl
Nanotechnology: Research and Perspectives} (B.C. {\sc Crandall} and
J.\ {\sc Lewis} eds.) (MIT Press, 1992).
        \bibitem{bridge} N. Margolus, ``A bridge of bits,''
in \cite[253--257]{physcomp92}.
	\bibitem{margolus-waterloo} N. Margolus, ``CAM-8: a
computer architecture based on cellular automata,'' in \cite[p.\ 
167--187]{waterloo} and {\tt comp-gas/9509001}.  See also {\tt
http://www.im.lcs.mit.edu/cam8.html}.
        \bibitem{margolus-fccm} N.  Margolus, ``An FPGA
architecture for DRAM-based systolic computations,'' in \cite[p.\
2--11]{fccm}.
        \bibitem{margolus-speed} N. Margolus and L. Levitin, ``The
maximum speed of dynamical evolution,'' in \cite[p.\
188--195]{physcomp96b} and {\tt quant-ph/9710043}. 
        \bibitem{margolus-hsc} N. Margolus, ``Crystalline
computation,'' to appear in {\em Proceedings of the Conference on High
Speed Computing} (Lawrence Livermore National Laboratory, 1998).
        \bibitem{physcomp92} D. Matzke (ed.), {\em Proceedings
of the Workshop on Physics and Computation---PhysComp '92} (IEEE
Computer Society Press, 1993).
        \bibitem{torino} R. Monaco (ed.), {\em Discrete Kinetic
Theory, Lattice Gas Dynamics and Foundations of Hydrodynamics} (World
Scientific, 1989).
        \bibitem{poly} B. Ostrovsky, M.~A. Smith and
Y. Bar-Yam, ``Simulations of polymer interpenetration in 2D
melts,'' in \cite[p.\ 931--939]{boston-lga}.
        \bibitem{fccm} K. Pocek and J. Arnold, {\em The
Fifth IEEE Symposium on FPGA-based Custom Computing Machines}
(IEEE Computer Society, 1997).
        \bibitem{pomeau-invariant} Y. Pomeau, ``Invariant in
cellular automata,'' {\em J. Phys.\ A \bf 17:}8, 415--418 (1984).
        \bibitem{preskill} J. Preskill, ``Fault-tolerant
quantum computation,'' in \cite{lo} and {\tt quant-ph/9712048}.
        \bibitem{princeton} Y.~H. Qian (ed.), {\em Discrete
Models for Fluid Mechanics}, special issue of {\em J. Stat.\ Phys.\ \bf
81}:1/2 (1995).
        \bibitem{ray1} T.~S. Ray, ``An approach to the
synthesis of life,'' in \cite[p.\ 371--408]{langton}.
        \bibitem{ray2} T.~S. Ray, ``Evolution of parallel
processes in organic and digital media,'' in \cite[p.\ 69--91]{waltz}.
        \bibitem{rothman-book} D. Rothman and S. Zaleski, 
{\em Lattice-Gas Cellular Automata---simple models of
complex hydrodynamics}, (Cambridge University Press, 1997).
        \bibitem{shor} P.~W. Shor, ``Algorithms for quantum
computation: discrete log and factoring,'' {\em Proceedings of the
$35^{\rm th}$ Annual Symposium on the Foundations of Computer
Science}, 124--134, (IEEE, 1994) and {\tt quant-ph/9508027}; reproduced in \cite{lloyd}.
        \bibitem{simons} N.~R.~S. Simons, M. Cuhaci,
N. Adnani and G.~E. Bridges, ``On the potential use of
cellular-automata machines for electromagnetic-field solution,'' {\em
Int.\ J. Numer.\ Model.\ Electron.\ N. \bf 8:}3/4, 301--312 (1995).
        \bibitem{sims} K. Sims, ``Evolving 3D morphology and
behavior by competition,'' in \cite[p.\ 28--39]{brooks}.
        \bibitem{smith-topol} M.~A. Smith, ``Representations of
geometrical and topological quantities in cellular automata,'' {\em
Physica D \bf 47}, 271--277 (1990).
        \bibitem{smith-thesis} M.~A. Smith,  ``Cellular
automata methods in mathematical physics,'' MIT Ph.D. Thesis (1994).
Reprinted as {\em Tech.\ Rep.\ MIT/LCS/TR-615}, MIT Lab.\ for Computer
Science (1994).
        \bibitem{takesue} S. Takesue, ``Boltzmann-type
equations for elementary reversible cellular automata,'' in
\cite[190-200]{paris}. 
        \bibitem{toffoli-universal} T. Toffoli, ``Computation and
construction universality of reversible cellular automata,'' {\em J.\
Comp.\ Syst.\ Sci.\ \bf 15}, 213--231 (1977).
        \bibitem{toffoli-pnc} T. Toffoli, ``Physics and
computation,'' in \cite[p.\ 165--175]{endicott}.
	\bibitem{cam5} T.  Toffoli, ``CAM: A
high-performance cellular-automaton machine,'' in \cite[p.\
195--204]{los-alamos}. 
	\bibitem{toffoli-pde} T. Toffoli, ``Cellular automata as
an alternative to (rather than an approximation of) differential
equations in modeling physics,'' in \cite[p.\ 117--127]{los-alamos}.
        \bibitem{cambook} T. Toffoli and N. Margolus,
{\em Cellular Automata Machines---a new environment for modeling}
(MIT Press, 1987).
        \bibitem{toffoli-four} T. Toffoli, ``Four topics in
lattice gases: ergodicity; relativity; information flow; and rule
compression for parallel lattice-gas machines,'' in \cite[p.\
343--354]{torino}.
        \bibitem{toffoli-cheap} T. Toffoli, ``How cheap can
mechanics' first principles be?''  in \cite[p.\ 301--318]{zurek}.
        \bibitem{ica} T. Toffoli and N. Margolus,
``Invertible cellular automata: a review,'' in \cite[p.\
229--253]{gutowitz}. 
        \bibitem{toffoli-nature} T. Toffoli, ``What are
nature's `natural' ways of computing?'' in \cite[p.\ 5--9]{physcomp92}.
        \bibitem{physcomp96} T. Toffoli, M. Biafore and
J. Le\~ao (eds.), {\em PhysComp96} (New England Complex Systems
Institute, 1996).  Also online at {\tt http://www.interjournal.org}.
        \bibitem{physcomp96b} T. Toffoli and M. Biafore (eds.), 
{\em PhysComp96}, special issue of {\em Physica D\ \bf 120}:1/2
(1998). 
        \bibitem{ulam} S. Ulam, ``Random processes and
transformations,'' {\em Proceedings of the International Congress on
Mathematics 1950, volume 2}, 264--275 (1952).
        \bibitem{vichniac} G. Vichniac, ``Simulating physics
with cellular automata,'' in \cite[p.\ 96--116]{los-alamos}.
        \bibitem{waltz} P.~D. Waltz (ed.), {\em Natural and
Artificial Parallel Computation} (SIAM Press, 1996).
        \bibitem{vieri-thesis} C. Vieri, ``Reversible computer
engineering,'' MIT Ph.D. Thesis (1998).
        \bibitem{wolfram-book} S. Wolfram, {\em Cellular
Automata and Complexity} (Addison-Wesley, 1994).
        \bibitem{yakhot} V. Yakhot and S. Orszag,
``Reynolds number scaling of cellular-automaton hydrodynamics, {\em
Phys.\ Rev.\ Lett. \bf 56}, 1691--1693 (1986).
	\bibitem{yepez-crystal} J. Yepez, ``Lattice-gas
crystallization,'' in \cite[p.\ 255--294]{princeton}.
	\bibitem{yepez-blobs} J. Yepez, ``A lattice-gas with
long-range interactions coupled to a heat bath,'' in \cite[p.\
261--274]{waterloo}. 
	\bibitem{yepez-thesis} J. Yepez, ``Lattice-gas
dynamics,'' Brandeis Physics Ph.D. Thesis (1997).  Available as {\em
Air Force Research Laboratory Tech.\ Rep.\ PL-TR-96-2122(I),
PL-TR-96-2122(II) and PL-TR-96-2122(III)}, AFRL/VSBE Hanscom AFB, MA
01731 (1996).
        \bibitem{younis-knight} S.~G. Younis and T.~F. Knight, Jr., 
``Practical implementation of charge recovering
asymptotically zero power {CMOS},'' in {\em Proceedings of the 1993
Symposium on Integrated Systems}, 234--250 (MIT Press, 1993).
        \bibitem{younis-thesis} S.~G. Younis, ``Asymptotically
zero energy computing using split-level charge recovery logic,'' MIT
Ph.D. Thesis (1994).
        \bibitem{zurek} W. Zurek (ed.), {\em Complexity,
Entropy, and the Physics of Information} (Addison-Wesley, 1990).

\end{thebibliography}
\end{document}